\documentclass[aps,prb,preprint,onecolumn,amsmath,amssymb,superscriptaddress,eqsecnum,floatfix,scrartcl]{revtex4-1}
\pdfoutput=1      
\usepackage{amssymb}  
\usepackage{color,soul} 
\usepackage{graphicx}  
\usepackage[normalem]{ulem}
\usepackage[tight]{subfigure}  
\usepackage{mathrsfs}
\usepackage[pdftex,pdfusetitle,bookmarks=true,colorlinks=true,citecolor=blue,urlcolor=blue,linkcolor=magenta]{hyperref}
\usepackage{hypcap}

\definecolor{OliveGreen}{cmyk}{0.64, 0, 0.95, 0.40}

\begin{document}
%\title{Level repulsion in the chaotic wavefunction}    
\title{Universal Spectral Correlations  in the Chaotic Wave Function,
and the  
%Onset
Development of Quantum Chaos}
\author{Xiao Chen}   
\email{xchen@kitp.ucsb.edu}
\affiliation{Kavli Institute for Theoretical Physics, University of California at Santa Barbara, CA 93106, USA}

\author{Andreas W.W. Ludwig}
%\email{ludwig@physics.ucsb.edu}
\affiliation{Department of Physics, University of California at Santa Barbara, CA 93106, USA}

\date{\today}

\begin{abstract}   
  We investigate the appearance of quantum chaos in a {\it single}  many-body wave function by analyzing
the statistical properties of
% level repulsion statistics between 
the  eigenvalues of 
%the 
its reduced density matrix ${\hat  \rho}_A$ of  a spatial  subsystem   $A$.  
We find  that  (i):  the spectrum of the density matrix is described by so-called {\it Wishart random matrix theory},  which
%and
%(ii): that it 
(ii): exhibits besides {\it level repulsion}, {\it spectral rigidity} and  {\it universal spectral correlations} between eigenvalues separated 
by distances
ranging from one  up to  many mean level spacings,   which we investigate.
%exhibits {\it universal spectral correlations} 
%%``{\it spectral rigidity}'',
%%, a rigid structure of its eigenvalues, 
%%a universal property reflecting repulsion 
%between eigenvalues separated by distances
%% ranging 
%which range from one  up to  many mean level spacings,
%and  (ii): that  these  universal features are  described by 
%%the 
%so-called {\it Wishart random matrix theory}.
%We propose that such level repulsion statistics (and concomitant spectral rigidity)
%%  spectral rigidity 
%is a universal property of the spectrum of the density matrix on  scales not too far from   the level spacing, 
%%at low-energy differences,  and  is 
%%can be 
%and that it is described by that of 
%the so-called Wishart random matrix ensemble. 
We use these universal spectral characteristics of the reduced density matrix
as a definition of chaos in the wave function.
% this spectral statistics 
A simple and  precise characterization
%A hallmark 
of  such universal
correlations
%spectral rigidity of 
in a spectrum
%level repulsion in a spectrum 
is  a segment of  strictly  linear growth at sufficiently  long times, 
%sometimes 
recently 
%also 
called the ``{\it ramp}'',
%\cite{CotlerEtAl1611.04650}
% referred to as 
of the {\it spectral form factor} which is the Fourier transform of the
% level-level
{\it  correlation function between a pair of eigenvalues}. It turns out that Wishart  and standard random matrix theory  have the same
universal  ``ramp''.
Specifically, here numerical results for the spectral form factor
of the density matrix of
generic non-integrable many-body systems, such as 
 one-dimensional quantum Ising 
%as well as
and   Floquet spin models,  are  found to exhibit an universal  ``ramp'' identical to that appearing for a {\it ``random pure state''} (``{\it Page state}'', or ``Haar state''). The density matrix of the latter
%whose density matrix
 is precisely the Wishart random matrix, the reduced density matrix of a
%an ``ideal'' quantum chaotic 
completely random wave function.
% in ``random pure state'', the typical state for both non-integrable Floquet spin model and static Ising model. In all these examples, we observe a linear ra%mp which characterize the level repulsion statistics. 
In addition,  we study the {\it  
%onset 
development of chaos}  in the wave function by  letting an initial direct product state  evolve under the unitary  time evolution. We find that
% spectral rigidity 
%and  level repulsion 
 the universal spectral correlations as manifested by the ``ramp'' set in  
%only after 
as soon as the entanglement entropy
% has already  saturated,
begins to 
grow, 
%increase, 
and first develop for the eigenvalues
% in the middle
at the top
 of the spectrum of the density matrix ${\hat \rho}_A$, subsequently spreading
%out  to 
over the entire  spectrum at  later times. 
%In addition,  
Finally, we study
a {\it prethermalized regime}  described by a {\it generalized Gibbs ensemble}, which develops in 
 a rapidly driven Floquet model at intermediate times.
% with a stable 
%prethermalization 
%prethermalized regime 
%described by a
%approximated by 
%generalized Gibbs ensemble. 
We find that the prethermalized regime exhibits no chaos, as evidenced by the absence of a ``ramp'' in
the spectral form factor of the density matrix, 
% there is no chaos in this regime
while 
%and 
% spectral rigidity  starts
the universal spectral correlations start  to develop when the
% prethermal
prethermalized regime finally
% state 
relaxes at late times to the 
%full thermalization
fully thermalized (infinite temperature)  chaotic regime.
% of the Floquet system.

\end{abstract}

\maketitle   
\tableofcontents

\section{Introduction}
\label{SectionIntroduction}
The characterization of chaos in quantum mechanical systems has a long history,
%\footnote{See e.g. Ref.s \onlinecite{HaakeBook},
%\onlinecite{StoeckmannBook}, or
%\onlinecite{GuhrMuellerGroelingWeidenmuellerPhysRpts1998} for a review.}
and chaos plays a key role in  the process of  thermalization, i.e. relaxation to equilibrium in generic  isolated many-body quantum 
systems\cite{Srednicki1994, Deutsch1991}. 
(See e.g. Ref.s \onlinecite{HaakeBook},
\onlinecite{StoeckmannBook}, 
\onlinecite{GuhrMuellerGroelingWeidenmuellerPhysRpts1998}, \onlinecite{DAlessio2014} for a review.)
It also plays an important role for the quantum nature of black holes.\cite{Shenker2013b,Shenker2013a,Shenker2014, Maldacena2015, Kitaev2014,Cotler2016}
% hole  dynamics.\cite{Shenker2013b,Shenker2013a,Shenker2014, Maldacena2015, Kitaev2014,Cotler2016}
An important milestone in the study of quantum chaos has been the so-called 
{\it Bohigas, Giannoni and Schmidt conjecture}\cite{BohigasGiannoniSchmidtConjecture}, which states that chaos manifests itself in the spectral properties of the Hamiltonian of a quantum system by exhibiting
% whose 
universal  features which are the same as those of the spectrum of a random Hamiltonian matrix  in the same symmetry class.
 Such {\it universal}
features include, besides
%are the
 level repulsion statistics between {\it adjacent}  spectral levels, spectral rigidity and more generally
the correlation function between two levels which is universal for  levels separated by
% as well as  so-called ``{\it spectral rigidity}'' which quantifies 
%{\it repulsion} betweenlevels
%which
%that  are  {\it separated by energy scales
% ranging 
energy scales 
%which
that  range from the mean level spacing 
%all the way
%up  
to energy differences which can be
much 
%orders of magnitude 
larger,  up to scales
at which model-dependent (``ultraviolet'') features set in. 
% level repulsion characteristics which are universal (independent of details of the systems), 
%on energy  scales comparable to the mean level spacing, and which coincide on those scales with those
%of a corresponding random matrix theory. 
The corresponding  universality classes are solely determined by the action of the anti-unitary time-reversal operator, giving rise to the  three
possible symmetry classes of spectral statistics
depending on whether time reversal symmetry is absent (``GUE'', Dyson index $\beta=2$), or is present and squares to
the identiy operator (``GOE'', Dyson index $\beta=1$) or squares to minus the identity operator (``GSE'', Dyson index $\beta=4$).

\begin{figure}[hbt]
\centering
\includegraphics[width=.6\textwidth]{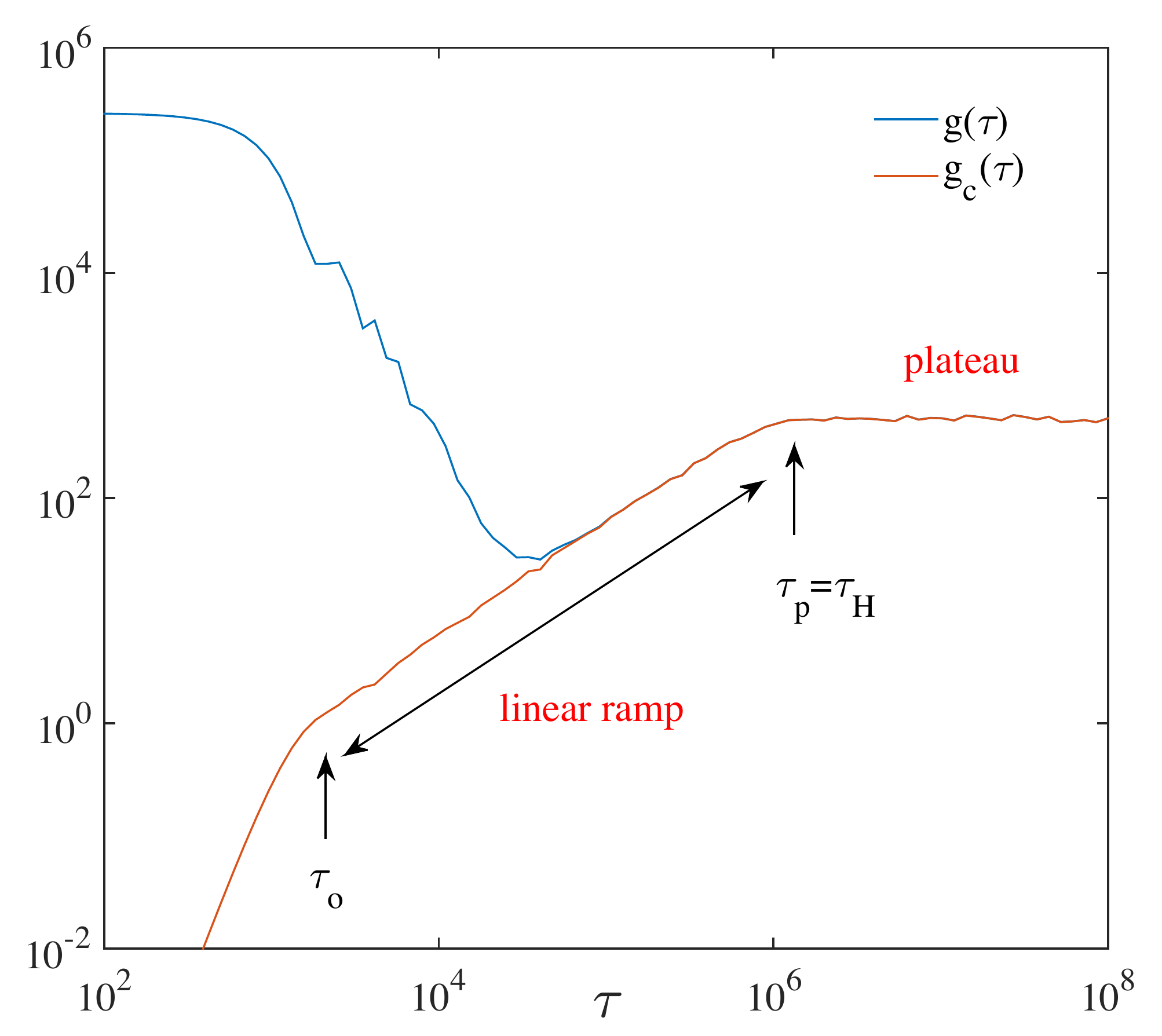}
\caption{Typical
%Generic 
structure\cite{Note2} of the linear universal  ``ramp''  in the spectral form factor $g(\tau)$ as well as of the connected spectral form factor $g_c(\tau)$, which exhibits a {\it longer} ``ramp'' ranging from a microscopic short  time scale $\tau_0$ below which non-universal effects set in, up to
the Heisenberg time $\tau_H$ (also called plateau time $\tau_p$).
}
\label{fig:fig-sketch-ramp}
\end{figure}

Spectral  characteristics of a discrete spectrum of levels $E_i$ are  conveniently
% reflected in
described by the so-called {\it spectral form factor}
\footnote{See e.g. Ref.s \onlinecite{GuhrMuellerGroelingWeidenmuellerPhysRpts1998},
\onlinecite{Cotler2016}}, which is  the {\it Fourier transform} of the
 {\it correlation function} between two levels, and  can be written in the form
\begin{eqnarray}
\label{IntroSpectralFormfactor}
g(\tau)\equiv  \langle\sum_{i,j} e^{-i \tau (E_i-E_j)}\rangle, \qquad  \qquad {\rm ({\it Spectral \ Form \ Factor})}.
\end{eqnarray}
Here $\tau$ denotes  an {\it auxiliary}  real time ({\it not} to be confused with  an `Euclidean' or `imaginary' time coordinate, often denoted by the same symbol), and $\langle ... \rangle$ 
% is
stands for  a certain  average,
to be described in detail below, whose sole  purpose is to remove non-universal rapid temporal fluctuations  (in $\tau$) 
from the signal which originate from  (non-universal) 
high frequency components corresponding to large energy differences $(E_i-E_j)$.
In random matrix theory, considering here  the  simplest case where time-reversal symmetry
is absent  (``GUE-type'' statistics),
% the {\it hallmark of spectral rigidity}
a simple and precise characterization of 
%spectral rigidity
universal spectral correlations  is a {\it segment of strictly  linear  
growth}\footnote{For the other two  symmetry classes the detailed shape  of this growth segment has similar features, but the details are slightly different} 
 in time $\tau$, recently called\cite{Cotler2016}
 the  ``{\it ramp}'',  of  the spectral form factor $g(\tau)$ at  sufficiently long   times
 up to the
so-called Heisenberg time
$\tau_H$ (defined to be $2\pi$ times the inverse of the mean level spacing), 
% ``plateau time'' $\tau_p$,
where it  suddenly becomes completely flat, reaching its long-time ``plateau'' value, as 
sketched\footnote{The plotted data are actually for
the reduced density matrix of
a typical one-dimensional Floquet many-body system to be discussed in Section \ref{SubSectionFloquetModel} and in   Fig. \ref{fig:Floq_Ising}
of the main text, but this is not relevant for the current discussion.}
 in Fig. \ref{fig:fig-sketch-ramp}.
[The Heisenberg time has  also been  called ``plateau time'' $\tau_p\equiv \tau_H$.]
More precisely, the {\it connected} spectral form factor $g_c(\tau)$ 
obtained\footnote{See detailed  discussions below  
(\ref{DEFFourierTransformEigenvalueDensity}), (\ref{DEFZZstar}), (\ref{DEFConnectedSpectralFormFactor}).}
 from (\ref{IntroSpectralFormfactor})
by subtracting a (non-universal) disconnected piece
$|\langle\sum_{i} e^{-i \tau E_i}\rangle|^2$,  turns out to exhibit a longer
segment of universal, strictly linear growth (``ramp'') for time scales $\tau$
larger than a shortest  time scale $\tau_0$ below which (in applications, e.g.  to spectra of Hamiltonians describing quantum chaos) possible
 non-universal features set in.
I.e., the region $\tau \lesssim \tau_0$ corresponds to differences of energies $ (E_i-E_j)$ which exceed the universal regime.
- In the (non-connected) spectral form factor
$g(\tau)$ from (\ref{IntroSpectralFormfactor}),  a portion of this universal segment of linear growth  in $g_c(\tau)$
turns out to be
hidden at small times larger than $\tau_0$ by  possible  non-universal features of the
disconnected part, and $g(\tau)$ typically only exhibits a shorter  part of the entire  universal linear ``ramp'',
as depicted  in Fig. \ref{fig:fig-sketch-ramp}.

%\cite{Note2}

%  (i.e. for  $\tau < \tau_H
%\equiv 2\pi \hbar{\bar  \rho}$, where 
%${\bar \rho}$ is the mean  level density
%and $1/{\bar\rho}$ the mean
%level spacing - here we set $\hbar=1$).
It was shown\cite{BerryProcRoySocLond1985,HannayDeAlmeidaJPhysA17-1984} 
many years ago that in chaotic  quantum systems with a small number of degrees of freedom whose classical
limit is ergodic,  the ``ramp'' for the energy spectrum of the Hamiltonian can be computed analytically in the semiclassical limit by making use of Gutzwiller's
Trace Formula\cite{GutzwillerBook}
 and  known properties of asymptotically  long classical  periodic orbits. In these cases, the time-scale $\tau_0$ characterizes
the onset of potential  non-universal contributions 
to $g_c(\tau)$ for $\tau \lesssim \tau_0$  arising from short 
%periodic 
orbits.
%{\color{blue} ``{\it COMMENT  by Andreas: Should we mention this semiclassical result?}'': XC: Yes. We can mention it here.}
%%, when the classical limit is ergodic.
On the other hand,  the spectral form factor has very recently  formed a topic
of extensive discussion  in the context 
%of the Hamiltonian  
 of the Sachdev-Ye-Kitaev (SYK)  model \cite{Sachdev1993,Kitaev2015,Maldacena2016}, a strongly chaotic quantum system, 
%which 
whose Hamiltonian   has been shown numerically  to exhibit a spectrum  possessing
 the expected ``ramp''. A recent  lucid  discussion of many aspects of the spectral form factor, with an emphasis on
 the Hamiltonian spectrum of the  SYK model, can be found
in Ref.\ \onlinecite{Cotler2016}.
-  In contrast, these universal  spectral correlations are   absent in an integrable system, where  the spectral form factor
exhibits no ``ramp'' (and the probability distribution for the spacing between adjacent levels is Poissonian).

In the present paper, we are going to show that
the universal spectral correlations manifested by a  strictly linear ``ramp''
% spectral rigidity 
% level repulsion statistics 
%can 
already appear  at the level of a single many-body wavefunction of a generic chaotic quantum system, without focusing
attention on the spectrum of the Hamiltonian of the system; 
we also discuss periodically driven Floquet systems.
%Our idea makes use of

For thermalizing (chaotic) systems whose time-evolution is governed by a time-independent Hamiltonian (i.e. not Floquet systems),
 our work can be motivated by
%Our work
% makes use of
%builds on 
 the connection
 between a typical state and the thermal ensemble, a notion  inherent  in the eigenstate thermalization hypothesis\cite{Srednicki1994,Deutsch1991}
 (ETH),
which we now  briefly summarize as follows: 
Let $|\psi\rangle$ be a state at finite energy density $e=E/V$ (i.e. 
$\langle \psi| {\hat H} | \psi \rangle=$  $E=$ $ e \ V$,
where $V$ is the volume), which can
either be  a highly excited  exact eigenstate of a chaotic Hamiltonian ${\hat H}$ in the spatial volume $V$, or just 
%an arbitrary
a typical  short-range entangled  initial state  (which is not an eigenstate of ${\hat H}$) 
acted on by the corresponding unitary quantum mechanical  time-evolution operator for a sufficiently long time. ETH states that the expectation value of a product of local operators in the state $|\psi\rangle$ equals the thermal expectation value of this product 
%of local operators
at a temperature determined by $e$ in the usual sense of microcanonical Statistical Mechanics.
For Floquet systems, these expectations
values in the analogous state $|\psi\rangle$ are at infinite temperature.

Here we  consider  the reduced density matrix in a spatial subregion $A$  ($B$ $={\bar A}$ is the complement of $A$) of such a typical  state,
\begin{eqnarray}
\label{DEFrhoA}
{\hat  \rho}_A = \mbox{Tr}_B \  |\psi\rangle  \langle \psi|.
\end{eqnarray}
We will show that the spectral form factor for the spectrum of  eigenvalues $\lambda_i$ of  the reduced density matrix  ${\hat  \rho}_A$, 
\begin{eqnarray}
\label{DEFSpectralFormFactor}
g(\tau)\equiv  \langle\sum_{i,j} e^{-i \tau (\lambda_i-\lambda_j)}\rangle, 
\end{eqnarray}
exhibits a ``ramp''.  As mentioned above, the presence of a ``ramp''  in the spectral form factor  demonstrates the presence of 
% level repulsion
universal
spectral 
correlations
%rigidity 
over a possibly large range of scales (determined by $\tau_0$ and $\tau_H$)  in the spectrum of eigenvalues of the density matrix. 
Thus, in  this paper we use
% the level
%repulsion between the eigenvalues 
the presence of these universal spectral correlations
%spectral rigidity of
in  the  spectrum of eigenvalues
of the reduced density matrix of a typical quantum state $|\psi\rangle$, as manifested by the presence of a ``ramp'' in the associated spectral
form factor, to define the notion of quantum chaos in the state (i.e. ``in the wavefunction'').
In particular, we will show at the technical level that spectral properties of the reduced density matrix ${\hat \rho}_A$ are described by
so-called {\it Wishart random matrix theory}\cite{loggas}. As it turns out, Wishart random matrix theory exhibits universal spectral correlations
idential to those appearing in
 standard (here\footnote{since we consider systems without time-reversal symmetry}  GUE) random matrix theory; in particular they have the same universal linear ``ramp'' (see Sect. \ref{SubSectionSpectralFormFactorReducedDensityMatrix}
and Appendix \ref{AppendixSectionDetailsWishartRandomMatrixTheory}).

For systems whose time-evolution is governed by a time-independent Hamiltonian,
we can look at this also from a  slightly different angle:
Instead of investigating the spectral statistics of the reduced density matrix, one may also  be inclined  to consider the spectral statistics of the
associated  entanglement Hamiltonian
${\hat H}_E$ defined  by
\begin{eqnarray}
\label{DEFEntanglementHamiltonian}
{\hat \rho}_A = {\cal N}_E^{-1} \exp\{- \beta_{eff} {\hat H}_E\}.
\end{eqnarray}
The spectral form factor for the entanglement Hamiltonian
is obtained
 from (\ref{DEFSpectralFormFactor}) by letting $\lambda_i \to$ $- \ln \lambda_i$.
% It  turns out, as
As discussed
% shown
 %briefly discussion 
in Appendix \ref{SectionAppendixSpectralFormfactorEntanglementHamiltonian},
%that
 the two spectral form factors,  of ${\hat \rho_A}$ and of ${\hat H}_E$, exhibit identical 
universal features in their respective   level statistics. In particular,  in a chaotic system
they 
%exhibit
 both exhibit  a linear ``ramp''.
Now, one may think
of the entanglement Hamiltonian 
% as follows:
and of (\ref{DEFEntanglementHamiltonian}) 
%may be viewed 
in the light of a strong version of ETH proposed in 
Ref.\ \onlinecite{Garrison2015}, which states that the reduced density matrix of the single  state $|\psi\rangle$ takes on a thermal form, ${\hat \rho}_A =$
${\cal N}_A^{-1}  \exp\{ - \beta {\hat H}_A \}$,
 where ${\hat H}_A$ is  the physical (chaotic)  Hamiltonian of the system, projected onto the region of subsystem $A$.
%(even if the spatial size of the subsystem is up to half that of the entire system).
%, which is presumably also chaotic.
Note that this strong version of ETH\cite{Garrison2015} is  a quite non-trivial statement  because even though ${\hat  \rho}_A$ is constructed from a
{\it  single}
 state [see (\ref{DEFrhoA})], 
this statement implies that ${\hat  \rho}_A$ contains the
knowledge of the {\it entire}  Hamiltonian of the system, or rather  at least of its projection onto $A$. -  Then, 
%given the validity of 
if one assumes the validity of the above-mentioned
strong version of ETH, one would naturally expect that 
%spectral rigidity 
the universal correlations in the spectrum of the  entanglement Hamiltonian ${\hat H}_E$ 
are  directly inherited from those  of the physical Hamiltonian ${\hat H}_A$ (which, 
according 
%owing 
to the 
Bohigas, Giannoni and Schmidt conjecture, is expected to  exhibit universal spectral correlations).
% to be   chaotic). 
Thus,
% if one assumes  the validity of this strong version of ETH, 
since we observe (as mentioned) that 
${\hat \rho}_A$ and ${\hat H}_E$ exhibit the same universal  features in their spectral form factors, 
it would be natural to expect the appearance
%spectral rigidity
%level repulsion 
the universal spectral correlations (and the ``ramp'') 
in the entanglement Hamiltonian of a single state $|\psi\rangle$. Put another way, for thermalizing (chaotic)  systems whose time-evolution is
governed by a time-independent Hamiltonian,  our results can thus also be viewed
as a confirmation of  the strong version of ETH proposed in Ref. \onlinecite{Garrison2015}.

In order to investigate explicitly
% the
% level repulsion
the presence of the mentioned
universal 
%spectral 
correlations
% spectral rigidity 
 in the spectrum of the reduced density matrix (\ref{DEFrhoA}) in many-body quantum chaos,
we numerically compute   the spectral form factor of the density matrix of a  typical 
{\it single} many-body  wave function $|\psi\rangle$  [as defined in the paragraph above (\ref{DEFrhoA})],  in two
generic  non-integrable  one-dimensional systems: a Floquet spin model, and a quantum Ising model in both transverse and logitudinal field. As will
be shown below
in the bulk of the paper, we clearly observe for  both
systems  a linear ``ramp'' in the spectral form factors of their density matrices,
%of these systems, 
confirming the 
%concomittant
corresponding universal spectral correlations
%spectral rigidity of
in  their  spectra of eigenvalues.
%level repulsion statistics. 
Furthermore, in
%In 
order to provide a generic, model-independent description of  the universal features of quantum chaos in a wave function, we consider a so-called  ``random pure state'',
or ``Page state'' (``Haar state'')\cite{Page1993}, 
\begin{align}
\label{DEFPageState}
|\Psi(\{\alpha_i\})\rangle=\sum_i\alpha_i \ |C_i\rangle,
\end{align}
in which the coefficients $\alpha_i$ of the state in a fixed 
 basis $\{ |C_i\rangle\}_i$ are random complex numbers subject solely to the normalization
constraint, with a probability distribution invariant under unitary basis changes. The set of coefficients  $\{\alpha_i\}_i$ can thus be considered a row (or column) vector of a 
%Gaussian 
unitary random matrix (distributed according to the Haar measure).
%\footnote{The Gaussian nature of the probability distribution is unimportant
%and can be modified without affecting the universal results discussed below.}. 
When we now form the 
reduced density matrix ${\hat  \rho}_A$ of the 
``random pure state''
%``Page state''
 (\ref{DEFPageState})
in a spatial region $A$, we obtain a random matrix which turns out to  belong to the well-studied ``Wishart random matrix ensemble''
(see Sect. \ref{SubSectionRandomPureStateWishart} below for a more detailed discussion).
The probability distribution for the eigenvalues of the Wishart random matrix and hence of the density matrix ${\hat  \rho}_A$
of the ``random pure state''  are  known analytically (as reviewed in Sect. \ref{SubSectionWishartRandomMatrixEnsemble}), 
  and the spectral form factor for ${\hat \rho}_A$ can be shown analytically (see 
Subsection \ref{SubSectionSpectralFormFactorReducedDensityMatrix} and Appendix \ref{AppendixSectionDetailsWishartRandomMatrixTheory})
 to exhibit a linear ``ramp'' in the limit of
large density matrices,  reflecting the presence of 
universal spectral correlations
%universal level repulsion statistics and
 %spectral rigidity 
in their spectra.
As already mentioned, the linear ``ramp'' in $g_c(\tau)$ for the eigenvalues of the Wishart random matrix turns out to be 
 identical to that of  standard
random matrix theory in the same symmetry class
(see Sect. \ref{SubSectionSpectralFormFactorReducedDensityMatrix} and Appendix \ref{AppendixSectionDetailsWishartRandomMatrixTheory}).
%, as we will see in
The spectral form factor for the ``random pure state'' is discussed in detail  in Sect. \ref{SubSectionRandomProductState} below.
% it exhibits 
%a ``ramp'' reflecting the presence of 
%level repulsion statistics
%spectral rigidity in its spectrum.
%of the density matrix.
%  (Sect. \ref{SubSectionRandomProductState} below). 
In Sect. \ref{SubSectionFloquetAndQuantumIsingModels}
%{SubSectionFloquetAndIsingModels}
we compare the numerically obtained spectral form factors
of the Floquet and the quantum Ising systems with that of the ``random pure state'' (for the same system sizes) and find full agreement
of the universal features. This means
that Wishart random matrix theory describes the  spectral 
correlations
% properties and the corresponding spectral rigidity 
 of the  reduced density matrix of a single many-body wave function
in typical chaotic systems of Hamiltonian and Floquet type in their universal regime, just as  ordinary random matrix theory is thought to describe
% describes 
the level statistics of a chaotic Hamiltonian (according to
the Bohigas, 
Giannoni and Schmidt conjecture\cite{BohigasGiannoniSchmidtConjecture}.) 
The  Floquet and quantum Ising systems considered  in this paper
%, they 
lack time-reversal symmetry and so the ``GUE" Wishart random matrix ensemble  will be
%expected to be 
appropriate.

Subsequently, we explore the important question of 
%onset
development of quantum chaos
 under quantum mechanical unitary time evolution.
 Recently, it has been  proposed that the ``out-of-time-ordered''  correlation function (OTOC) can probe the
% onset
development of chaotic dynamics
 and  scrambling of quantum information.\cite{Larkin1969,Shenker2013b,Shenker2013a,Shenker2014, Maldacena2015, Kitaev2014}
 At early times, 
%this quantity
the OTOC can exhibit 
an exponentially growing regime,   the growth rate of which  represents
 a quantum analog of the (classical)  Lyapunov exponent. In this paper, we will study instead as an indicator of the 
%onset
development  of quantum chaos
the  emergence  of
universal spectral correlations
and of the  corresponding  ``ramp'' in  the spectral form factor, 
% spectral rigidity
% level repulsion statistics 
in the 
%time-evolution of the 
spectrum of the  reduced  density matrix  ${\hat \rho}_A(t)$ as  a function of
% the
 time $t$
% increases 
 in a quantum quench problem. ETH states that  an (sufficiently general short-range entangled)  initial state 
%far from equilibrium
which is not an eigenstate,
relaxes under the quantum mechanical time evolution, after  a sufficiently long time, 
to a state which appears to be (in the sense of  ETH, as reviewed above) in 
thermal equilibrium 
(at infinite temperature for Floquet systems).\cite{Srednicki1994,Deutsch1991}
 Therefore,
 if we start with an  initial direct  product state, one expects that while
the  spectrum of the reduced density matrix ${\hat  \rho}_A(t)$ will initially exhibit no
%universal 
spectral correlations,
%spectral rigidity,
% level repulsion, 
%there is no level repulsion in ${\hat \rho}_A$ and 
under the unitary time evolution  
%level repulsion
a ``ramp''
%spectral rigidity
  will  
%appear 
emerge after a sufficiently long time $t$ in its spectral form factor.
We consider both (one-dimensional)  Floquet and quantum  Ising models and find that 
% level repulsion
a  ``ramp''
%spectral rigidity 
 starts to develop 
as soon as the entanglement entropy begins to grow.
%{\color{blue} ``{\it COMMENT by Andreas: We need to write this last sentence somehow differently, since as it
%stands it contradicts FIG. 5 and the corresponding discussion in *Sect. III. A* where we say at the beginning of the 2nd  paragraph
%"the ``ramp'' starts to emerge at time-step $n=11$... - Of course, as we say even in the introduction, spectral correlations at the top of the %spectrum as soon as the entanglement entropy starts to grow.}} 
% only after the 
%EE 
%entanglement entropy  has already reached its saturation value, where it exhibits   a volume law. 
%The repulsive log interaction first appears in the middle of the spectrum of ${\hat \rho}_A$ and then spreads out to the whole spectrum at a later time.
More precisely, it is interesting to note  that 
%Interestingly, 
%level repulsion
%spectral rigidity
universal spectral correlations are first   seen to appear 
%(i.e. as soon as the entanglement entropy begins to grow)
for the eigenvalues
% in the middle 
at the top
of the spectrum of the reduced density matrix ${\hat  \rho}_A$(t), and subsequently  spread
out over the entire spectrum at later times.

We  emphasize that there is no direct connection between the  appearance of a volume law in the entanglement entropy and  quantum chaos.
 In 
%many
 integrable systems,  initial direct  product states (as above)  are typically expected to
thermalize to a generalized Gibbs ensemble  (GGE)  after sufficiently  long unitary time evolution\cite{Rigol2007,Rigol2006,Calabrese2007, Cardy2015}.
%, they can thermalize to generalized Gibbs ensemble (GGE).\cite{Rigol2007,Rigol2006,Calabrese2007} 
Although the reduced density matrix ${\hat  \rho}_A$  for these  GGE states possesses  an  entanglement entropy exhibiting a volume law,
% can have volume law EE, 
the eigenvalues of this density matrix 
%exhibit no
are not expected to exhibit
the discussed universal spectral correlations,
%spectral rigidity,
in contrast to the reduced density matrix of chaotic (thermalizing) systems  discussed above.
% level repulsion.
%there is no level repulsion between the eigenvalues of ${\hat \rho}_A$. 
%To better illustrate this point, we further construct
To illustrate this point explicitly, we have constructed
 a rapidly  driven Floquet system whose time evolution, starting out from a direct product state, 
exhibits  a long, stable so-called  prethermalized regime\cite{Mori2016,Abanin2015,Abanin_2017,Else_2017} at intermediate times.
% with a stable prethermalization regime.\cite{Mori2016,Abanin2015,Abanin_2017,Else_2017} 
%This regime has long lifetime and can be well approximated by GGE.
This prethermalized regime, accompanied by a long plateau 
%(in time) of the 
in the time-dependence of the entanglement entropy  exhibiting a volume law,
will   be seen to be clearly  devoid of chaos as evidenced from the absence of 
the characterisic universal spectral correlations
%of a ``ramp''
% in the  spectral form factor
%level repulsion is
%spectral rigidity 
in the spectrum  of the density matrix ${\hat  \rho}_A(t)$,
which is manifested by the absence of a ``ramp'' in the corresponding spectral form factor.
%By computing spectral form factor in this regime, we can clearly see that there is no chaos in it. 
A  linear ``ramp'' in the spectral form factor is seen to develop
%appears
 only when the prethermalized regime eventually relaxes at very  late times to the fully thermalized  chaotic regime, in which the conservation laws
(approximately) present in the prethermalized regime cease to exist.
%are no longer present.
% the system move forward from prethermalization regime and relaxes to the full thermalization. 

We end the introduction by mentioning some  related work. Level repulsion statistics between adjacent levels of the density matrix 
of thermalizing systems with a 
main focus
on disordered systems has been  discussed  in the context of work investigating  Many-Body-Localization (MBL) in Ref.\ \onlinecite{ChamonPRB96-2017-020408}
and \onlinecite{GeraedtsNandkishoreRegnaultPRB93-2016-174202}.
Our work, in contrast, discusses spectral rigidity and, in particular,  the universal spectral correlations
and the ``ramp'' in the spectral form factor, focusing on non-random chaotic systems, and
it elucidates the origins of
% identifies
these spectral correlations 
%as arising from 
%of  these spectral correlations  of the density matrix
%as being  in 
 in the ``random pure state'' (Page state)  and  in  Wishart random matrix theory, both for Hamiltonian and Floquet systems. Furthermore,
we identify the 
%{\it onset}
{\it development  of chaos}
as the process of buildup
% building up  
of these spectral correlations in the density matrix under the unitary time-evolution.
We also discuss a prethermal regime, lacking chaos, and its late-time relaxation to a chaotic state.

The rest of the paper is organized as follows. In Sect. II, we first discuss the spectral form factor in the  ``random pure state'' (Page state)
 and
%exhibit 
its
% show that it has
 linear ``ramp''. Then, we discuss the spectral form factors of typical
 wavefunctions of  non-integrable  Floquet and quantum  Ising models
for the same system sizes,  and show that they both exhibit  the same universal  linear ``ramp''
as the ``random pure state''.
% have similar behaviors.
 In Sect. III, we discuss the 
%onset
development of chaos
 in these Floquet and Ising model wavefunctions by computing the time evolution of the spectral form factor.
% at different times. 
Moreover, 
%In particular, 
we explore the 
%onset
development  of chaos
 in  a Floquet model 
%exhibiting
which exhibits  a long prethermal regime at intermediate times. In Sect. IV, we compute the spectral form factor in ``random pure state'' 
(Page state) analytically by using some basic knowledge of the Wishart ensemble and compare the result with the numerical calculations  in Sect. II. We summarize and conclude in Sect. V.

%\section{$\langle ZZ^*\rangle$}

\section{Spectral Form Factor}

\subsection{General Discussion}
\label{SubSectionSpectralFormFactorGeneralDiscussion}

We 
%Let us 
decompose the Hilbert space of the total system of dimension $N$
 (of which the ``typical'' state $|\psi\rangle$ is an element) into a tensor product of
the Hilbert spaces of the two subsystems, system $A$ with Hilbert space dimension $N_A$,
and system $B$ with Hilbert space dimension $N_B$ (i.e. $N=N_A N_B$).
The spectral form factor $g(\tau)$  for  the $N_A$  eigenvalues $\lambda_i$ of the reduced density matrix ${\hat  \rho}_A$
%(where $N_A$ is the Hilbert space dimension of subsystem $A$), 
defined in
(\ref{DEFSpectralFormFactor}) above, can be conveniently expressed\footnote{In the present paper we often follow the notations used in Ref. \onlinecite{Cotler2016}.}
 in terms of the Fourier transform of the eigenvalue density
\begin{eqnarray}
\label{DEFFourierTransformEigenvalueDensity}
Z(\tau)\equiv \mbox{Tr} \ \exp(-i\tau {\hat \rho}_A)= \sum_{i=1}^{N_A} \exp(-i\tau
\lambda_i)
\end{eqnarray}
 as follows
\begin{eqnarray}
\label{DEFZZstar}
g(\tau) =  \langle\sum_{i,j=1}^{N_A} e^{-i \tau (\lambda_i-\lambda_j)}\rangle = \langle Z(\tau) Z^*(\tau)\rangle.
\end{eqnarray}
As seen from
(\ref{DEFZZstar}), at
%At
 $\tau=0$ the spectral form factor clearly takes on the value 
$g(\tau=0)= (N_A)^2$, while in the limit $\tau\to \infty$ only contributions with $\lambda_i=\lambda_j$ survive, which yields the smaller
value
$\lim_{\tau\to\infty} g(\tau)=$ $N_A$.
As we will see below,  the function $g(\tau)$ initially {\it decreases} starting from $\tau=0$ until it reaches a minimum (``dip''), then exhibits
 a segment
of linear rise  (``ramp''), until the curve suddenly becomes constant (at the Heisenberg time $\tau_H$) reaching its late-time
``plateau'' value [see e.g.  
%the light blue curve for $N_A=2^{12}$ 
%rising linearly over two decades  
%in the left panel (a) of  
Fig. \ref{fig:fig-sketch-ramp}]. 
%\ref{fig:page}.]
As we will review below, the initial {\it decrease} at early times is non-universal, whereas the  linear ``ramp'' is completely universal,
depending only  on the symmetry class.  We note that the presence of these  three distinct  regimes, 
the decrease until the ``dip'', the linear rise along the ``ramp'',
and the  flat plateau, was stressed in the context of the spectral form factor of the {\it  Hamiltonian} of the
SYK model  in the recent Ref.\ \onlinecite{Cotler2016} already mentioned above.

We will also consider the connected spectral form factor
\begin{eqnarray}
\label{DEFConnectedSpectralFormFactor}
g_c(\tau) =   \langle Z(\tau) Z^*(\tau)\rangle -  \langle Z(\tau)\rangle \ \langle Z^*(\tau)\rangle,
\end{eqnarray}
which 
%typically 
exhibits (as already mentioned) a longer and more pronounced ``ramp'' [compare e.g. Fig. \ref{fig:fig-sketch-ramp}]. 
%the light blue curve for $N_A= 2^{12}$ in the  right panel (b)  of  
%Fig.\ref{fig:page}].
Its analytic form for the ``random pure state'' and the Wishart random matrix ensemble 
is displayed
%shown
 in (\ref{``ramp''}) of
Sect. \ref{analytical}
 in the limit of a large density matrix.
In the context of  the spectrum of a  random (GUE, GOE or GSE)  Hamiltonian matrix, 
the connected spectral form factor $g_c(\tau)$ has been extensively discussed in the literature over many years.\footnote{See
e.g. Ref. \onlinecite{GuhrMuellerGroelingWeidenmuellerPhysRpts1998} for an extensive  review.}

%To summarize, we 
%%We 
%conclude 
%%by summarizing
 %that the presence of
%%which has been discussed frequently in the context of the spectrum of a random Hamiltonian matrix.
%%We will show that
%%As mentioned, 
%spectral rigidity
%%  level repulsion between 
%in  the spectrum of  eigenvalues $\lambda_i$ of the density matrix  is  reflected in a linear ``ramp'',  present  in both 
%$g(\tau)$ as well as $g_c(\tau)$. [We will explain this connection in more detail  in Sect. \ref{analytical}.]

%$\langle Z(\tau)Z^*(\tau)\rangle$ and $\langle Z(\tau)Z^*(\tau)\rangle-\langle Z(\tau)\rangle\langle Z^*(\tau)\rangle$.

As already mentioned in the Introduction (Sect. \ref{SectionIntroduction}), the purpose of the average $\langle ... \rangle$
in (\ref{DEFZZstar}) and (\ref{DEFConnectedSpectralFormFactor})
 is to remove non-universal rapid temporal fluctuations\cite{Prange1997} from the spectral form factor $g(\tau)$. 
In our work reported below, there will be a natural ensemble available  over which to perform the average as an {\it ensemble average}:
For the ``random pure state'' (Page state) discussed  in Sect. \ref{SubSectionRandomProductState} below,  this will be  an average over the statistical
ensemble of ``random pure states'', while for the Floquet and  quantum Ising models in Sect. \ref{SubSectionFloquetAndQuantumIsingModels}
%{SubSectionFloquetAndIsingModels} 
this will be an ensemble of initial direct product states.
Another way to remove the high-frequency fluctuations from the spectral form factor $g(\tau)$
is to coarse grain the latter in time  $\tau$  by convolution with
a temporal  ``smearing function'' which eliminates high frequencies components from  the signal.  (For example, see Refs.\ \onlinecite{Cotler2016,Balasubramanian2017}.) Since the ensemble averages were more convenient
for us, we did not  use the coarse graining approach in the present work to remove the high frequency fluctuations.

%\subsection{Random product state}
\subsection{Random pure state}
\label{SubSectionRandomProductState}
Since, as already mentioned in the Introduction,
 this will turn out to provide a model-independent description of the universal properties of quantum chaos in a wave function,
we
first study the spectral form factor
% in
of  the ``random pure state'' (Page state), discussed in (\ref{DEFPageState}) and the paragraph below that equation.
%\begin{align}
%|\Psi(\{\alpha_i\})\rangle=\sum_i\alpha_i|C_i\rangle,
%\end{align}
%in which the
%coefficients $\alpha_i$ of the state in a fixed basis $|C_i\rangle$ are random complex numbers subject to the normalization constraint, with a %probability distribution invariant under (unitary) basis changes. The set of coefficients can thus be considered a row (or column) 
%vector of a 
%%Gaussian U
%unitary random matrix.
%, i.e., in short ``it is in the GUE random matrix ensemble".
The reduced density matrix for a ``random pure state''  is also a random matrix and
it turns out to belong to the so-called
(unitary) 
Wishart ensemble with Dyson index
%at 
$\beta=2$, in which the spectral density satisfies
the so-called
Marchenko-Pastur distribution.\cite{loggas}
(See Sect. \ref{analytical}
for a review.)

 Using this property, Page showed\cite{Page1993}
 that the ensemble-averaged (von Neumann) entanglement entropy [EE] of the reduced density matrix for subsystem A
of the ``random pure state''  is equal to
\begin{align}
\label{EntanglementEntropyRandomPureState}
\langle S_A\rangle=\log N_A-\frac{N_A}{2N_B}.
\end{align}
(Recall that $N_A$ and $N_B$ are  Hilbert space  dimensions for subsystem A and its complement B, respectively, and we have assumed 
 $N_A\leq N_B$
without loss of generality.)

Since the Hilbert space dimension $N_A$ grows exponentionally with the volume of subsystem $A$,
the  entanglement entropy of the random product state 
exhibits
according to (\ref{EntanglementEntropyRandomPureState})
a volume law (as expected).
For example, for
 the Ising-type systems considered in the present paper which have a local (onsite) Hilbert space dimension of two,
we have $N_A=2^{L_A}$ where $L_A$ is the number of lattice sites of subsystem $A$.
We thus seen from   (\ref{EntanglementEntropyRandomPureState})
that the entanglement entropy of the ``random pure state'' exhibits a volume law of
 maximal possible value (given the dimension of the onsite Hilbert space), up to 
a small subleading 
%constant 
term which depends on the ratio of the Hilbert space dimensions of subsystems $A$ and $B$,
which we denote by $\alpha\equiv N_A/N_B$. The latter 
%constant
subleading term in  (\ref{EntanglementEntropyRandomPureState})
takes on its  maximal value $1/2$ at $\alpha=1$,
 and approaches zero as  $N_A\ll N_B$.

\begin{figure}[hbt]
\centering
 \subfigure[]{\label{fig:page_z2} \includegraphics[width=.4\textwidth]{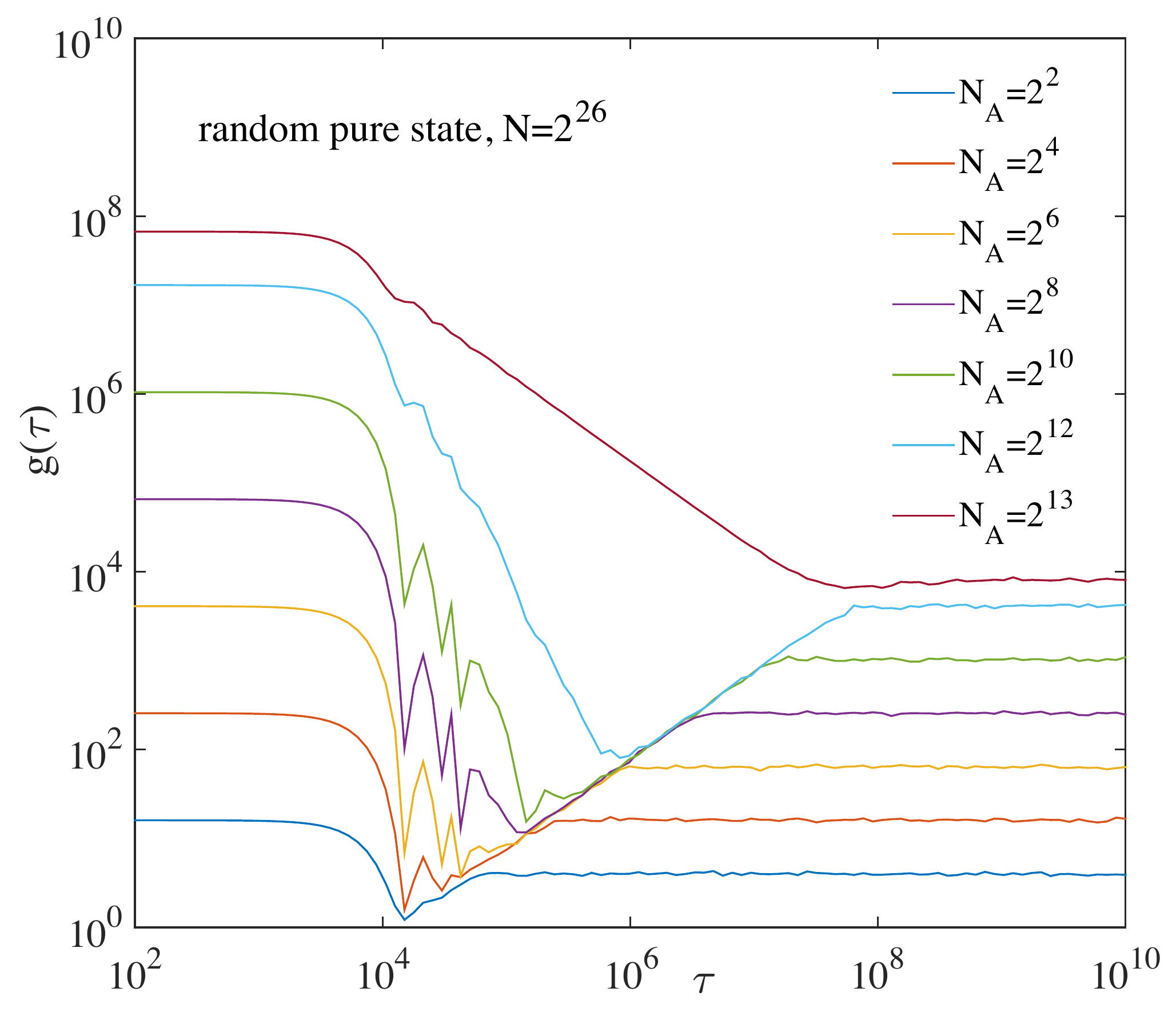}}
 \subfigure[]{\label{fig:page_z2_sub} \includegraphics[width=.4\textwidth]{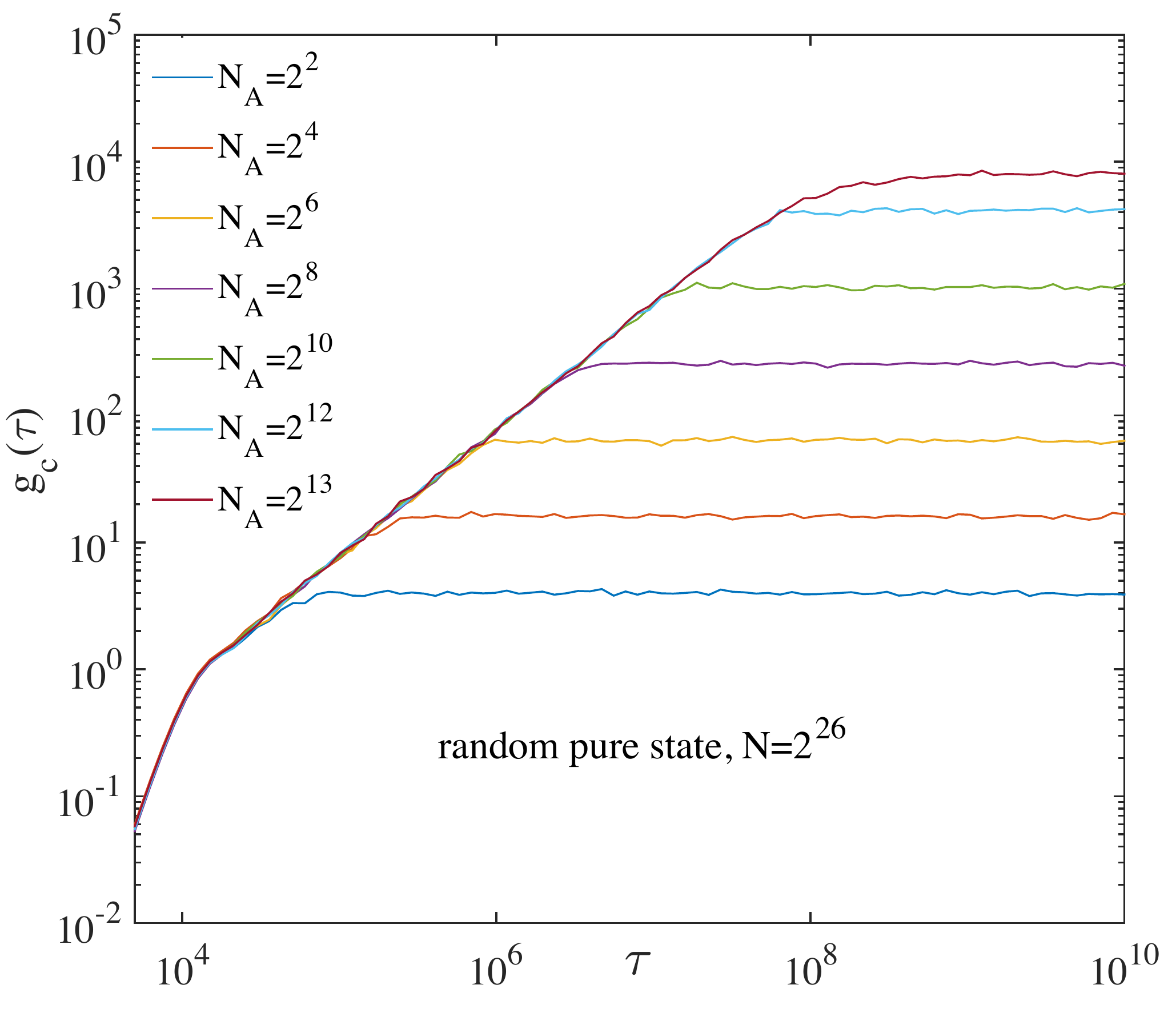}}   
\caption{
(a) Spectral form factor $g(\tau)=$  $\langle ZZ^*\rangle$ for the ``random pure
state'' (Page state) with fixed value of $N=2^L$ and 
different values of  $N_A=2^{L_A}$, where $L$  is the total number of lattice sites, and $L_A$ denotes the number of lattice sites in subsystem $A$.
 The curve is obtained by taking the  disorder average over 1000 states.
 (b) Connected spectral form factor $g_c(\tau)=$ $\langle ZZ^*\rangle-\langle Z\rangle\langle Z^*\rangle$ for the
`` random pure
state'' with fixed
values of $N$ and different values of $N_A$. 
Again, the curve is obtained by taking the  disorder average over 1000 states.
}
\label{fig:page}
\end{figure}

It it known analytically (as reviewed in Sect.
\ref{analytical} below)
 that the eigenvalues of the Wishart random matrix
exhibit the same universal spectral correlations as
those of the Hamiltonian of the GUE  random matrix ensemble, which manifest themselves, as already
mentioned, in the {\it connected} spectral form factor $g_c(\tau)$.
%two eigenvalues separated by energy scales 
%ranging  from the mean level spacing up to a scale that can be orders of magnitudes larger
%, similar to  the eigenvalues of the Hamiltonian of the GUE Hamiltonian matrix ensemble,
% also exhibitspectral rigidity, i.e. they repel when .
%even on scales much larger than the mean level spacing.}
% level repulsion, i.e.
%}
%the probability that neigboring eigenvalues are close to each other is rare. 
We have computed numerically the (non-connected)
spectral form factor $ g(\tau)=$$\langle Z(\tau)Z^*(\tau)\rangle$ for the eigenvalues of
the 
%resulting 
Wishart random matrix, describing the reduced density matrix of the ``random pure state''. 
The results are plotted  in Fig.~\ref{fig:page_z2}
which shows that
when $\alpha= N_A/N_B<1$, there is 
an intermediate  linear ``ramp'' where $g(\tau)=$ $\langle Z(\tau)Z^*(\tau)\rangle$ grows linearly with
% the
 time $\tau$.
The presence of the ``ramp''
% indicates
demonstrates the presence of the mentioned universal spectral correlations, as discussed
%shows  that there is
% level repulsion between eigenvalues
%spectral rigidity in the spectrum,
%a statement that 
% as we 
%will  be proven 
 analytically  in Sect.~\ref{analytical} below (compare also Fig. \ref{fig:fig-sketch-ramp}).

Continuing with $\alpha=N_A/N_B <1$,
we also observe  in Fig.~\ref{fig:page_z2}
an early time regime where $g(\tau)=$ $\langle Z(\tau)Z^*(\tau)\rangle$ drops down quickly to a minimum value.
It turns out 
%(see below) 
that at early times,
$\langle Z(\tau)Z^*(\tau)\rangle$  
factorizes into $\langle Z(\tau)\rangle\langle Z^*(\tau)\rangle$ and is  therefore  determined by the Fourier 
%transformation
transform  of the 
average of the eigenvalue density, $\langle Z(\tau)\rangle$, defined in (\ref{DEFFourierTransformEigenvalueDensity}).
%One infers
%We see  from
One  can determine from
Fig.~\ref{fig:page_z2}, 
where
$g(\tau)=$ $\langle Z(\tau)Z^*(\tau)\rangle$ is plotted versus $\tau$ for
 $N_A=2^{12}$ (and $N=2^{26}$, i.e. when  $\alpha=1/4$), 
that it scales as $1/\tau^3$ in this early time regime. 
Moreover, we observe in  the plots 
%in 
shown in the same figure  for smaller values of $\alpha$,
% closer to zero, 
that
there are large oscillations in 
this  early time regime\footnote{These oscillations can be systematically reduced by averaging over larger samples.}, 
but  with an envelope function  that is  still close to $1/\tau^3$, when  compared to the $\alpha=1/4$ case.

This power law decay behavior of the spectral form factor at early times 
% is determined by 
originates from the eigenvalue distribution function 
$\langle Z(\tau)\rangle$
%for
of  the Wishart matrix which will be analytically computed in Sect.~\ref{SubSectionWishartRandomMatrixEnsemble}.
%{analytical}.
For a generic chaotic system with Hamiltonian ${\hat H}$, the 
details of the eigenvalue 
%density
distribution function 
%for
of  the density matrix ${\hat \rho}_A$ for a typical wavefunction
will in general be different from that of the Wishart matrix, and will
not be universal.
In particular,  in the
early time regime where the spectral  form factor $g(\tau)=$  $\langle Z(\tau)Z^*(\tau)\rangle$ factorizes into
$\langle Z(\tau)\rangle \langle Z^*(\tau)\rangle$, it  will be 
model-dependent,
in contrast to the regime of  intermediate  $\tau$
%time regime 
where it exhibits
% exhibiting 
a  universal ``ramp'', whose presence depends solely  on the universal
spectral 
%rigidity
correlations in  the spectrum of eigenvalues.

% level repulsion between eigenvalues on a scale of the distance between individual levels
%( much smaller than the total width of the spectrum). 
%}

As already discussed,
at  late times $\tau$ larger than the Heisenberg time $\tau_H$,  the spectral form factor $g(\tau)=$  $\langle Z(\tau)Z(\tau)^*\rangle$ will saturate to a constant value $N_A$, which is coming from the terms with  $\lambda_i=\lambda_j$
[see (\ref{DEFZZstar})].
 Since, as  has also been  mentioned,  the saturation value is much smaller than the initial value $N_A^2$ attained at $\tau=0$ [see again (\ref{DEFZZstar})], 
we plot  $g(\tau)=$  $\langle Z(\tau)Z^*(\tau)\rangle$ on a log-log scale so that the behavior of $\langle ZZ^*\rangle$
at the different time scales $\tau$
can be seen clearly. 
%Similar  to the case of the spectral form factor of a chaotic  Hamiltonian discussed in (\ref{CotlerEtAl1611.04650}),
The 
%There are
three time-regimes
mentioned in Sect.
(\ref{SubSectionSpectralFormFactorGeneralDiscussion})
are separated by two typical time scales:
The time where the ``dip'' occurs (``dip time") $\tau_d$, and the 
time where the plateau begins (``plateau time'', or ``Heisenberg time'') $\tau_p=\tau_H$.
We find that the
%The
 dip time $\tau_d$
%is found to scale
scales
 as $\sqrt{N_A N}$,
 while $\tau_p$
is found to scale as $N_A\sqrt{N}$. (Recall $N=N_A N_B$.)
  [Both statements are obtained analytically in Sect.~\ref{analytical}, and have also been checked
numerically.]
This is analogous to the three regimes  observed in Ref.\ \onlinecite{Cotler2016} for  the spectral form factor for a $M \times M$  random matrix in the GUE ensemble (as compared
to the Wishard random matrix ensemble discussed here),  where  $\tau_d\sim \sqrt{M}$ and $\tau_p\sim M$.
%,  where $M$ is the dimension of GUE matrix.

We finally discuss a subtlety occuring
%In constrast, 
when $\alpha= N_A/N_B =1$.
In contrast to the case where $\alpha <1$ discussed above, we see from
Fig.~\ref{fig:page_z2} [top curve, $N_A=2^{13}$, where the total Hilbert space dimension is $N=2^{26}$]
 that for $\alpha=1$  the intermediate ``ramp''  in $g(\tau)$ disappears. 
 The difference between $\alpha=1$ and $\alpha<1$ is caused by
the different behavior of
$\langle Z(\tau)\rangle$:
At early times, where  $g(\tau)=$ $\langle Z(\tau)Z^*(\tau)\rangle \sim$ $ \langle Z(\tau)\rangle\langle Z^*(\tau)\rangle$
factorizes, for $\alpha=1$ 
the spectral form factor
%this quantity 
scales as $1/\tau$ and then directly transits to the  plateau. 
However, the
absence of the ``ramp'' does not mean that
% there is
% no level repulsion 
universal spectral correlations are 
%rigidity 
%is 
absent
%present 
 in the spectrum  of
%between 
eigenvalues when $\alpha=1$. Rather, the different
behavior of
$\langle ZZ^*\rangle$
%caused by the slow decay of $\langle Z(\tau)\rangle$ 
%which turns out  to just
just  turns out to  {\it hide} the ``ramp'' due to the slow decay of 
the disconnected part $\langle Z(\tau)\rangle\langle Z^*(\tau)\rangle$.
The effect of the slowly decaying $\langle Z(\tau)\rangle$ can be removed if we 
%compute 
consider instead the connected  spectral form factor
$g_c(\tau)=$
$\langle ZZ^*\rangle-\langle Z\rangle\langle Z^*\rangle$,
which is plotted in
Fig.~\ref{fig:page_z2_sub}. In the latter figure  we observe a long ``ramp'' {\it even}  for $\alpha=1$.
% i.e. a signature of level repulsion between a large number of eigenvalues.
Actually, for the
other
% rest 
curves with $\alpha<1$ in the same figure, the linear ``ramp''
in $g_c(\tau)$ starts at an earlier time (denoted earlier by $\tau_0$ - see Fig. \ref{fig:fig-sketch-ramp})
than the ``ramp'' in $g(\tau)$,
where part of the longer linear ``ramp'' in $g_c(\tau)$
% which 
is
in fact covered up by $\langle Z\rangle\langle Z^*\rangle$ as  depicted in Fig.~\ref{fig:page_z2}. 
Since small values of $\tau$ correspond to large eigenvalue differences
on the scale of the mean level spacing,
the early-time part $\tau_0 \leq \tau \ll \tau_H$
%this part 
of  the ``ramp'' in $g_c(\tau)$  describes
%describes 
the
% repulsion between 
%presence of 
spectral correlations
% rigidity 
of eigenvalues  separated by an energy scale of many times the mean level spacing.
Eventually, as $\tau$ is close to $\tau_0$,  the universal behavior of $g_c(\tau)$ reflected in the linear ``ramp''  will be limited
%at small $\tau$
 by model-dependent (``ultraviolet'') effects at large separations of eigenvalues, leading to deviations from the linear ``ramp'' at 
yet smaller 
values of $\tau \lesssim \tau_0$.

The length of the ``ramp'' in $g(\tau)$  increases
%turns out to grow 
with the number of eigenvalues that 
%have 
exhibit 
%level 
universal spectral 
%rigidity.
correlations.\cite{Cotler2016}
% repulsion between them.\cite{CotlerEtAl1611.04650}
We clearly see from
Fig.~\ref{fig:page_z2_sub}
that both the length of ``ramp'', and the position of $\tau_p$, are linearly proportional (on a log scale) to $\log N_A$. 

We finally
want to mention that the reduced density matrix ${\hat \rho}_A$ studied in this section belongs to the Wishart random matrix ensemble 
{\it lacking time-reversal symmetry}, described by Dyson index  $\beta=2$ (i.e. the ``GUE-type'' version of the Wishart random matrix ensemble).
% In principle we 
We can also consider a density matrix  ${\hat \rho}_A$ described by  a Wishart ensemble with Dyson index $\beta=1, 4$ (the ``GOE'' and the ``GSE'' version of the Wishart random matrix ensemble), in which the details of 
the universal spectral
% rigidity 
correlations are slightly different. In the spectral form factor, these modified spectral correlations
%rigidity of 
between the eigenvalues  are  reflected in a similar but slightly  more complicated  universal ``ramp''.\cite{loggas} While a straightforward extension,
we will not discuss details of these cases 
%them 
explicitly in this paper. 

%
%%%%%%%%%%%%%%%
%\begin{figure}%[hbt]
%\centering
%\includegraphics[width=.6\textwidth]{Page_Z_2.pdf}
%\caption{$\langle ZZ^*\rangle$ for random product state with fixed $N$ and different $N_A$. The curve is taking disorder average over 1000 states.}
%\label{fig:page_z2}
%\end{figure}
%%%%%%%%%%%%%%% 

%%%%%%%%%%%%%%%%%%%%%%
%%%%%%%%%%%%%%%%%%

\subsection{Floquet and  Quantum Ising models}
\label{SubSectionFloquetAndQuantumIsingModels}

For a generic chaotic system with a time-independent Hamiltonian ${\hat H}$, we expect that for the states
$|\psi\rangle$ with energy expectation value $E$ ($=\langle \psi |{\hat H} | \psi \rangle = e V$, where $e$ is the energy density)
% $E =
%\footnote{i.e. with expectation values of the Hamiltonian $\langle \psi |{\hat H} | \psi \rangle=$}
%\langle \psi |{\hat H} | \psi \rangle=$}
% $E = e V$.
%}
 in the middle of the spectrum of ${\hat H}$, the resulting density matrix ${\hat \rho}_A$ of the subsystem is well thermalized and
its entanglement entropy exhibits a volume law.
As already  mentioned in the introduction,  one 
%expects
might expect  based on 
%ideas of
notions from  ETH that the spectrum of
this reduced density matrix exhibits universal spectral correlations.
%spectral rigidity. 
%More specifically, one might conjecture that
%the spectra of the reduced density matrices of these states $|\psi\rangle$  have the same universal properties as those of the random
%pure states (Page states), and that their reduced density matrix is in fact close to the Wishart random matrix, which is the reduced
%density matrix of the ``random pure state''.
%%These 
%states are analogous to the ``random pure state'', and that
%the spectrum of their reduced density matrix 
%%is %One may expect that these
%expected to 
%exhibits
%% level repulsion.
%spectral rigidity.
%
In this section, we will 
show more specifically 
for both, one-dimensional  Floquet and quantum Ising models, that 
the spectra of the reduced density matrices of the above-mentioned states $|\psi\rangle$  
have the same universal properties as those of the ``random
pure states''  (Page states), and that the universal features of their reduced density matrix 
%is
are  in fact 
%close to
those of  the Wishart random matrix, which is the reduced
density matrix of the ``random pure state'', the density matrix of a completely random wave function.
%verify
% this statement 
%these statements for both, one-dimensional  Floquet and Ising models. 

%At a practical level,
For practical reasons,
 instead of diagonalizing the Hamiltonian (or the Floquet operator) to
investigate the 
%presence of
 spectral statistics
%studying 
%the level repulsion
%in the spectrum of 
the reduced density matix obtained for the subsystem for a typical {\it eigenstate}\cite{footnote_1},
we will choose 
%some
a set of
initial product states $|\Psi_0\rangle$
(which thus are not entangled)
and let them evolve under the unitary evolution governed by  the Floquet operator or Ising Hamiltonian, i.e., 
\begin{eqnarray}
\label{TimeEvolvedInitialState}
|\Psi_0(t)\rangle=U(t)|\Psi_0\rangle.
\end{eqnarray}
 This is actually  a quantum quench problem.  For a generic non-integrable system,
the initial wave function $|\Psi_0\rangle$
% this single wavefunction
will eventually, at long times $t$,  thermalize under 
its  own dynamics and 
the reduced density matrix ${\hat \rho}_A(t)=\mbox{Tr}_B|\Psi_0(t)\rangle\langle\Psi_0(t)|$ will
% be very close to
approach the reduced
density matrix of  a generic eigenstate of the Hamiltonian.\cite{Srednicki1994,Deutsch1991}
% in the sense that the difference (in norm)  between 
%these two density matrices will become exponentially small.\cite{Srednicki1994,Deutsch1991}
Universal spectral correlations
%rigidity
% level repulsion
 will develop in the spectrum of the  reduced density matrix starting from 
such
%these 
unentangled  initial states, so that the final state obtained
 after sufficiently  long 
%enough 
time evolution will be fully thermalized. The advantage of this method is that we can work with relatively
large  systems (the time evolution operator simply has to be applied for a long time).
%In particular,
Specifically,  we will consider below
one-dimensional Floquet and Ising models with  $L=20$ lattice sites.

\subsubsection{Floquet model}
\label{SubSectionFloquetModel}
We first consider a Floquet model.
It is known
that Floquet systems can thermalize very 
%fast
rapidly  due to the absence of any  conservation laws.\cite{Kim_Huse_2014, DAlessio2014,Lazarides2014,PONTE2015,Zhang2015}
The properties of 
such periodically driven systems  are determined by the unitary time evolution operator over one period, i.e., the Floquet operator. Following Ref.\ \onlinecite{Kim_Huse_2014}, 
we consider the following Floquet operator
\begin{align}
\hat{U}_F=\exp[-i t_0 \hat{H}_z]\exp\left[-it_0\hat H_x\right] ~,
\label{flo_op}
\end{align}
where
\begin{align}
&\hat{H}_x=\sum_{j=1}^{L}g\hat\sigma_j^x\nonumber\\
&\hat{H}_z=\sum_{j=1}^{L-1}\hat\sigma_j^z\hat\sigma_{j+1}^z+\sum_{j=1}^Lh\hat\sigma_j^z ~,
\label{floq}
\end{align}
and ${\hat \sigma}^x_j$ and ${\hat \sigma}^z_j$ are standard Pauli matrices acting on lattice site $j$.
This model is a one-dimensional periodically driven system with period $T=2t_0$. In the numerical calculations
discussed below
we choose open boundary conditions and typical system parameters
 $(g,h,t_0)=(0.9045, 0.8090, 0.8)$.

We choose a set of 
%take the 
initial states which are
%to be 
random product states (hence unentangled,  having vanishing EE)
with  the direction of the spin 
at each  lattice site chosen independently from a uniform distribution on the Bloch sphere.

 Since we are considering a Floquet model, the evolution time $t$ is 
%chosen as 
an integer multiple of $T$, i.e., $t=nT$ with $n\in\mathbb{Z}_+$. For the parameters we are considering here, it only takes a  small number of
time-steps to
% realize
achieve  thermalization. The details  of the  thermalization process itself and of  the 
%onset
development of chaos
 will be discussed in Sect.~\ref{onset_chaos} below. 
Here we discuss the properties of the fully thermalized state that the system takes on after  sufficiently long time-evolution.
Note that since
%Since 
for a Floquet system energy is not conserved, 
one expects that the subsystem ($L_A\leq L/2$) will always thermalize 
%with the
to a  state  at  infinite temperature close to the  ``random pure state'' discussed above.\cite{Zhang2015}

\begin{figure}[hbt]
\centering
 \subfigure[]{\label{fig:floq_ising_z2} \includegraphics[width=.4\textwidth]{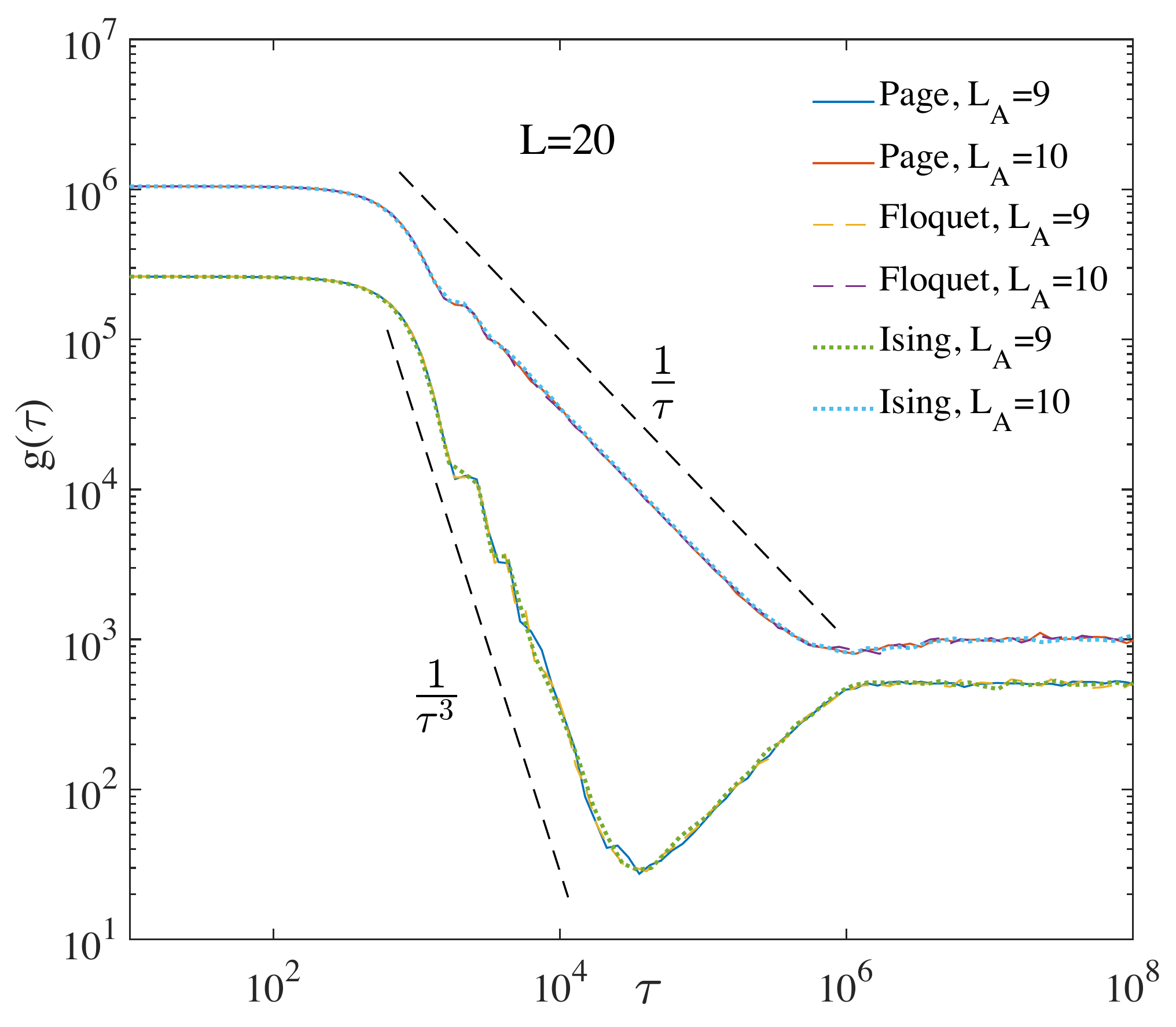}}
 \subfigure[]{\label{fig:floq_ising_z2_sub} \includegraphics[width=.4\textwidth]{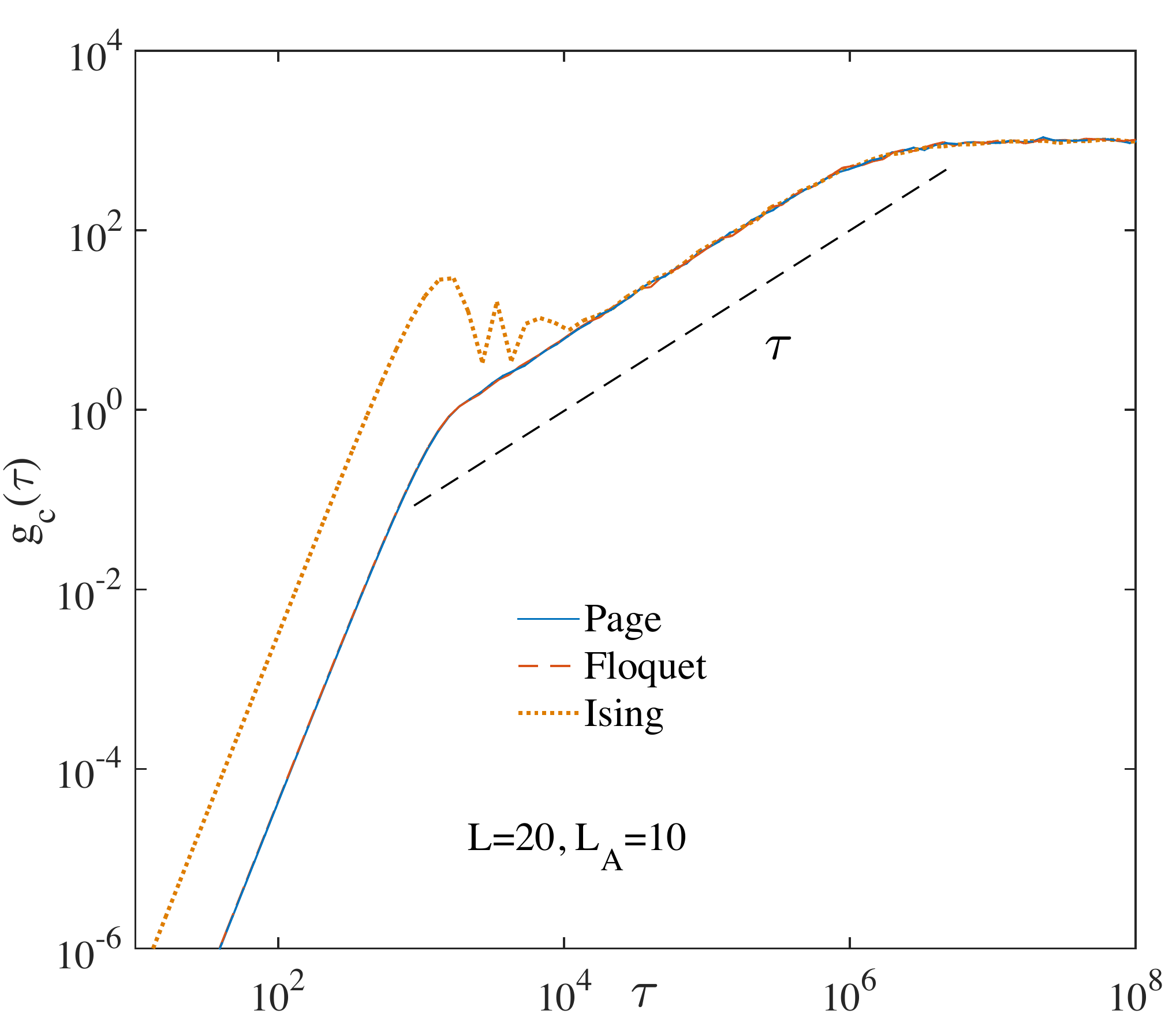}}   
\caption{
(a) $g(\tau)=$ $\langle ZZ^*\rangle$ for  Page state, Floquet and quantum Ising models. 
%over 1000 states.
 (b) $g_c(\tau)=$ $\langle ZZ^*\rangle-\langle Z\rangle\langle Z^*\rangle$ for Page state, Floquet and quantum  Ising models.  - For both (a) and (b), we average over an ensemble containing over 1000 states.
} 
\label{fig:Floq_Ising}
\end{figure}

In   Fig.~\ref{fig:Floq_Ising} we
present numerical  results for the spectral form factor $g(\tau)=$  $\langle Z(\tau)Z^*(\tau)\rangle$ and for
the connected spectral form factor $g_c(\tau)=$  $[\langle Z(\tau)Z^*(\tau)\rangle-\langle Z(\tau)\rangle\langle Z^*(\tau)\rangle]$ 
at  time-step $n=30$, when the system
is fully thermalized.
We see from
 Fig.~\ref{fig:Floq_Ising}
that both, 
for  $L_A=9$ and for $L_A=10$ [here $L=20$], both quantities
% $\langle ZZ^*\rangle$
$g(\tau)$  and  
%$[\langle ZZ^*\rangle-\langle Z\rangle\langle Z^*\rangle]$ 
$g_c(\tau)$
are 
%the same as 
indistinguishable from those for the ``random pure state''.
 The ``ramp''  in
%for 
$g(\tau)=$ $\langle Z(\tau)Z^*(\tau)\rangle$ 
is absent  when  $L_A=10=L/2$, but
becomes visible once we subtract the disconnected part
 $\langle Z(\tau)\rangle\langle Z^*(\tau)\rangle$ to obtain the connected spectral form factor
$g_c(\tau)$  (as discussed above
% for 
in the context of the ``random pure state'').

%{\color{blue}

\begin{figure}%[hbt]
\centering
\includegraphics[width=.5\textwidth]{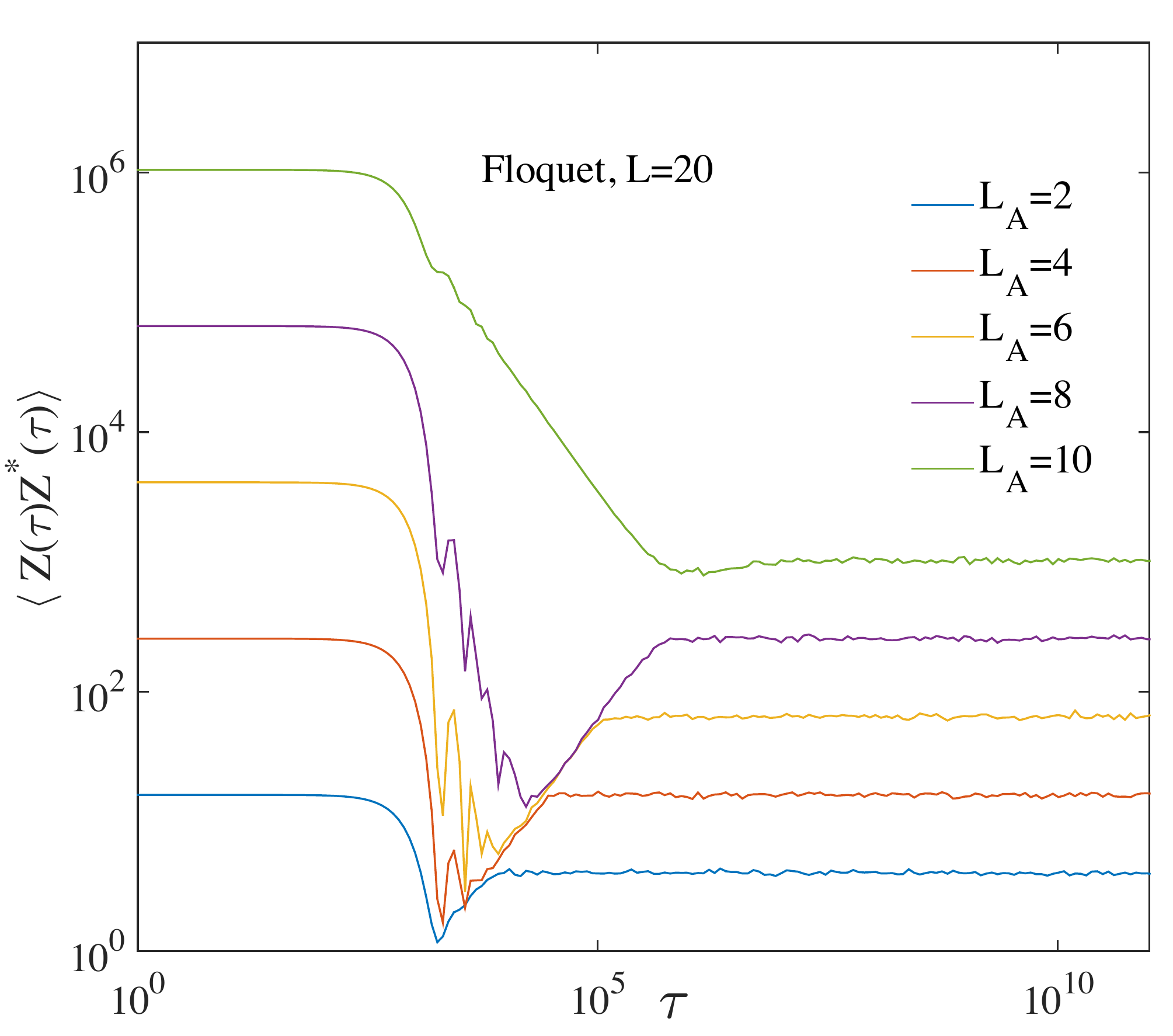}
\caption{Spectral form factor averaged over a time-series of 1000 states $|\Psi_0(t_m)\rangle$ generated from a single initial product state
 (see \eqref{TimeSeriesOfStates})
at times $t_m=T_0 + m \  \delta t$, where $T_0=40$ is chosen sufficiently large to ensure the initial state has already thermalized.
}
\label{fig:floq_Z2_time}
\end{figure}

We  also note that there is another way to generate an ensemble that can be used to perform the  average: We can pick a {\it fixed
initial state} $|\Psi_0\rangle$, but consider an ensemble of states whose members 
%are
consist of  the {\it time-series of
states} originating from the time-evolution of this {\it fixed} state by different amounts of time
$t_m \equiv T_0 + m  \ \delta t$,  where $T_0$ is a large time ensuring that the initial
state has ``thermalized'', $\delta t$ is some time-step ($=T$ in the Floquet case),
and $m=1, 2, ..., M$, i.e.
\begin{equation}
\label{TimeSeriesOfStates}
|\Psi_0(t_m)\rangle = {\hat U}(t_m) |\Psi_0\rangle, \qquad (m=1, ..., M).
\end{equation}
In this situation the time-average over the set of states
(\ref{TimeSeriesOfStates}) at  times $t_m$ then generates the ensemble-average
of the spectral form factor. The resulting averaged spectral form factor is displayed in Fig.~\ref{fig:floq_Z2_time},
and seen to exhibit the same universal linear ``ramp'' as that arising  from averaging over  the ensemble of initial states $|\Psi_0\rangle$
displayed in Fig.\ref{fig:Floq_Ising}.

%%%%%%%%%%%%%%%%%%%%%%
%%%%%%%%%%%%%%%%%%

%%%%%%%%%%%%%%
%%%%%%%%%%%%%% 

\subsubsection{Quantum Ising model}
\label{SubSectionSpectralFloquetAndQuantumIsingQuantumIsingModel}

In this section we
study the transverse field quantum  Ising Hamiltonian
with a longitudinal field. The Hamiltonian is 
\begin{equation}
\hat H=\sum_i\hat\sigma^z_i\hat\sigma^z_{i+1}+h_x\sum_i \hat\sigma^x_i+h_z\sum_i \hat\sigma^z_i ~.
\label{ising_H}
\end{equation}
The system parameters are
$(h_x,h_z)=(1.05, 0.5)$.\cite{Banuls2011} This model is 
far from integrable
 due to the large longitudinal field. 
%Although  the
The  reduced density matrix of
% the projection operator on
 the initial direct product state will eventually thermalize under the time-evolution generated
by the time-independent Hamiltonian (\ref{ising_H}), the total energy 
%will 
always 
%be 
being conserved.
We choose the  initial states to be  random product states with Hamiltonian expectation values $E$
 within a small energy 
%window
interval  $E \in [-0.1, 0.1]$ 
%(the energy is 
(close to the middle of the spectrum of ${\hat H}$), 
and study the spectral correlations and the emergence of a ``ramp'' in
% the level repulsion 
%spectral rigidity of 
the spectrum of
% for the 
eigenvalues of the reduced density matrix
at a  sufficiently long time 
$t=100$, when the system is fully thermalized. 
The results are presented  in Fig.~\ref{fig:Floq_Ising}. We see
that the spectral form factor $g(\tau)=$ $\langle Z(\tau)Z^*(\tau)\rangle$ 
is indistinguishable from that
%very close to 
%the same quantity 
computed for the ``random 
%product
pure  state",
as well as from  that computed  for the  Floquet model,  for times $\tau$
ranging from close to two orders of magnitude below the Heisenberg time scale all the way up to the plateau
%at times larger than that time scale,
and it exhibits a linear  ``ramp'' in that range of times $\tau$.
On the other hand, 
there is some difference
%with
in  the
% in 
{\it connected} spectral
form factor $g_c(\tau)=$  $[\langle Z(\tau)Z^*(\tau)\rangle-\langle Z(\tau)\rangle\langle Z^*(\tau)\rangle]$ 
displayed in Fig.~\ref{fig:floq_ising_z2_sub}:  The length of the ``ramp'' for the  Ising model is shorter than that for ``random pure state'', and 
that for the Floquet model, 
%showing
and shows an overshoot at early $\tau$. This suggests that for the Ising model,
whose time-evolution is  
constrained by the energy conservation law, the subsystem is ``less chaotic'' in the sense that model-dependent features appear in the connected
spectral form factor
% at sufficiently
already at (small) times, here  $\tau \approx \tau_0= 10^4\approx 10^{-2} \tau_H$, see  Fig.~\ref{fig:floq_ising_z2_sub},
reflecting deviations from  universal spectral
% rigidity 
correlations for eigenvalues $\lambda_i$ at  correspondingly large  separations.
We will discuss this issue in more detail in the next section.

\section{The 
%onset
development of chaos and thermalization
 in Floquet and Ising model}
\label{onset_chaos}
In this section, we study the 
%{\it onset}
{\it development of quantum chaos }
in  the many-body wave function. Starting (as before) with an initial product state $|\Psi_0\rangle$, 
the time-evolved  reduced density matrix 
${\hat \rho}_A(t)$
of 
the subsystem,
% is equal to
\begin{align}
{\hat \rho}_A(t)=\mbox{Tr}_B \left[{\hat U}(t)|\Psi_0\rangle\langle \Psi_0|{\hat U}^{\dag}(t)\right],
\end{align}
%It
will eventually thermalize
under the unitary time evolution 
operator ${\hat U}(t)$ of a 
generic non-integrable system, 
 and
its spectrum will in the process
 develop universal spectral correlations,
% rigidity.
%level repulsion between eigenvalues in ${\hat \rho}_A$.
%As we showed in the last section, this
% level repulsion
%spectral rigidity  will be 
manifested by
% lead to 
a linear   ``ramp'' in the corresponding spectral form factor. 
In this section we ask: At what times $t$, under the quantum mechanical time evolution, does the ``ramp'' 
%appear, 
emerge, and how does it evolve in time $t$ until
it reaches its final fully thermalized regime  at long times?
That is, we will
be studying the 
%onset
development of chaos
 in the density matrix.
To answer these questions, we will study the spectral form factor at different  times $t$ before 
${\hat \rho}_A(t)$ has
fully thermalized. 

\subsection{Floquet System}
\begin{figure}[hbt]
\centering
 \subfigure[]{\label{fig:floq_z2} \includegraphics[width=.4\textwidth]{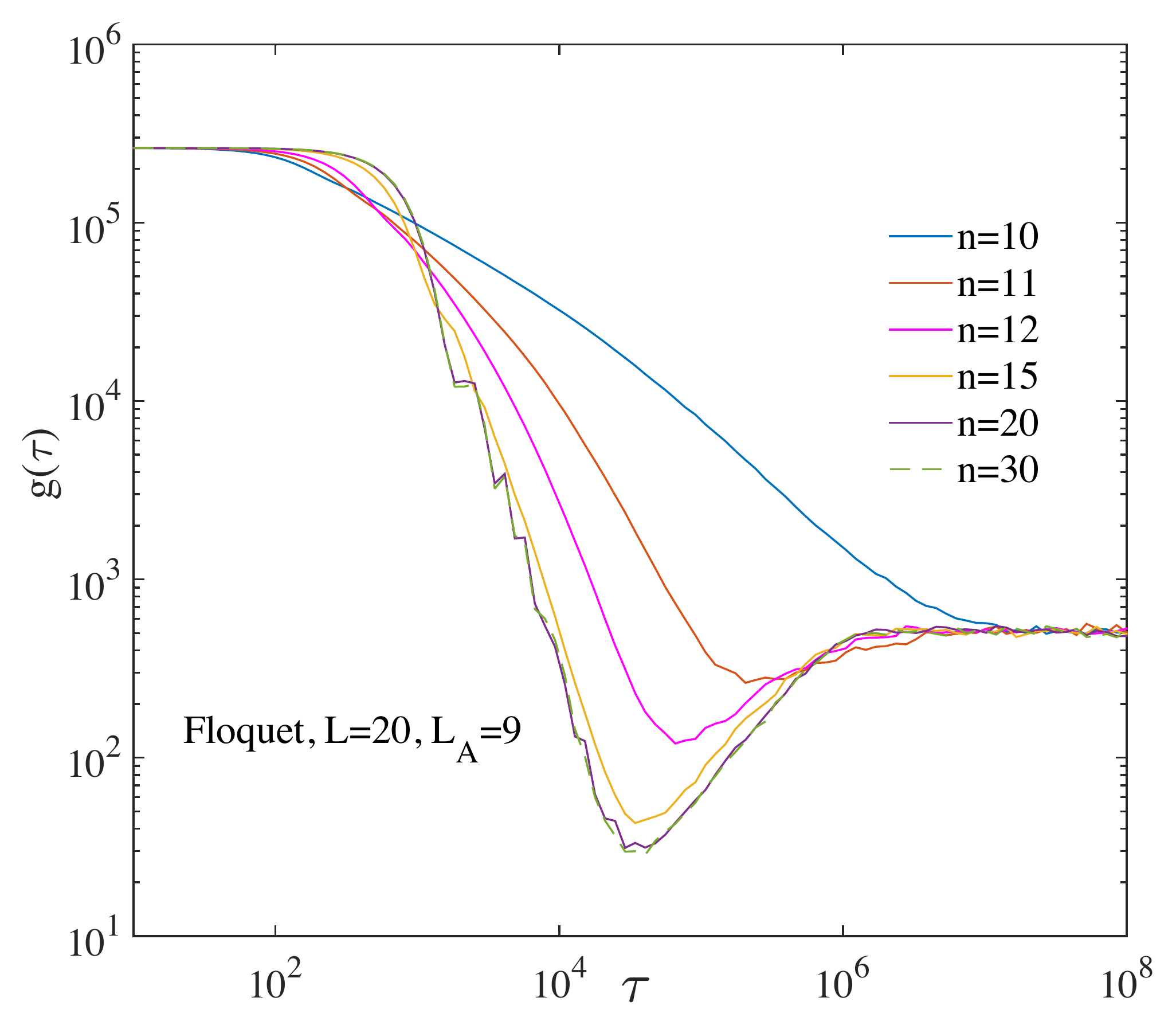}}
 \subfigure[]{\label{fig:floq_z2_sub} \includegraphics[width=.4\textwidth]{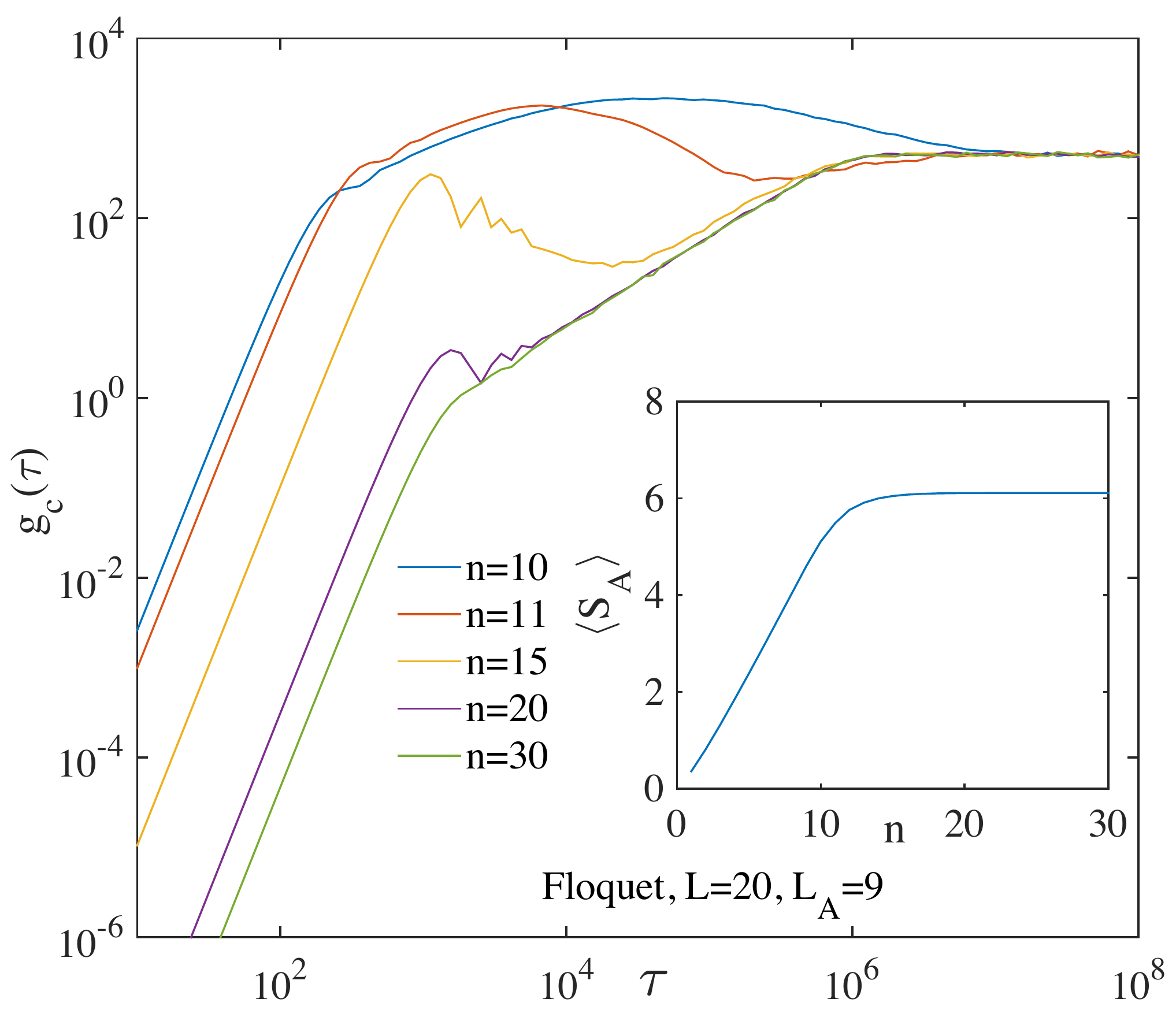}}   
\caption{
(a) $g(\tau)=$ $\langle ZZ^*\rangle$ versus $\tau$ for the Floquet model defined in \eqref{floq} at different time-steps $n$,  averaged over 1000 samples. (b) $g_c(\tau)=$  $\langle ZZ^*\rangle-\langle Z\rangle\langle Z^*\rangle$ for the same model at different time-steps $n$. 
The inset shows
% is 
the averaged 
%von Neumann 
entanglement entropy (EE) as a function of time-step $n$.
 } 
\label{fig:floq}
\end{figure}

We first study the
Floquet system defined in 
(\ref{flo_op}) and (\ref{floq}).\cite{Kim_Huse_2014, DAlessio2014,Lazarides2014,PONTE2015,Zhang2015}
%%\eqref{flo_op},
%where the {\it entanglement entropy} (EE)  approaches
%equilibrium very rapidly due to the absence of any conserved quantities.\cite{Kim_Huse_2014, DAlessio2014,Lazarides2014,PONTE2015,Zhang2015}
 As shown in the inset of Fig.~\ref{fig:floq_z2_sub}, when the subsystem size is $L_A=9$ (total system size $L=20$), the EE grows linearly with time 
$t=n T$
%when
for time-steps  $n\leq 10$, and then 
%simply
quickly  saturates exponentially in the  time-step $n$  to the Page value. 
%with an 
%exponentially 
%with
As is clearly seen from
the inset  of Fig.~\ref{fig:floq_z2_sub},
at time-step $n=30$  the 
deviation of the EE from its Page value is  negligible.

When we consider the 
%quantity
{\it spectral form factor}
$g(\tau)=$ 
 $\langle Z(\tau)Z^*(\tau)\rangle$
for  the same  density matrix ${\hat \rho}_A(t=n T)$,
 the ``ramp'' starts to emerge at time-step  $n=11$:
 In Fig.~\ref{fig:floq_z2}, we can clearly observe that as the time-step $n$ increases
beyond $n=11$,
% further, 
the dip in $g(\tau)$  becomes much deeper,
and  at the same time
the  ``ramp'' is getting longer.
At time-step $n=15$, the length of the  ``ramp'' in 
$g(\tau)=$ $\langle Z(\tau)Z^*(\tau)\rangle$  is already very close to that observed at time-step $n=30$.
In Fig.~\ref{fig:floq_z2_sub} we plot   the connected
spectral form factor
$g_c(\tau)=$ 
$\langle Z(\tau)Z^*(\tau)\rangle-\langle Z(\tau)\rangle\langle Z^*(\tau)\rangle$, 
which is seen to exhibit a ``ramp'' whose length continues to increase 
beyond time-step $n=15$ until the fully thermalized regime at time-step $n=30$ is reached.
These plots also show an overshoot  at the low-$\tau$ end of the ``ramp'' in $g_c(\tau)$, which
 is however  suppressed as the time-step $n$ increases further, 
 and at $n=30$  the overshoot  has basically disappeared 
(and $g_c(\tau)$   is  the same as that for the  ``random pure state''
- compare  Fig. \ref{fig:Floq_Ising} (b)),
indicating that
at this time-step chaos has fully developed in the subsystem. All these time scales depend on the length $L_A$ of
the  subsystem and become smaller as the subsystem size $L_A$  is decreased.

\begin{figure}[hbt]
\centering
 \subfigure[]{\label{fig:floq_H_z2} \includegraphics[width=.3\textwidth]{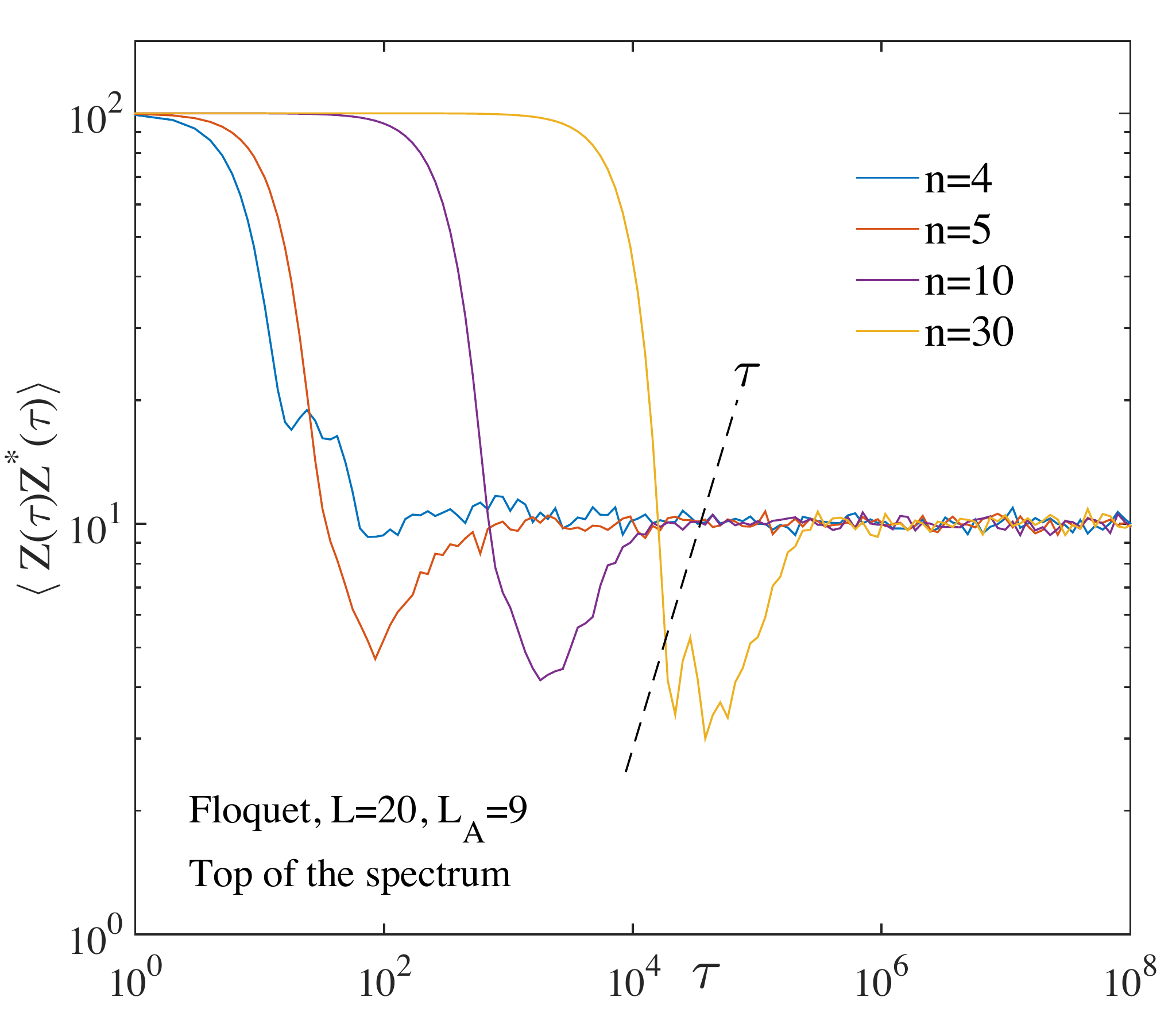}}
 \subfigure[]{\label{fig:floq_L_z2} \includegraphics[width=.3\textwidth]{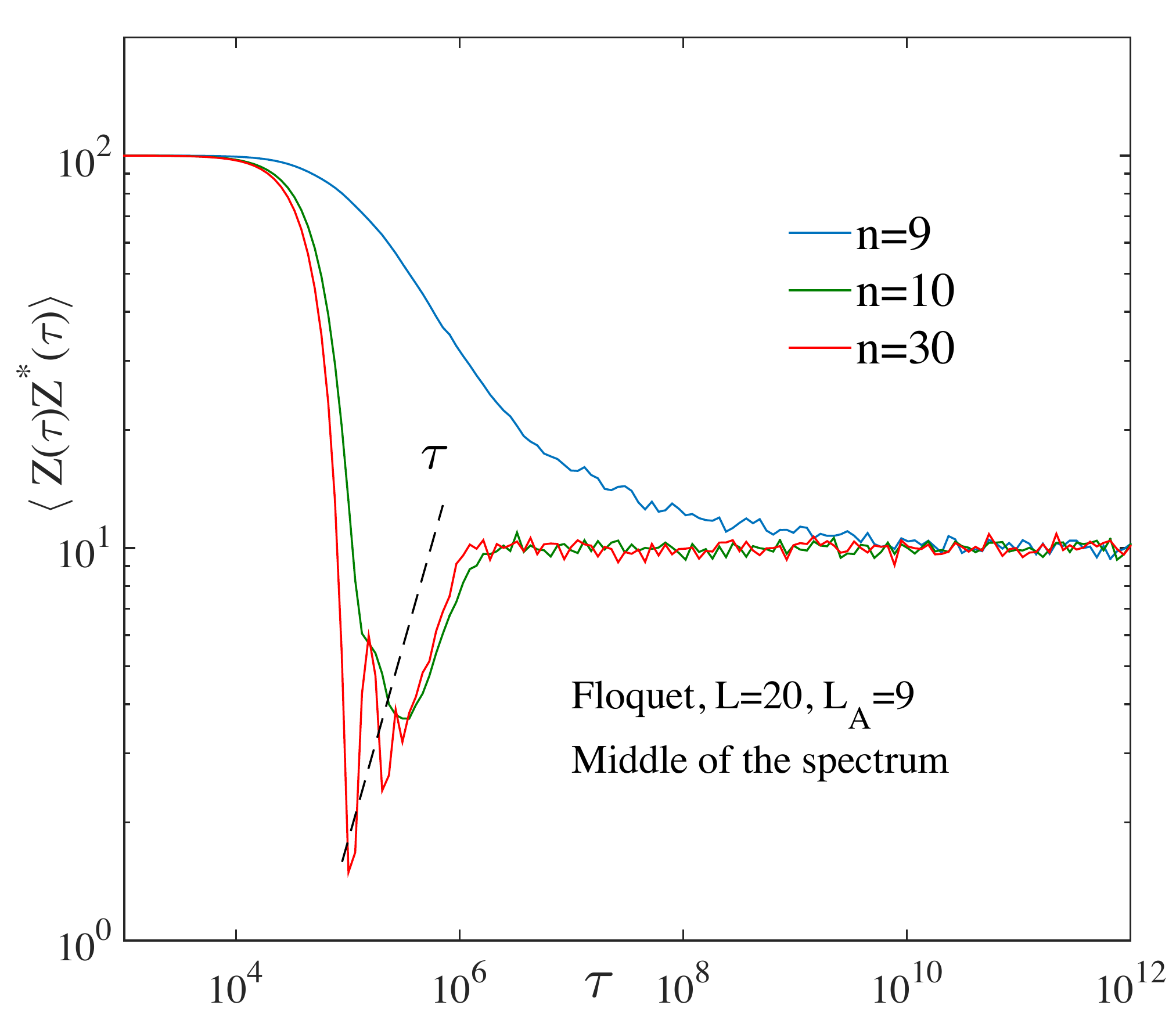}}  
 \subfigure[]{\label{fig:floq_L_z2} \includegraphics[width=.3\textwidth]{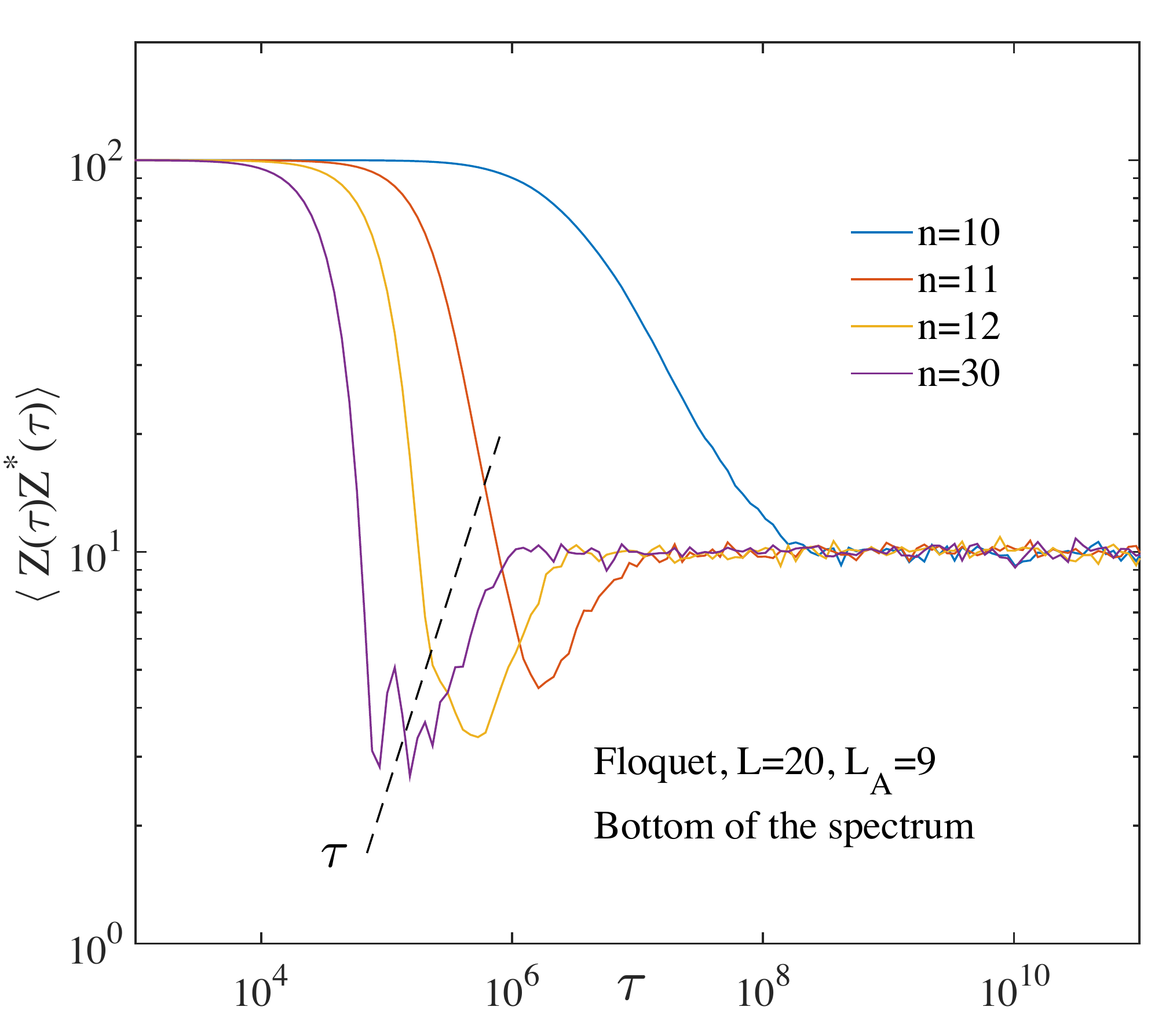}}   
\caption{
(a) $g(\tau)=$  $\langle ZZ^*\rangle$ versus $\tau$ for a subset (window) of 10 eigenvalues at the top of the spectrum of ${\hat \rho}_A$ for the  Floquet model defined in \eqref{floq} at different time-steps,  averaged over 1000 samples. 
(b)  $g(\tau)=$ $\langle ZZ^*\rangle$ for a subset of 10 eigenvalues in the middle of the spectrum of the same model. 
(c) $g(\tau)=$ $\langle ZZ^*\rangle$ for
a subset of 10 eigenvalues at the lower edge of the spectrum of the same model.
} 
\label{fig:floq_ME}
\end{figure}

We note that we have obtained the above results
% in the above calculations, we  computed 
upon computing the  spectral form factor by using in  \eqref{DEFSpectralFormFactor}
or (\ref{DEFZZstar}) 
%for
{\it all}  the eigenvalues 
%in 
of the reduced density matrix ${\hat \rho}_A(t=n T)$.
Actually, 
in order to gain additional insight,
it is useful to
%we can
 limit the eigenvalues
used to compute the spectral form factor in \eqref{DEFSpectralFormFactor} 
or (\ref{DEFZZstar}) 
to 
%values 
a subset lying in a  
%an energy 
window around a fixed 
%energy
eigenvalue, and to compute the spectral form factor
by only using 
%for the energies 
the eigenvalues of the density matrix in this window.
This procedure can then detect 
%the ``local level repulsion''
%the 
``local universal spectral correlations''
% rigidity" 
characterizing the correlations  amongst the 
%between the 
eigenvalues in this
% energy 
window.
 In Fig.~\ref{fig:floq_H_z2}, we present results for the
% quantity
 spectral form factor $g(\tau)=$ $\langle Z(\tau)Z^*(\tau)\rangle$ for a window of 10 consecutive eigenvalues at the top of 
spectrum\footnote{
We choose here the 10 largest eigenvalues, of a total of approx. $N_A=500$ levels of ${\hat \rho}_A$.
}
 in ${\hat \rho}_A$.
We  notice the appearance of a linear ``ramp'' as early as at time-step $n=5$. 
As we move the window of eigenvalues
 away from the top to the bottom of the spectrum of the density 
matrix\footnote{We choose here the 10 smallest eigenvalues, of a total of approx. $N_A=500$
levels of ${\hat \rho}_A$.
}, 
%{\color{blue} \it `` COMMENT by Andreas: Xiao, do you agree with the previous footnote?''}
we find that the linear ``ramp'' develops only  at  later time-steps
- here at $n=11$  (Fig.~\ref{fig:floq_L_z2}).
 This result demonstrates that in the Floquet model, as time $t$  evolves,
the universal spectral  correlations
%rigidity 
first  emerge  at the top of spectrum of ${\hat \rho}_A(t=nT)$ and 
subsequently spread over the
entire  spectrum at  later times $t$. 
This behavior, i.e. the fact that not the entire spectrum of the density matrix develops
the spectral correlations
% rigidity 
uniformly in time $t$,  is also responsible for the  shallowness of the dip that appears,
when the time-step  $n$ is between 11 and 15,
in  the plot in Fig.~\ref{fig:floq_z2}
of  the spectral form factor 
$g(\tau)=$ $\langle Z(\tau)Z^*(\tau)\rangle$ 
which uses the entire
%for the whole
spectrum as input.

Finally, we would like to discuss the connection between the linear growth of the  EE and the development
of universal  spectral correlations.
% rigidity.
When we 
%We 
look at the
% values
magnitudes  of  the eigenvalues  of $\rho_A(t)$ at early times, we
% and 
find that there are only
a few of them which
%which  
are appreciably different from zero;
%  it is  these
% several
%some  of them with nonzero values;
%these 
%which 
%which 
and it is them that
%they 
exhibit the spectral correlations
%level repulsion 
and are also responsible for the observed  value of the EE. 
Actually, they also give rise to the volume law  in the EE for  smaller subsystem sizes.
% with the same entanglement cut.
 %--- ``{\it  COMMENT by Andreas: What does * with the same entanglement cut* mean here?}'' ---
As time evolves, more and more eigenvalues become appreciably different from zero.
% nonzero. 
They develop spectral
% rigidity
correlations and lead to the linear growth of the EE. This is in contrast with 
an
%the 
integrable system, where the linear growth of the  EE is due to the ballistic propagation of quasiparticles\cite{Calabrese2007}
 and there are no spectral correlations
% is no level repulsion 
between the eigenvalues and hence there is  no ``ramp'' in the spectral form factor.\footnote{See the last paragraph
before Sec. \ref{LabelSectionFloquetPrethermal} and Fig. \ref{fig:ising_h_0}.} 
In Sec. \ref{LabelSectionFloquetPrethermal}
we will show an example of such a phenomenon within a Floquet system.

%%%%%%%%%%%%%%%%%%%%%%
%%%%%%%%%%%%%%%%%%

%%%%%%%%%%%%%%%%%%%%%%
%%%%%%%%%%%%%%%%%%

%%%%%%%%%%%%%%%%%%%%%%
\begin{figure}[hbt]
\centering
 \subfigure[]{\label{fig:Ising_z2} \includegraphics[width=.4\textwidth]{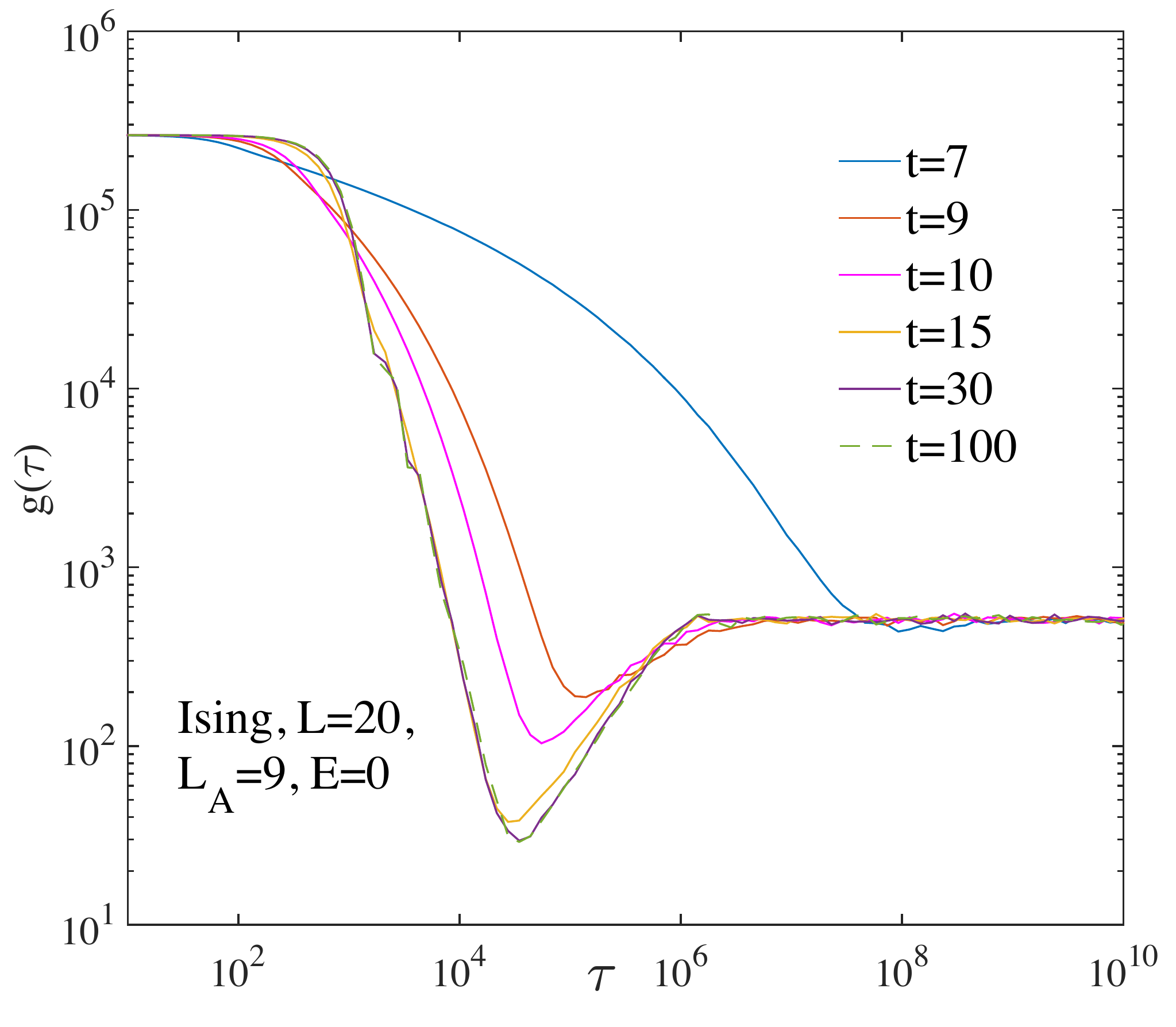}}
 \subfigure[]{\label{fig:Ising_z2_sub} \includegraphics[width=.4\textwidth]{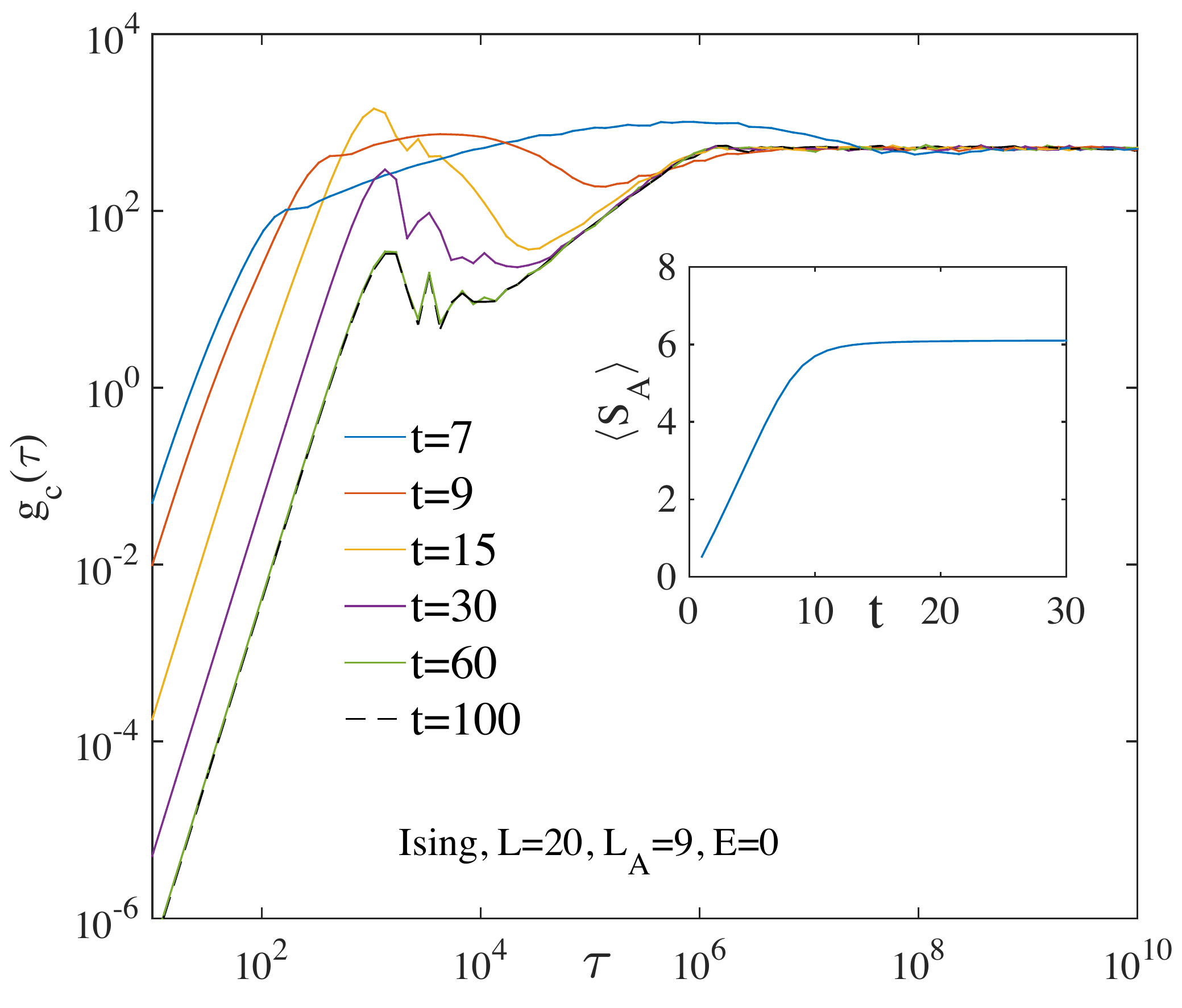}}   
\caption{
(a) $g(\tau)=$ $\langle Z(\tau)Z^*(\tau)\rangle$ versus $\tau$
of the spectrum of ${\hat \rho}_A$ 
%of the quantum Ising model
 for states at ``high energy'' expectation values  $E\in [-0.1, +0.1]$ of the quantum Hamiltonian defined in (\ref{ising_H}), which has support in the interval $[-26, +33]$. Different curves correspond to  different times $t$, and  averages were taken over 1000 samples. 
 %for Ising model defined in \eqref{floq} at different times $t$, averaged over 1000 samples.
 (b) $g_c(\tau)=$  $\langle Z(\tau)Z^*(\tau)\rangle-\langle Z(\tau)\rangle\langle Z^*(\tau)\rangle$ 
for the same model at different times $t$, under otherwise identical conditions.
 The inset
% is
shows  the averaged 
%von Neumann EE 
entanglement entropy (EE) as a function of time $t$.
} 
\label{fig:ising}
\end{figure}
%%%%%%%%%%%%%%%%%%

%%%%%%%%%%%%%%%%%%%%%%
\begin{figure}[hbt]
\centering
 \subfigure[]{\label{fig:Ising_T_z2} \includegraphics[width=.3\textwidth]{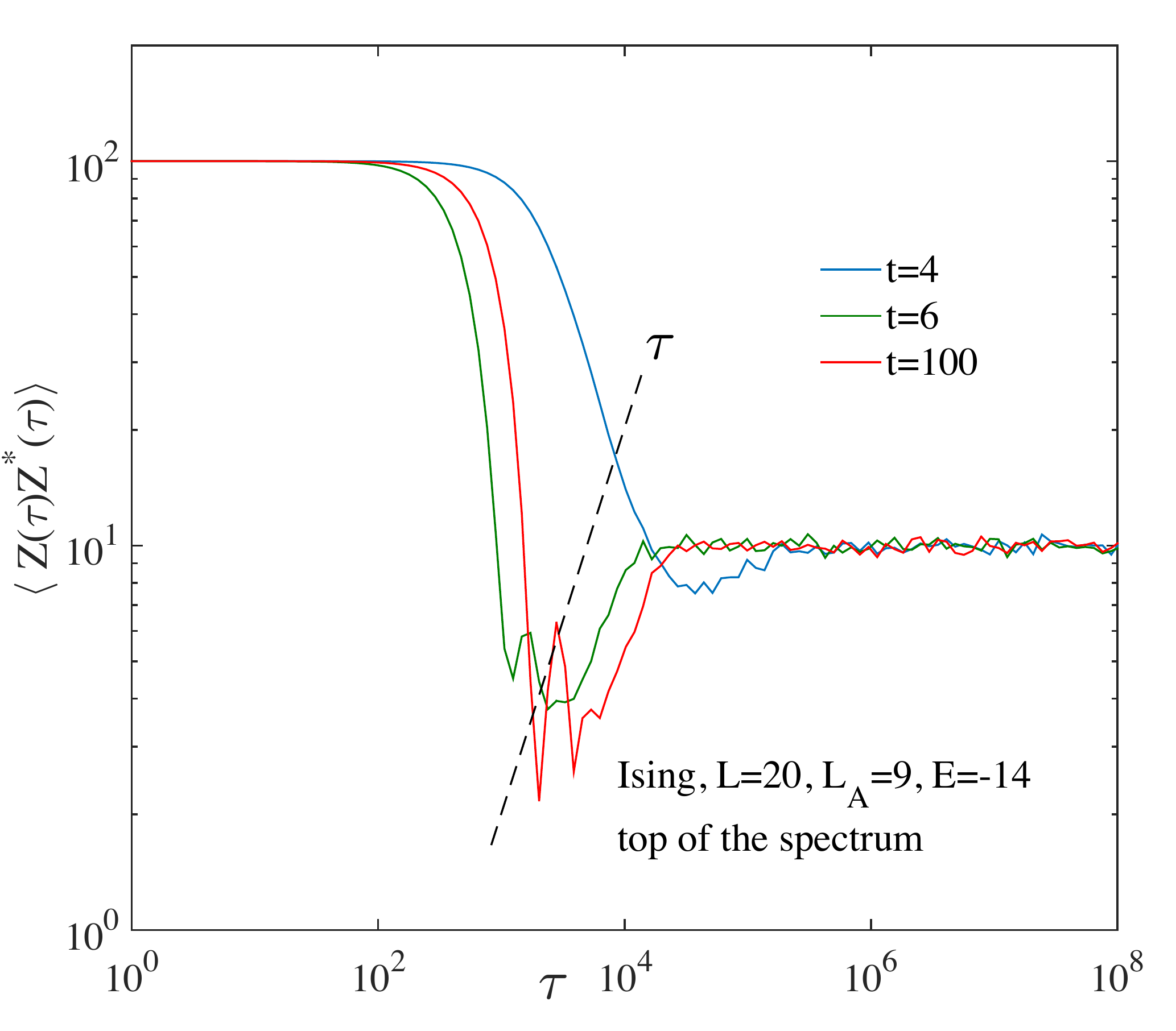}}
 \subfigure[]{\label{fig:Ising_M_z2} \includegraphics[width=.3\textwidth]{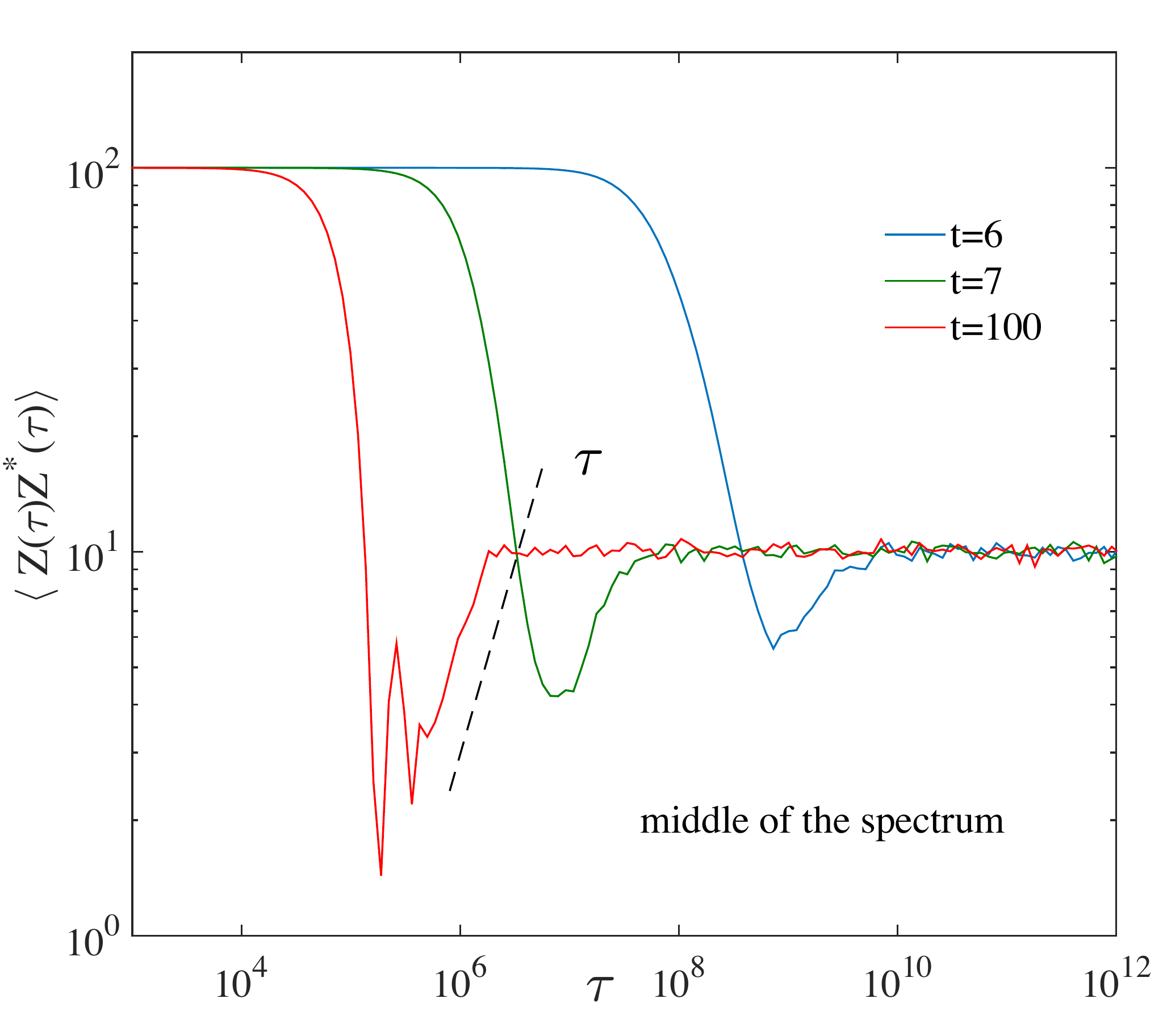}}
 \subfigure[]{\label{fig:Ising_E_z2} \includegraphics[width=.3\textwidth]{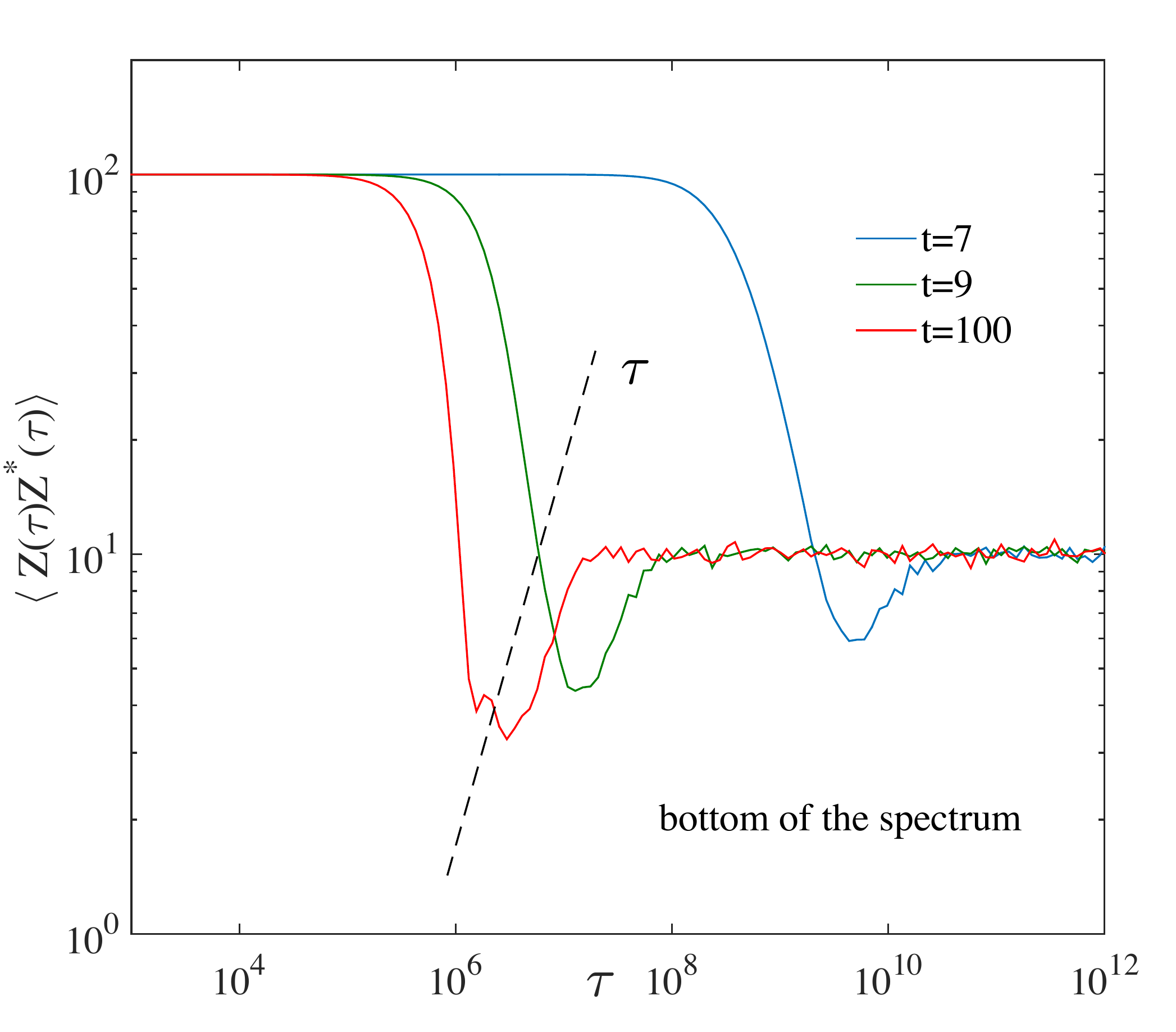}}   
\caption{
(a) $g(\tau)=$ $\langle ZZ^*\rangle$ versus $\tau$ for a subset (window)  of 10 eigenvalues  close to the upper edge 
%top 
of the spectrum of ${\hat \rho}_A$ of the quantum Ising model for states at ``low energy''  expectation values  $E\in [-14.1, -13.9]$ of the Hamiltonian defined in
(\ref{ising_H}), which has support in the interval $[-26, +33]$. Different curves correspond to  different times $t$, and  averages were taken over 1000 samples. (b)  $g(\tau)=$ $\langle ZZ^*\rangle$ for a subset of 10 eigenvalues in the middle of the spectrum of the same model, and otherwise identical conditions. (c)  $g(\tau)=$ $\langle ZZ^*\rangle$ for a subset of 10 eigenvalues close to the lower edge of the spectrum of the same model, and otherwise identical conditions.
} 
\label{fig:Ising_ME}
\end{figure}
%%%%%%%%%%%%%%%%%%

%%%%%%%%%%%%%%%%%%%%%%%
%\begin{figure}[hbt]
%\centering
% \subfigure[]{\label{fig:Ising_M_z2} \includegraphics[width=.4\textwidth]{Ising_low_Z2_M.pdf}}
% \subfigure[]{\label{fig:Ising_E_z2} \includegraphics[width=.4\textwidth]{Ising_low_Z2_E.pdf}}   
%\caption{
%(a) $g(\tau)=$ $\langle ZZ^*\rangle$ versus $\tau$ for a subset (window)  of 20 eigenvalues in the middle of the spectrum of ${\hat \rho}_A$ for the quantum Ising model defined in Eq. \eqref{floq} at different times,  averaged over 1000 samples. (b)  $g(\tau)=$ $\langle ZZ^*\rangle$ for a subset of 20 eigenvalues at the lower edge of the spectrum of the same model.
%} 
%\label{fig:Ising_ME}
%\end{figure}
%%%%%%%%%%%%%%%%%%%

\subsection{Ising Model}
In this section we investigate the 
%onset
development of chaos  in a quantum 
%For the 
Ising model, where the thermalization process is slower due to the
presence of the energy conservation law.

{\it High energy states}:
 For an ensemble of
% the
 initial random direct product states
$|\Psi_0\rangle$
 with Hamiltonian expectation value in the small  energy interval 
 $E\in[-0.1,0.1]$
 (i.e. $E=\langle \Psi_0| {\hat H}| \Psi_0\rangle$
in the middle of the spectrum of the Ising Hamiltonian ${\hat H }$ in (\ref{ising_H})),  
the EE grows linearly with time until 
% $t\leq 7$ 
$t \approx 7$ and  then saturates
exponentially to
the final  volume law 
%exponentially
 at  longer times around $t=100$ [see inset in Fig.~\ref{fig:ising}(b)].
Since this initial  state has energy $E$ close to zero
(middle of the spectrum of ${\hat H}$), the effective temperature is high and the saturation value of the EE is
only  slightly smaller than the Page value.  
%We notice that
In particular,  when $L_A=9$ (total system size $L=20$), the  difference between 
the two values of the EE
 is seen to be around 0.0045, which is less than $0.1\%$ of the  EE of the Page state.
%{\color{blue} {\it "We need to write this out $(S_I-S_P)/S_P$ ??}
%}

In Fig.~\ref{fig:ising}, we present
results for 
the spectral form factors $g(\tau)=$ $\langle Z(\tau)Z^*(\tau)\rangle$ and 
$g_c(\tau)=$ $\langle Z(\tau)Z^*(\tau)\rangle-\langle Z(\tau)\rangle\langle Z^*(\tau)\rangle$ at 
different 
%values of time 
times $t$.
%various times. 
In Fig.~\ref{fig:Ising_z2}, the ``ramp''  in $g(\tau)$ starts to
%appear
emerge at around $t=9$. As $t$ increases further, the dip becomes deeper and shifts to 
%an 
earlier times. At $t=15$, a linear ``ramp'' has fully  developed and
 remains almost  unchanged until the system is fully thermalized at  $t=100$.
In Fig.~\ref{fig:Ising_z2_sub} we plot the
% ``connected'' observable
connected spectral form factor $g_c(\tau)$, which is seen to exhibit a ``ramp'' that continues to grow further in length, even beyond time $t=15$.
%In
However, in contrast to  the Floquet model,  the
overshoot appearing at the low-$\tau$ end of the ``ramp''
cannot be fully suppressed and is always 
present even after a very
% sufficient
 long time evolution 
( - compare also
Fig. \ref{fig:Floq_Ising}(b)).
This   indicates that the energy conservation law makes the Ising model ``less chaotic'' than the Floquet model,
in the sense that universal spectral 
%rigidity does 
correlations do not extend to pairs of eigenvalues $\lambda_i$ as far separated as in the Floquet model.
We have also computed the 
%quantity
spectral form factor  using only a subset  of eigenvalues of ${\hat \rho}_A(t)$ in a  window around an eigenvalue 
%in the middle of its 
at the top, in the middle and at the bottom of its
spectrum
and we find similar behavior as 
 in the Floquet model:
%level repulsion 
Universal spectral 
%rigidity 
correlations first 
%develops 
emerge at the top
%in the middle 
of the spectrum and then spread over the rest of spectrum.

%%%%%%%%%%%%%%%%%%%%%%%
%\begin{figure}[hbt]
%\centering
% \subfigure[]{\label{fig:Ising_low} \includegraphics[width=.4\textwidth]{Ising_low_Z2.pdf}}
% \subfigure[]{\label{fig:Ising_sub_low} \includegraphics[width=.4\textwidth]{Ising_low_Z2_sub.pdf}}   
%\caption{
%(a) $\langle ZZ^*\rangle$ versus $\tau$ for Ising model defined in Eq.\eqref{ising_H} at different time averaged over 1000 samples. The initial product state is close to Neel state with lower energy between -13.9 and 14.1. (b) $\langle ZZ^*\rangle-\langle Z\rangle\langle Z^*\rangle$ for the same model at different time. The inset is the averaged von Neumann EE as a function of time.
%} 
%\label{fig:ising_low}
%\end{figure}
%%%%%%%%%%%%%%%%%%%

{\it Lower energy states}:
 Since for  the 
Ising model energy is conserved,
we can also study 
the 
% level repulsion 
spectral correlations
%rigidity 
%characteristics 
of ${\hat \rho}_A(t)$ which arise  upon time evolution starting
 from an initial state
% arising from an initial state
with  a lower energy $E$  (= Hamiltonian expectation value $=\langle \psi_0| {\hat H} | \psi_0\rangle$)
corresponding to properties of the quantum Ising Hamiltonian (\ref{ising_H}) at relatively low temperatures. (Though,
 $E$ is  separated
from the ground state by many levels. The spectrum of ${\hat H}$ has support in an interval which is approximately $[-26,+33]$.)
In particular, we
consider an initial direct  product state close to the Neel state with
% within 
%the
%and 
an  energy  in the narrow
%window
interval  $E\in[-14.1, -13.9]$ rather than the  random product state
in the middle of the spectrum of ${\hat H}$, which  has high excitation energy $E$, considered above. 
Furthermore, we introduce some randomness into this
ensemble of  initial states so that we can perform a 
 disorder (ensemble) average over them. Under
the unitary time evolution evolution, the EE is found to initially grow
linearly with  time $t$, and to saturate as expected
to a smaller volume law after long time evolution
as compared to the case of a random initial state with energy $E$  in the middle of the spectrum
of ${\hat H}$.
%,  considered above.
For $L_A=9$ (total system size $L=20$), 
we start to observe  a dip in
the spectral form factor $g(\tau)=$  $\langle Z(\tau)Z^*(\tau)\rangle$ at times around $t=9$.
% (Fig.~\ref{fig:Ising_M_z2}).
As before, we also compute the spectral form factor $g(\tau)=$ $\langle ZZ^*\rangle$ by only using
a subset of eigenvalues of ${\hat \rho}_A(t)$ locally in a window around
a fixed eigenvalue of the density matrix.
% (Fig.~\ref{fig:Ising_E_z2}).
For a window of 10 consecutive
% the 
eigenvalues 
%in 
%at the very  
%near 
close to the top of the spectrum
%\footnote{
%\color{blue}
%Here  we choose a window of  10 consecutive eigenvalues, the largest being separated by 10 eigenvalues from the top of the spectrum, of a total %of approx. $N_A=500$ levels of ${\hat \rho}_A$.
%}
of the density matrix ${\hat \rho}_A$
%{\color{blue} \it  ``COMMENT by Andreas: Xiao, do you agree with the previous footnote?''}
we find,
% a linear ``ramp'' similar to that found  for
similar to the  Floquet model,   a  ``ramp''
%and for 
%the  Ising model  at higher energy $E\approx 0$ (Fig.~\ref{fig:Ising_T_z2}),
%but a ``ramp'' appears to emerge already
 already at an 
%earlier
early  time $t=4$ which becomes linear at $t=6$ (Fig.~\ref{fig:Ising_T_z2}). 
On the other hand, as we move the window of 10 consecutive eigenvalues
close  to the 
%very 
bottom of the 
spectrum
%{\color{blue} \it  ``COMMENT by Andreas: Xiao, do you agree with the previous footnote?''}
 of the density matrix
 (Fig.~\ref{fig:Ising_E_z2}),
%level repulsion
the spectral
correlations 
% rigidity 
%also 
emerge only  
%also appears 
at 
%a 
later times  $t$  as compared to the case where the window is 
%in the center
at the top  of the spectrum.
%\footnote{\color{blue} For this window the ``ramp'' appears to be not precisely linear, 
%which we believe is a consequence of numerical issues arising %from the smallness of the eigenvalues near the bottom of the spectrum  of ${\hat %\rho}_A$.
%Different curves correspond to  different times $t$, and  averages were taken over 1000 samples. }
This is analgous to what was observed in the Floquet case.

% except that the ``ramps" generated from the windows of the very top 10 and the very  bottom 10 levels of the spectrum of a total of approx. %$N_A=500$ levels are not linear (the ramp generated from the window of 10 levels in the middle
%of the spectrum {\it is}  linear). This is likely due to effects arising at the extreme edges of the spectrum, including  the fact that the level spacing %between the top 10 levels is far from uniform - and similar for the bottom 10 levels -,  and we have not ``unfolded'' these two  spectra to ones with %the same level spacing between all 10 levels, before computing the ``ramp''.

%{\color{blue} \it  ``COMMENT by Andreas: Xiao, do you agree?''}

%{\color{red}Most properties are similar to what were observed in the Floquet case. One small difference is that in the Ising model for the state with %low energy, the ``ramp" is  linear only in the middel of the spectrum. }
%%
%%{\color{blue}
% %``{\it We could mention spont. symmetry breaking in this phase; mention Fratus+Srednicki work. Could say that there are no effects
%of many-body physics on spectral rigidity/chaos, since this is just about stat mech, valid in all phases (of course excluding any MBL phas%%es)}''
%%}

\begin{figure}%[hbt]
\centering
\includegraphics[width=.5\textwidth]{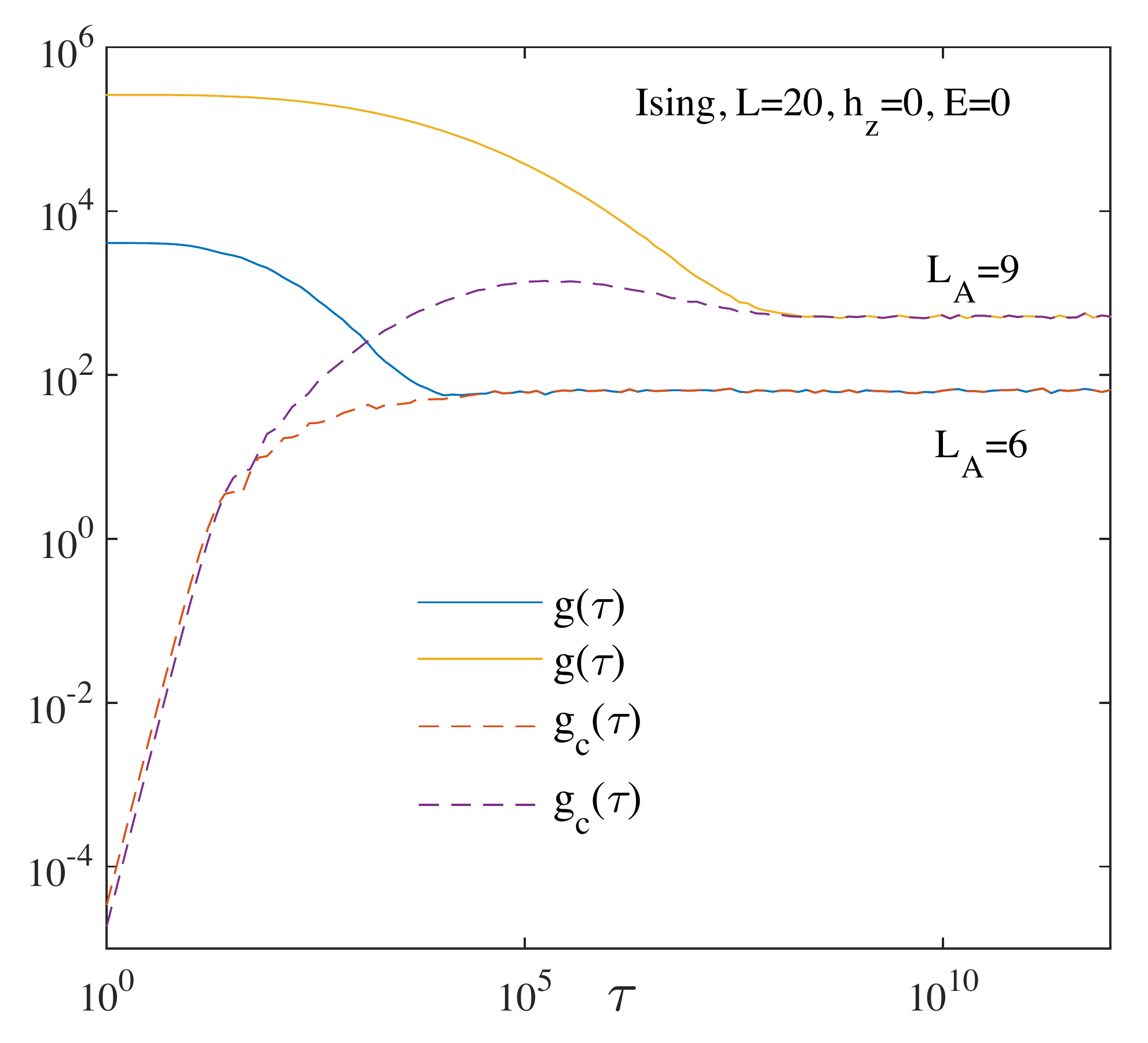}
\caption{
$g(\tau)$ (solid curves) and $g_c(\tau)$ (dashed curves) versus $\tau$ for the quantum  Ising model defined in Eq.\eqref{ising_H} for the integrable case where the longitudinal field  vanishes,  $h_z=0$, averaged over 1000 samples of  initial random direct product states
% in the energy 
with energy expectation values in the interval  $E\in [-0.1, + 0.1]$, i.e. $E$ is close to zero.
%(a) $g(\tau)=$ $\langle ZZ^*\rangle$ versus $\tau$ for the quantum  Ising model defined in Eq.\eqref{ising_H} for the integrable case
%where the longitudinal field  $h_z=0$,  
%%{\color{blue} {\it Comment:} at different times, ?}
% averaged over 1000 samples. 
%%The longitudinal field is $h=0$, and
%% with
%Average oven an ensemble of 1000 samples of  initial random states in the energy interval
%% has energy 
%$E$  between -0.1 and 0.1.
% (b) $g_c(\tau)=$  $\langle ZZ^*\rangle-\langle Z\rangle\langle Z^*\rangle$ 
%versus $\tau$  for the same model and under the same conditions.
%{\color{blue}
%The inset 
%%is
%shows  the averaged 
%%von Neumann 
%entanglement entropy (EE) as a function of time. 
%``{\it COMMENT by Andreas: There is no inset in this figure.}''
%}
}
\label{fig:ising_h_0}
\end{figure}

Finally, we turn off the longitudinal field $h_z$
in the quantum Ising Hamiltonian (\ref{ising_H}),
 so that the model becomes integrable. 
An
%For the 
initial random direct product state
equilibrates after the quantum quench
to a thermal state described by the generalized Gibbs ensemble (GGE) with an extensive number of conserved quantities.\cite{Rigol2007,Rigol2006,Calabrese2007} As shown in Fig.~\ref{fig:ising_h_0}, we do not observe any ``ramp'' in
the spectral form factor $g(\tau)=$  $\langle Z(\tau)Z^*(\tau)\rangle$, 
indicating the absence of chaos
reflected in the absence of universal spectral correlations 
% level repulsion
%spectral rigidity  and chaos 
in the reduced density matrix  ${\hat \rho}_A$.

%%%%%%%%%%%%%%
%%%%%%%%%%%%%% 

\subsection{Floquet system with prethermal regime}
\label{LabelSectionFloquetPrethermal}

As  discussed before, for a generic Floquet system
% model
the reduced  density matrix of a general  
short-range entangled  initial state 
 will reach a steady state 
%of
at  infinite temperature
after a sufficiently long time evolution, since energy is not conserved.\cite{Kim_Huse_2014,DAlessio2014,PONTE2015,DALESSIO2013}
How thermal equilibrium and chaos emerge
in the wavefunction  is model-dependent. Recently, it has been shown that 
 a rapidly driven system 
may exhibit 
an
%a long 
 intermediate prethermal regime of long duration
 in which the system reaches a thermal equilibrium state governed by an approximate time-independent Hamiltonian with the effective temperature set by the initial energy.\cite{Mori2016,Abanin2015,Abanin_2017,Else_2017,Machado2017} This regime can have an
exponentially long lifetime (in units of inverse frequency, and other parameters of the system).
%{\it ``COMMENT by Andreas: Do we still want to talk about an *exponentially long lifetime*  here?"}
We   note that a prethermalized regime in a system with  a time-independent (as opposed to Floquet)  Hamiltonian
has also been considered recently in the context of non-integrable perturbations of integrable many-body systems\footnote{See e.g. 
Ref. \onlinecite{BertiniEsslerGrohaRobinsonPRB2016} and references therein.}, but we do not discuss these situations here.

In this section, we are going to explore a Floquet model 
that exhibits
%which can exhibit
% a
such prethermalization to a thermal state 
%approximated by 
which is close to  that of a nearly-integrable system.
 The Floquet operator that we use to achieve such
a prethermalized regime takes the following form,
\begin{align}
\hat{U}_F=\exp[-i t_0 \hat{H}_1]\exp\left[-it_0\hat H_2\right] ~,
\label{flo_op_pre}
\end{align}
where
\begin{align}
&\hat{H}_1=-\sum_{j=1}^{L-1}\hat\sigma_j^z\hat\sigma_{j+1}^z-h_x\sum_{j=1}^L\hat\sigma_j^x \nonumber\\
&\hat{H}_2=-h_y\sum_{j=1}^{L}\hat\sigma_j^y ~.
\label{floq_pre}
\end{align}
This model is a one-dimensional periodically driven system with period $T=2t_0$. In the numerical calculations, we choose open
boundary conditions and system parameters $(h_x,h_y)=(1, 1)$.
 The period $T=0.2$ is chosen to be very small
in order to realize a long prethermal regime.

\begin{figure}%[hbt]
\centering
\includegraphics[width=.5\textwidth]{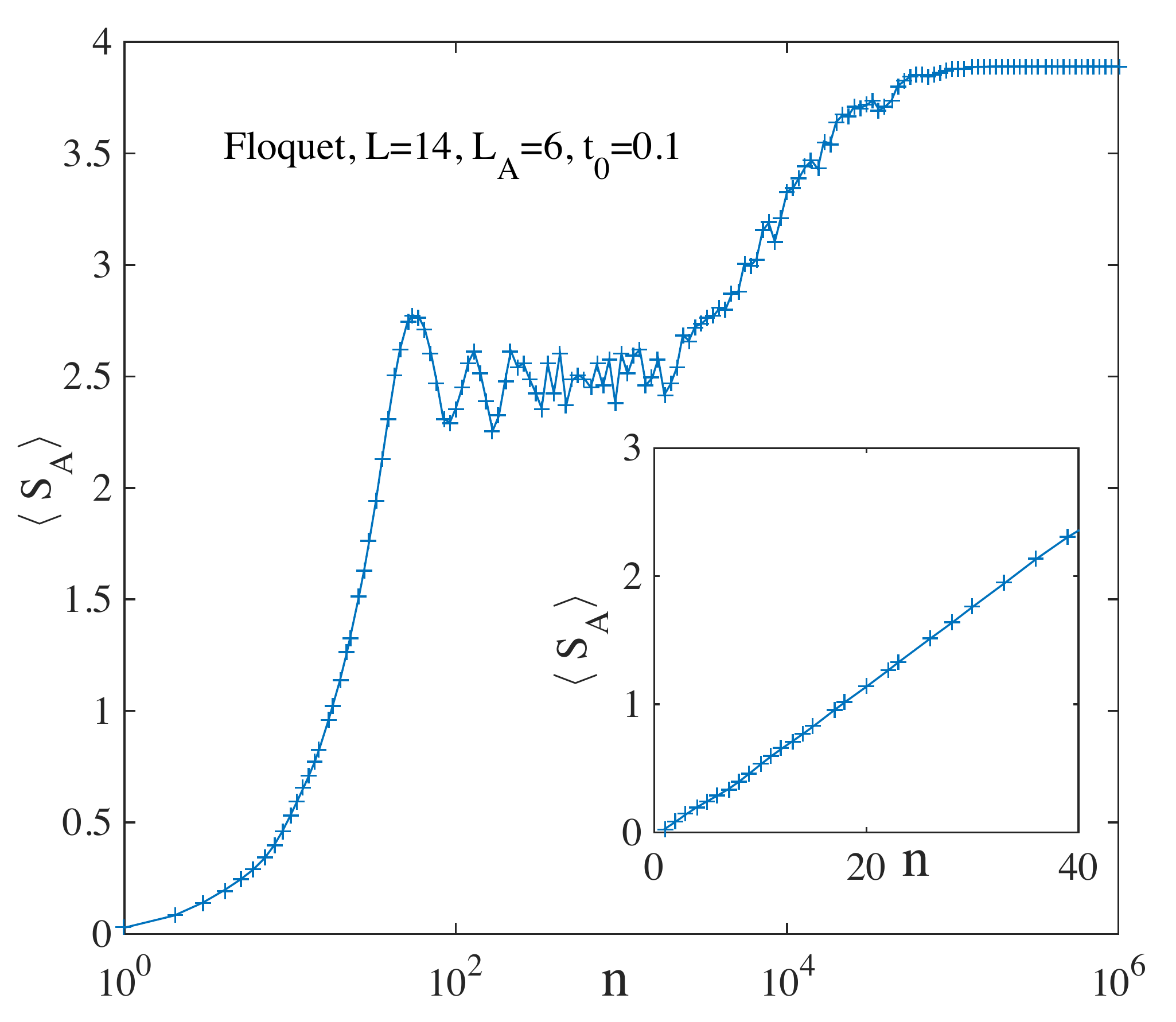}
\caption{Time dependence of the entanglement entropy
%The EE dynamics for
of  the Floquet system described by \eqref{floq_pre} on 
%the 
a semi-log scale.
The period of $T= 2 t_0 = 0.2$.
The result is averaged over an ensemble of 400 wavefunctions.
 The inset
% is EE
shows the entanglement entropy  at early times on the linear scale, 
exhibiting linear growth as expected.
 }
\label{fig:prethermal_EE}
\end{figure}

\begin{figure}%[hbt]
\centering
\includegraphics[width=.8\textwidth]{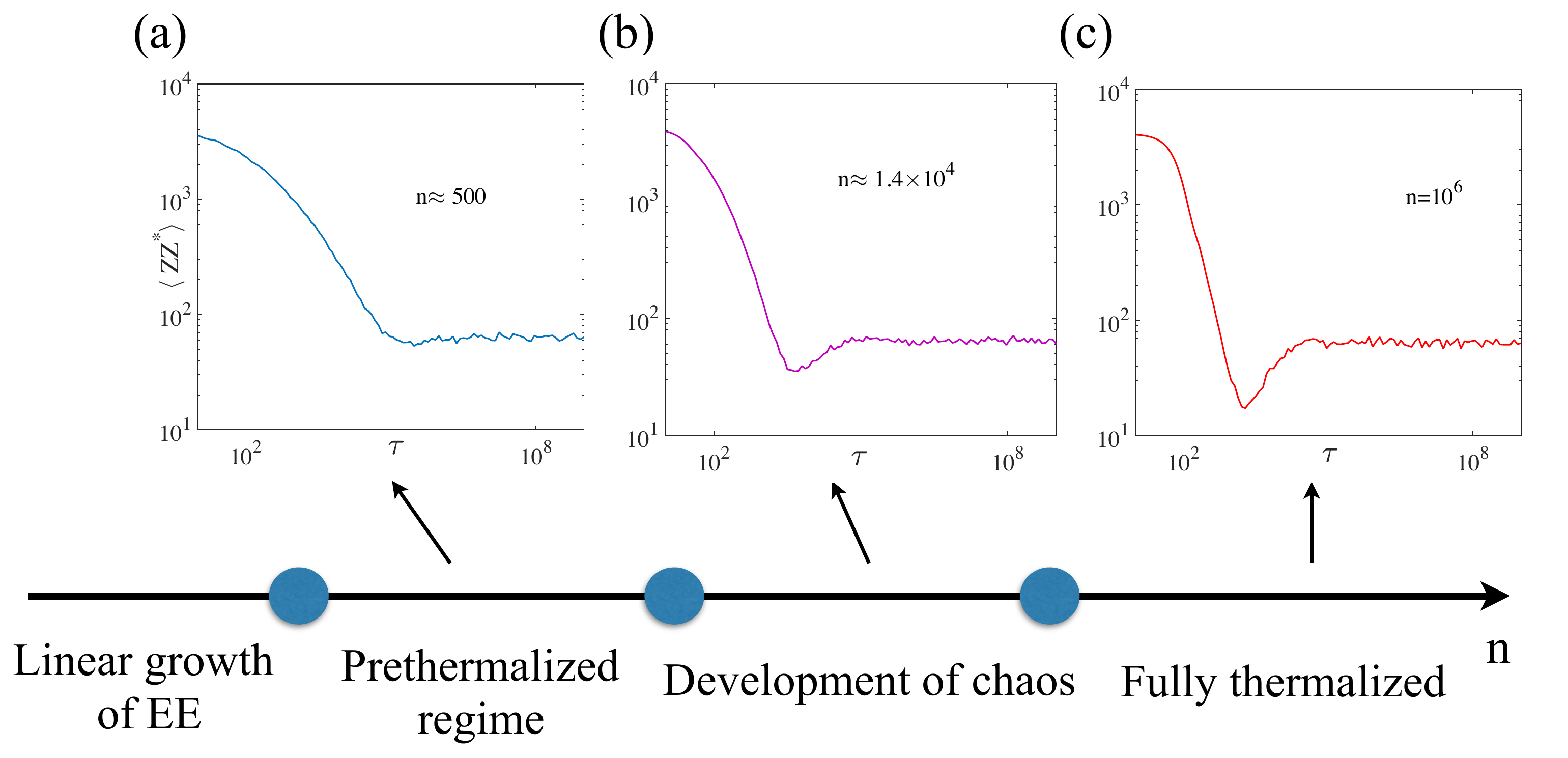}
\caption{Spectral form factor $g(\tau)=$ $\langle ZZ^*\rangle$ for the Floquet model in \eqref{floq_pre} at different stages in the time-evolution
(time-step $n$). 
Each curve 
%is
represents the average over an ensemble of 400 wavefunctions.} 
\label{fig:prethermal_z2}
\end{figure}

Since the period $T$ is very small, it takes a 
large number  of time-steps for the Floquet system to relax to 
its ultimate,  fully thermalized (chaotic)  state.
The previous method used above for rapidly
% fast
 thermalizing Floquet systems 
that
%which 
simply amounted to applying the time-evolution operator many times 
to an initial state, 
which did not 
require
%involve  
diagonalizing the Floquet operator,  is no longer useful here due to the large number of required time-steps.
Here we will instead consider
a smaller system size with $L=14$ so that we can diagonalize the Floquet operator explicitly  and study the long time dynamics
by applying that operator for any  length of time to the initial state.
We start with a  random  direct product state and evolve it under the Floquet operator.  The result for the time evolution of the  EE is shown in Fig.~\ref{fig:prethermal_EE}, where we clearly observe a long  intermediate plateau which corresponds to the 
prethermalized regime. 
We further have computed the spectral form factor 
$g(\tau)=$  $\langle Z(\tau)Z^*(\tau)\rangle$ 
in this (prethermalized)  regime and we do not observe any dip or ``ramp'',
demonstrating that
% there is no 
%level repulsion between
 the eigenvalues of the reduced density matrix ${\hat \rho}_A(t=nT)$
do not exhibit any universal spectral 
%rigidity
correlations  in this regime. 
The lack of the spectral 
%rigidty 
correlations shows that chaos is absent in this regime, and we expect
that it is
%We expect that this
% regime, exhibiting a long plateau of the EE, is 
described by a Generalized Gibbs Ensemble (GGE).
%, and by absence of chaos.
On the other hand,  chaos starts to
appear at yet longer times where the 
EE increases further
and eventually relaxes to the Page value;
% where 
at those longer times, a small ``ramp'' is seen to  develop in the spectral form factor depicted  in Fig.~\ref{fig:prethermal_z2} (b). 
A more pronounced linear ``ramp'' can be observed 
(see Fig. \ref{fig:prethermal_z2}) once the state reaches full thermalization at still  larger
time-steps $n$.
% times $t$.

Therefore, in the present  Floquet model, we can separate the long time evolution into four stages 
 (see Fig.s~\ref{fig:prethermal_EE} and
\ref{fig:prethermal_z2}): 
(1) a regime of linear growth of the EE,
which also  appears in all of the previous models (see inset of Fig.~\ref{fig:prethermal_EE}), (2) the prethermal regime described by GGE and absence of chaos,
which is reflected by a plateau in the time-evolution of the EE,
 (3) the regime of 
% onset
development  of chaos
where
% level repulsion
universal spectral 
%rigidity 
correlations start to develop 
%in the middle 
at the top of the spectrum  of the density matrix ${\hat \rho}_A$,
%  starts to develop in the middle of its spectrum,
and (4) the fully thermalized regime, where the
initial state has time-evolved into a state whose reduced density matrix exhibits a spectrum indistinguishable from that of the density
matrix of a
% is the 
featureless ``random pure state''.

%%%%%%%%%%%%%%
%%%%%%%%%%%%%% 

%%%%%%%%%%%%%%
%%%%%%%%%%%%%% 

\section{Analytical calculations for the  ``random pure state''}
\label{analytical}

\subsection{Random pure state and Wishart-Laguerre ensemble}
\label{SubSectionRandomPureStateWishart}

In this subsection we briefly review the connection between  the reduced density matrix of the ``random pure state'' (Page state)
and  the Wishart random matrix ensemble.
First, we
decompose the total Hilbert space into
a tensor product of the Hilbert spaces of
% those of
 the two subsystems $A$ and $B$
 with dimensions $N_A$ and $N_B$, respectively
(assuming $N_A \leq N_B$ without loss of generality), and write the ``random pure state'' defined in   (\ref{DEFPageState})
 in a direct (tensor) product basis
\begin{align}
|\Psi\rangle=\sum_{i=1}^{N_A} \sum_{J=1}^{N_B} X_{iJ} \ |\Psi_A^i\rangle\otimes|\Psi_B^J\rangle
\end{align}
where the coefficients $X_{iJ}$ are complex Gaussian random variables,  and form a rectangular $N_A\times N_B$ random matrix $X$ subject to the normalization constraint $\mbox{Tr}(XX^{\dag})=1$. After tracing out subsystem B, we obtain from this wavefunction
 the reduced density matrix  ${\hat \rho}_A=XX^{\dag}$ for subsystem A, which is a $N_A\times N_A$ square matrix.

In order to make contact with the Wishart-Laguerre random matrix ensemble, we consider a (unconstrained) 
$N_A\times N_B$ complex  random matrix $Y=\{Y_{iJ}\}$ whose statistically independent complex  matrix elements are drawn from a Gaussian 
probability distribution
\begin{eqnarray}
\label{WishartGaussianDistribution}
P(\{Y_{iJ}\})
=
{\cal N}^{-1}
\exp  \{ - {\beta\over 2} \  N_B \mbox{Tr}(Y Y^\dagger) \}.
\end{eqnarray}
The $N_A\times N_A$ matrix $W\equiv YY^\dagger$ is then a random matrix belonging to what is known as the $\beta=2$ 
(``GUE-type'') Wishart
random matrix ensemble. Consequently, the density matrix for the ``random pure state'', discussed above, can  be expressed in terms of the Wishart
random matrix as follows
\begin{eqnarray}
\label{DEFDensityMatrixNormalizedWishart}
{\hat  \rho}_A \equiv {Y Y^\dagger
\over \mbox{Tr} (Y Y^\dagger)}.
\end{eqnarray}
We finally note that the denominator on the right hand side of (\ref{DEFDensityMatrixNormalizedWishart}) has
expectation value
\begin{eqnarray}
\label{ExpectationValueTraceW}
\langle \mbox{Tr} (Y Y^\dagger)\rangle = 
\sum_{i=1}^{N_A}\sum_{J=1}^{N_B} \ 
\langle |Y_{iJ}|^2\rangle
={(N_A N_B)\over N_B} = N_A.
\end{eqnarray}
Thus,
 in the limit where both $N_A$ and $N_B$ tend to infinity while the ratio $\alpha\equiv N_A/N_B$ remains fixed, the relative fluctuations $f$
of the random variable $ \mbox{Tr} (Y Y^\dagger)=$ $N_A (1+ f)$ about its mean $N_A$ vanish, and we can replace 
\begin{eqnarray}
\label{LimitTrYYdagNA}
\mbox{Tr} (Y Y^\dagger) \to N_A.
\end{eqnarray}
Owing to (\ref{DEFDensityMatrixNormalizedWishart})
the eigenvalues $\lambda_i$ of the reduced density
matrix ${\hat \rho}_A$ are thus  related in the  limit of large $N_A$ and $N_B$ to the eigenvalues $\mu_i$ of the Wishart matrix $W$
via
\begin{eqnarray}
\label{DEFEigenvaluesOfRhoA}
\lambda_i = {\mu_i\over N_A}.
\end{eqnarray}

Clearly, the above-described relationship 
 immediately extends to the other two universality  classes of GOE ($\beta=1$) and GSE ($\beta=4$) ``random pure state''s and Wishart random matrix ensembles.

% and belongs to $\beta=2$ Wishart-Laguerre ensemble with a fixed trace constraint.

\subsection{Eigenvalue statistics of the Wishart-Laguerre ensemble}
\label{SubSectionWishartRandomMatrixEnsemble}
Here we first briefly review some important results for the Wishart-Laguerre random matrix ensemble.\cite{loggas}
For more details, see Appendix \ref{AppendixSectionDetailsWishartRandomMatrixTheory}.

 In general, for a Wishart matrix $W=YY^{\dag}$ with $Y$ being a $N_A\times N_B$ matrix with real $(\beta=1)$, complex $(\beta=2)$ or quaternion $(\beta=4)$ Gaussian entries drawn from the joint distribution
as in (\ref{WishartGaussianDistribution}),
% $P(Y_{ij})\propto \exp\left[ -\frac{\beta}{2}\mbox{Tr}YY^{\dag} \right]$, 
the joint probability probability  distribution for the $N_A$ eigenvalues  $\mu_i$ of 
% the Wishart matrix
 $W$ is known to be\cite{loggas} 
\begin{eqnarray}
\label{log_gas}
&&
\qquad \qquad  P[\{\mu_i\}]=
%C_{N_A,N_B}e^{-\frac{\beta}{2}\sum_{i=1}^{N_A}\mu_i}\prod_{i=1}^{N_A} \mu_{i}^{\nu\beta/2}\prod_{1\leq j<k\leq N_A}|\mu_j-\mu_k|%^{\beta}
 {\cal {\tilde  N}}^{-1}
 \exp[-\beta E(\{\mu_i\})], 
\\ \nonumber
&& {\rm where} \\ \nonumber
%&&
%{\rm where} \ \  \\ \nonumber
&&
\qquad 
\qquad E[\{\mu_i\}]=\frac{1}{2}\sum_i^N
%V_i
\left [ 
V(\mu_i) -\frac{1}{2}\log|\mu_i-\mu_j|
\right ],
\quad \mu_i>0,
\end{eqnarray}
%where $\nu=(1+N_B-N_A)-2/\beta$ and $C_{N_A,N_B}$ is a normalization factor.
%where
and
$V(\mu)=(\mu-\kappa \log \mu)$ with
$\kappa=(1+N_B-N_A)-2/\beta$;
%and
 \  ${\cal {\tilde  N}}^{-1}$ is a normalization factor.
% This equation can be written in standard Boltzmann form, $P[\{\mu_i\}]\propto \exp[-\beta E(\{\mu_i\})]$, where
%\begin{align}
%E[\{\mu_i\}]=\frac{1}{2}\sum_i^N
%%V_i
%\left [ 
%V(\mu_i) -\frac{1}{2}\log|\mu_i-\mu_j|
%right ]
%\label{log_gas}
%\end{align}
The weight
$E[\{\mu_i\}]$
can be thought of 
%can be identified 
as the energy of a one-component Coulomb gas of charges with 
%self-
logarithmic interaction 
%and
in an  external potential 
$V(\mu)$.

 In the limit $N_A,N_B\to\infty$, the average of the spectral density
\begin{eqnarray}
\label{DEFSpectralDensityMu}
{\hat \nu} (\mu) \equiv \sum_{i=1}^{N_A} \delta(\mu - \mu_i),
\qquad \ \  {\rm satisfying} \quad \int d \mu \  {\hat \nu}(\mu) = N_A,
\end{eqnarray}
 of the matrix $W=Y Y^\dagger$
can be calculated via the saddle point approximation and is 
found to be equal to the so-called Marchenko-Pastur (MP) distribution\cite{loggas},
\begin{eqnarray}
\bar{n}(\mu)\equiv
%f(w)\equiv
{\langle
{\hat \nu} (\mu) 
\rangle\over N_A}
%\underset{N_A\to\infty}{\lim} \left\langle\frac{1}{N_A}\sum_{i=1}^{N_A}\delta(w-\mu_i)\right\rangle
=\frac{1}{2\pi\alpha \mu}\sqrt{(\mu-\alpha_-)(\alpha_+-\mu)},
\qquad  \qquad \int_{\alpha_-}^{\alpha_+} d\mu \  {\bar n}(\mu) =1,
\label{mp_dis}
\end{eqnarray}
%\end{align}
where $\alpha_-\leq \mu \leq \alpha_+$ with $\alpha_{\pm }=(1\pm \sqrt{\alpha})^2$, $\alpha=N_A/N_B$ and $N=N_A N_B$. This distribution is independent of the Dyson  index $\beta$.
% and satisfies $\int_{\alpha_-}^{\alpha_+} d\mu \  {\bar n}(\mu) =1$ by construction.
Note that $\mu$ has support in the finite interval $\alpha_- \leq \mu \leq \alpha_+$ of $N_A$-independent length
$(\alpha_+-\alpha_-)= $ $4 \sqrt{\alpha}$. Since all $N_A$ eigenvalues lie in this interval, the average level spacing is
\begin{eqnarray}
\label{AverageLevelSpacingMu}
\overline{\left (\Delta \mu\right )} 
=
{4 \sqrt{\alpha}
\over N_A},
\qquad ({\rm average \ level \ spacing \ of \ eigenvalues} \ \mu_i).
\end{eqnarray}

%For a ``random pure state'', if we decompose the total Hilbert space into two subsystems A and B with dimensions $N_A$ and $N_B$ respectively (with %assumption $N_A\leq N_B$), the wavefunction can be rewritten as a linear combination
%\begin{align}
%|\Psi\rangle=\sum_{i,j}X_{ij}|\Psi_A^i\rangle|\Psi_B^j\rangle
%\end{align}
%where the coefficient $X_{ij}$ is a Gaussian complex variable and forms a rectangular $N_A\times N_B$ random matrix $X$ subjected to the %normalization constraint $\mbox{Tr}(XX^{\dag})=1$. For this wavefunction, after tracing out the subsystem B, we obtain the reduced density %matrix  ${\hat \rho}_A=XX^{\dag}$ for subsystem A, which is a $N_A\times N_A$ square matrix and belongs to $\beta=2$ Wishart-Laguerre ensemble %with a fixed trace constraint.

Consider now the spectral density 
%of states 
of the reduced density matrix ${\hat \rho}_A$ (eigenvalues $\lambda_i=\mu_i/N_A$,
and $\lambda = \mu/N_A$),
\begin{eqnarray}
\label{DEFDensityOfStates}
{\hat \nu}(\lambda) \equiv \sum_{i=1}^{N_A} \delta(\lambda-\lambda_i) = N_A  \ {\hat \nu}(\mu),
\qquad \ \  {\rm satisfying} \quad \int d \lambda \  {\hat \nu}(\lambda) = N_A.
\end{eqnarray}
%Considering 
In view of (\ref{DEFDensityMatrixNormalizedWishart})   and (\ref{LimitTrYYdagNA}),
% of (\ref{LimitTrYYdagNA}) 
valid
in the limit $N_A,N_B \to \infty$ which we are currently considering, it follows from 
(\ref{mp_dis}) that the averaged spectral density 
% of states   
of the   density matrix ${\hat  \rho}_A$ satisfies
%The fixed trace constraint can be simply realized by writing ${\hat \rho}_A$ in this way,
%\begin{align}
%{\hat \rho}_A=\frac{YY^{\dag}}{\mbox{Tr}(YY^{\dag})}
%\end{align}
%Since $\mbox{Tr}(YY^{\dag})=N_A$, the spectral density for ${\hat \rho}_A$ is
\begin{align}
%\tilde{f}(\tilde{w})
\bar{n}(\lambda)
\equiv
{\langle{\hat \nu}(\lambda)\rangle \over N_A}
=
N_A \ \bar{n}(\mu)=  
%f(w)=
\frac{N_A}{2\pi\alpha \lambda}\sqrt{(\lambda-\frac{\alpha_-}{N_A})(\frac{\alpha_+}{N_A}-\lambda)},
\ \  \  ({\rm where} \ \lambda=\mu/N_A).
\label{mp_dis_cons}
\end{align}
%where we used the condition $f(w) d w = {\tilde f} (\tilde w)  d {\tilde w}$.
%where $\tilde{w}=w/N_A$. 
It follows from
%(\ref{mp_dis})
(\ref{mp_dis_cons})
 that
$\bar{n}(\lambda)$ is defined on the interval  $\lambda \in [\alpha_-/N_A,\alpha_+/N_A]$, satisfying  
$\int_{\scriptscriptstyle \alpha_-/N_A}^{\scriptscriptstyle \alpha_+/N_A} d\lambda \ 
%\tilde{f}(\tilde{w})
\bar{n}(\lambda)=1$
by construction.
%Note from (\ref{MarchenkoPasturDistributionNindependent}, \ref{DEFEigenvaluesOfRhoA}, \ref{AverageLevelSpacingMu}) 
%It follows from (\ref{mp_dis_cons})  that the  
%Thus, 
Since the $N_A$ eigenvalues $\lambda_i$ of the reduced density matrix lie in the interval
$\alpha_-/N_A \leq \lambda_i \leq \alpha_+/N_A$ (which becomes small when $N_A$ becomes large),  their average
level spacing is
\begin{eqnarray}
\label{AverageLevelSpacingLambdai}
\overline{\left (\Delta \lambda\right )} 
=
{4 \sqrt{\alpha}
\over (N_A)^2},
\qquad ({\rm average \ level \ spacing \ of \ eigenvalues} \ \lambda_i).
\end{eqnarray}

Starting from  the average spectral density
% $\tilde{f}$
$\bar{n}(\lambda)$
 in (\ref{mp_dis_cons}), we can 
%compute 
write the Fourier transform (\ref{DEFFourierTransformEigenvalueDensity}) of the expectation value of the eigenvalue density of the 
%Wishart matrix
reduced density matrix ${\hat \rho}_A$  as
%$\langle Z(\tau)\rangle$ as
\begin{align}
\langle Z(\tau)\rangle=
\int \ d\lambda \ 
\langle{\hat \nu}(\lambda)\rangle 
\ e^{-i\lambda \tau}\ =
N_A\int d\lambda  \
\bar{n}(\lambda) \ 
%\tilde{f}(\lambda)
e^{-i\lambda\tau}.
\end{align}
%where we have replaced for notational 
%simplicity
%reasons ${\tilde w} \to \lambda$.
%, the eigenvalue of reduced density matrix ${\hat \rho}_A$.
% [using(\ref{LimitTrYYdagNA}), (\ref{DEFDensityMatrixNormalizedWishart})].
When $\alpha=N_A/N_B<1$,  the density $\bar{n}(\lambda)$ vanishes at both edges, 
%and scales 
scaling 
as $\sqrt{|\lambda-\alpha_{
a}/N_A|}$
as $\lambda \to \alpha_a$ where $a=\pm$ [see (\ref{mp_dis}) above]. These two edges 
$\lambda=\alpha_{\pm}$ dominate the expectation value $\langle Z(\tau)\rangle$ and contribute 
\begin{align}
|\langle Z_{\alpha_{\pm}}(\tau)\rangle|=\frac{1}{(1\pm\sqrt{\alpha})^2}\frac{N_A^{\frac{5}{2}}}{2\sqrt{\pi}}\alpha^{-\frac{3}{4}}\frac{1}{\tau^{3/2}}.
\end{align}
Assuming $\alpha = N_A/N_B \ll 1$, we have
\begin{align}
\label{DecayOneOverThreeOverTwo}
|\langle Z(\tau)\rangle|^2=\frac{N_A^2N^{3/2}}{\pi \tau^3}.
\end{align}
As mentioned, 
%At early time, 
the spectral form factor factorizes at early times $\tau$, 
%implying
where it thus reads 
 $\langle Z(\tau)Z^*(\tau)\rangle\approx|\langle Z(\tau)\rangle|^2\sim 1/\tau^3$. This is 
%consistent
in agreement fs with the numerical results shown  in Fig.~\ref{fig:page} and Fig.~\ref{fig:Floq_Ising}.

On the other hand, when $\alpha=N_A/N_B=1$, the lower edge for MP distribution (\ref{mp_dis})
%(\ref{MarchenkoPasturDistribution})
 is pushed to $\alpha_-=0$ and the spectral density has  a 
%$1/w^{1/2}$ 
$1/\lambda^{1/2}$-divergence at this edge. This divergence will lead to a different 
behavior of $\langle Z(\tau)\rangle$, namely
\begin{align}
\langle Z(\tau)\rangle=\frac{N_A^2}{2\pi}\int_0^{4/N_A}d\lambda\sqrt{\frac{4/N_A-\lambda}{\lambda}}e^{-i\lambda\tau}=N_A\left[J_0(\frac{2\tau}{N_A})+iJ_1(\frac{2\tau}{N_A})\right]e^{-\frac{2i\tau}{N_A}},
\end{align}
where $J_\alpha(z)$ is the Bessel function of the first kind,
which behaves
% and
 in the limit $z\gg 1$
%, it has
as  $|J_\alpha(z)|\sim 1/\sqrt{2\pi z}$. This  leads
%gives rise
 to 
\begin{align}
\label{ZtauModSquaredForAlphaEuqalZero}
|\langle Z(\tau)\rangle|^2\sim\frac{N_A^3}{\pi\tau},
\end{align}
which 
%It 
decays much slower than  in the case   $\alpha <1$ displayed in (\ref{DecayOneOverThreeOverTwo}). 
% thant that for $\alpha<1$. 
This is also 
%consistent 
in agreement with the numerical results in Fig.~\ref{fig:page} and Fig.~\ref{fig:Floq_Ising}.

%It is easy to verify that
%\begin{align}
%& \int_{\alpha_-/N_A}^{\alpha_+/N_A}\tilde{f}(\tilde{x})d\tilde{x}=1\nonumber\\
%& \int_{\alpha_-/N_A}^{\alpha_+/N_A}\tilde{f}(\tilde{x})\tilde{x}d\tilde{x}=\frac{1}{N_A}
%\end{align}

\subsection{Spectral form factor for the  reduced density matrix 	from Wishart Random Matrix Theory}
\label{SubSectionSpectralFormFactorReducedDensityMatrix}
For any  random matrix ensemble with the joint probability density described by \eqref{log_gas}, the level-level correlation function 
(``pair correlation function'') is universal and only depends\cite{Nagao1992,BREZIN1993}
 on the symmetry type, although the spectral density depends on the explicit form of the potential $V(\mu_i)$ defined in  (\ref{log_gas}). Specifically,
% considering the density of states
%\begin{eqnarray}
%\label{DEFDensityOfStates}
%{\hat \nu}(\lambda) \equiv \sum_i \delta(\lambda-\lambda_i)
%\end{eqnarray}
the connected  correlation function
of the 
spectral density ${\hat \nu}(\lambda)$,
% of states,
 defined in (\ref{DEFDensityOfStates}),
% in the middle of the spectrum 
takes  for  Dyson index $\beta=2$ (``GUE-type'' class)   in the large $N_A$ limit  the following universal form which
can be expressed in terms of the celebrated so-called sine-kernel\cite{Fox1964,Nagao1992,BREZIN1993},
% For $\beta=2$ case we are interested in, the pair correlation function for the bulk spectrum in the large N limit includes the celebrated sine kernel \cite{Fox1964,Nagao1992,BREZIN1993},
\begin{align}
\langle {\hat \nu}(\lambda){ \hat \nu}(\lambda')\rangle-\langle {\hat \nu}(\lambda)\rangle \ \langle{\hat \nu}(\lambda')\rangle
=\langle {\hat \nu}(E)\rangle \ \delta(\omega)-\langle{\hat \nu}(\lambda)\rangle \ \langle{\hat \nu}(\lambda')\rangle \ 
\frac{\sin^2[\pi\langle{\hat \nu}(E)\rangle \omega]}{[\pi\langle{\hat \nu}(E)\rangle \omega]^2},
\label{pair_corr}
\end{align}
where 
\begin{eqnarray}
\label{DEFOmegaE}
\omega=\lambda-\lambda', \qquad  E=(\lambda+\lambda')/2.
\end{eqnarray}
(For more details see Appendix \ref{SubSectionAppendixDetailsOnTheSpectralFormFactorWishart}.)
We also  recall  $\langle{\hat \nu}(\lambda)\rangle=N_A \ \bar{n}(\lambda)$
from (\ref{mp_dis_cons}).
% is the density of states.
Here, in order to obtain a universal expression, the argument of the sine function was  rescaled by the non-universal factor
 $\langle{\hat \nu}(E)\rangle$ that determines the local mean level
 spacing.\footnote{From (\ref{DEFDensityOfStates}) the number $\Delta N$ of levels $\lambda_i$  that lie in the interval $[\lambda-\Delta\lambda/2,
\lambda+\Delta \lambda/2]$ is $\Delta N =$ 
$\int_{\lambda-\Delta \lambda/2}^{\lambda+\Delta\lambda/2}<{\hat \nu}(\lambda')>
d\lambda'\approx$ $<{\hat \nu}(\lambda)> \ \Delta \lambda$, so that the local density of levels and thus the inverse of the local  mean
level spacing is $<{\hat \nu}(\lambda)>=$ $1/(\Delta \lambda/\Delta N)=$  $1/ \ \overline{(\delta \lambda)}$,
where $\overline{(\delta \lambda)}=$  $(\Delta \lambda/\Delta N)$
denotes the local level spacing near the eigenvalue $\lambda$.}

The spectral form factor 
\begin{eqnarray}
\nonumber
g(\tau)&=&  \langle Z(\tau)Z^*(\tau)\rangle
=
\left [
\langle Z(\tau)Z^*(\tau)\rangle - \langle Z(\tau)\rangle \langle Z^*(\tau)\rangle\right]+
\langle Z(\tau)\rangle \langle Z^*(\tau)\rangle \\ 
\label{Connected-NonConnnectedSpectralFormfactor}
&=&g_c(\tau)+\langle Z(\tau)\rangle \langle Z^*(\tau)\rangle
\end{eqnarray}
 is related 
%with
to  the level-level correction function
 $\langle {\hat \nu}(\lambda){ \hat \nu}(\lambda')\rangle$ through Fourier transformation, 
%i.e., 
\begin{eqnarray}
g_c(\tau) 
%&=& 
%\langle Z(\tau)Z^*(\tau)\rangle
%-
%langle Z(\tau)\rangle \langle Z^*(\tau)\rangle
 %= \\ \nonumber
%&=&
=\int d\lambda d\lambda' \ 
\left [ \langle {\hat \nu}(\lambda) { \hat \nu}(\lambda')\rangle 
-
 \langle {\hat \nu}(\lambda)\rangle \langle { \hat \nu}(\lambda')\rangle
\right ]
\  e^{-i(\lambda-\lambda')\tau},
\end{eqnarray}
and we focus here on the connected function $g_c(\tau)$ as the disconnected part has already been discussed in
Sect. \ref{SubSectionWishartRandomMatrixEnsemble}. 
Taking the  Fourier transform of the sine
 kernel in (\ref{pair_corr}) which  is
determined by the following elementary integral ($a$ is any real parameter)
% equal to 
\begin{align}
\label{IntegralOfSine}
\int_{-\infty}^{\infty}e^{-i\omega \tau} \ \frac{\sin^2[\pi a \omega]}{\pi^2\omega^2}d\omega=
\begin{cases} a-\frac{|\tau|}{2\pi}, & |\tau|<2\pi a \\ 0, & |\tau|\geq 2\pi a \end{cases},
\end{align}
%for any real parameter $a$, 
%{\color{blue}
we  obtain (for more details see
Appendix
\ref{SubSectionAppendixDetailsOnTheSpectralFormFactorWishart})
\begin{eqnarray}
\label{``ramp''}
%\label{gConnectedResultAppendix}
g_c(\tau)=
\begin{cases}  \   {2 \over \pi} {1\over \sqrt{N}} \ |\tau|, & |\tau| <   \tau_H \\ 
N_A, &   |\tau| >   \tau_H
 \end{cases},
\qquad \quad {\rm where} \  \ \ \tau_H=
(2\pi/ \overline{\left (\Delta \lambda\right )} )=
{\pi\over 2} N_A \sqrt{N},
\end{eqnarray}
%\begin{align}
%g_c(\tau)= \langle Z(\tau) Z^*(\tau)\rangle-|\langle Z(\tau)\rangle|^2=\begin{cases} \frac{2\sqrt{\alpha}|\tau|}{\pi N_A}, & |\tau|<\frac{\pi %N_A\sqrt{N}}{2} \\ N_A, & |\tau|\geq \frac{\pi N_A\sqrt{N}}{2} \end{cases},
%\abel{``ramp''}
%\end{align}
where  we recall that 
%$\alpha = N_A/N_B$ and 
$N=N_AN_B$. 
%In the above calculation, we 
%have assumed that $\langle{\hat \nu}(E)\rangle$ does not change much over the 
%%entire
%relevant part of the  spectrum. 
The regime of linear growth  with $\tau$ is universal and reflects the 
%level repulsion
universal 
spectral
% rigidity 
correlations present   in the spectrum, which are represented by the sine-kernel 
on the right hand side of  (\ref{pair_corr}).
Note also that the prefactor of the linear growth term in (\ref{``ramp''})  is independent of subsystem size $N_A$. This is the origin of the fact
that
the  linear ``ramps'' appearing  for different subsystem sizes $N_A$ all lie on top of each other - see e.g. Fig\
\ref{fig:page}. The spectral form factors of the entanglement Hamiltonian,
discussed in Appendix  \ref{SectionAppendixSpectralFormfactorEntanglementHamiltonian}
and   depicted in Fig. \ref{fig:SpectralFormFactorEntanglementHamiltonian} do not show this feature, but are instead
shifted with respect to each other by a $N_A$-dependent constant (on a log-log plot), reflecting a $N_A$-dependent
coefficient of the term linear in $\tau$.
%}

In view of (\ref{DecayOneOverThreeOverTwo}),(\ref{ZtauModSquaredForAlphaEuqalZero}),(\ref{Connected-NonConnnectedSpectralFormfactor}),(\ref{``ramp''}), the disconnected part in $g(\tau)$ hides
the early-time $\tau$ part of
% at early times $\tau$ 
the universal  linear ``ramp''  (\ref{``ramp''}) appearing in $g_c(\tau)$. This effect
gives rise to the ``dip'' (minimum) in $g(\tau)$, and allows one to estimate the ``dip-time'' $\tau_d$ as follows:
Equating $|\langle Z(\tau)\rangle|^2$ and the ``ramp'' in \eqref{``ramp''} gives the dip time in $\langle Z(\tau)Z^*(\tau)\rangle$. 
Based on this logic, 
%When $\alpha<1$, 
we expect  the dip time $\tau_d$ to be around $(N_AN)^{1/2}$
when $\alpha<1$, which is consistent with the numerical results. On the other hand, when $\alpha=1$,
this logic yields
% we have 
$\tau_d\sim N_A^2$. However this 
is a time scale of the order of
%this has the same time scale as 
the plateau time $\tau_p=$ $\tau_H$, and therefore we cannot observe the ``ramp'' in $g(\tau)=$ $\langle Z(\tau)Z^*(\tau)\rangle$ when $\alpha=1$, consistent with our numerical findings reported above.

%As we discussed in this section, the log interaction between the eigenvalues in the reduced density matrix is responsible for the linear ``ramp'' in the spectral form factor. In general, the eigenvalue of the reduced density matrix forms a one-dimensional gas of charge with energy
%\begin{align}
%E[\{\lambda_i\}]=\frac{1}{2}\sum_i^NV_i-\frac{1}{2}|\lambda_i-\lambda_j|^{1-\delta}
%\end{align}
%where $0<\delta<1$. The $|\lambda_i-\lambda_j|^{1-\delta}$ interaction will give rise to a kernel with the form $1/(\lambda_i-\lambda_j)^{1+\delta}$. In the spectral form factor, this will lead to a power law ``ramp'' as $\tau^{\delta}$. 
%%At late time, $\langle Z Z^*\rangle$ is determined by the level repulsion between the spectrum, we have sine kernel as GUE random matrix. 
%%Fourier transformation for it gives
%%\begin{align}
%%\int dx \frac{\sin^2[\pi\langle{\hat \rho}(E)\rangle x]}{[\pi x]^2}e^{ix \tau}
%%= \begin{cases} \langle {\hat \rho}(E)\rangle-\frac{\tau}{2\pi}, &  \tau< 2\pi\langle {\hat \rho}(E)\rangle \\ 
%%0, &   \tau\geq 2\pi\langle {\hat \rho}(E)\rangle \end{cases}
%%\end{align}
%%where we assume $\langle{\hat \rho}(E)\rangle$ takes a uniform distribution between $\alpha_-/N_A$ and $\alpha_+/N_A$ and equal to $N_A^2/(4\sqrt{\alpha})$. We expect that $\langle ZZ^*\rangle$ takes this form
%%\begin{align}
%%\langle ZZ^*\rangle
%%\sim \begin{cases} \frac{2\sqrt{\alpha}\tau}{\pi N_A}, &  \tau< 2\pi\langle {\hat \rho}(E)\rangle \\ 
%%N_A, &   \tau\geq 2\pi\langle {\hat \rho}(E)\rangle \end{cases}
%%\end{align}

\section{Discussion and Conclusion}

In conclusion, we  have explored 
%level repulsion statistics 
the presence of universal spectral correlations in
%spectral rigidity of 
the spectrum of 
%in 
the reduced density matrix ${\hat \rho}_A$ of a many-body wavefunction and used 
%this property  
the presence of these correlations to define quantum chaos at the level of 
a single many-body  wavefunction. To detect 
%this level repulsion statistics
these spectral 
correlations,
%rigidity, 
we constructed the spectral form factor $g(\tau)$ for ${\hat \rho}_A$ and
% argue that there is 
identified the presence of 
%an intermediate 
a  ``ramp'' as a hallmark of 
the spectral
correlations.
% rigidity.
% characterizing the level repulsion statistics.
% To support this argument, we 
We explicitly considered  three wavefunctions: the ``random pure state'', a  typical state 
%for
of a  Floquet spin model, and 
%for the 
of a quantum Ising model in both transverse and longitudinal fields, both in one spatial dimenstion.
 In all three cases, we numerically found the presence of the universal
%  a 
linear   ``ramp'' in the spectral form factor.
% In particular, for 
For the ``random pure state'', we also analytically computed the spectral form factor by using Wishart random matrix theory and found
 %that it is consistent
agreement  with  our numerical results. 

Moreover, we discussed  how
% does level repulsion statistics
universal
spectral
% rigidity
correlations  develop  in a quantum quench problem from an initial product state lacking
% without 
chaos. We found  that
% level repulsion 
the spectral 
correlations
%rigidty 
first 
%appear 
emerge
% in the middle
at the top  of spectrum of the reduced density matrix  ${\hat \rho}_A$,
 and then spread over the entire spectrum at  later times. We verified this statement numerically in both,
the  Floquet and quantum Ising models. Finally we 
studied a rapidly  driven Floquet system 
%with
which posseses  a  long prethermalized regime 
%which has
exhibiting  an ``EE plateau",
on which the system
% and 
can be well approximated by a GGE. 
For times when the system is on that EE  plateau, we
%We 
don't observe any ``ramp''
% for
in the  spectral form factor,
% in this regime, 
which is consistent with the absence of chaos in the GGE. 
We found that 
%level repulsion
universal
spectral  correlations (and a ``ramp'') 
%rigidity
 in the density matrix develop only when the wavefunction starts to relax to the fully thermalized regime at  late times.

%For the quantum quench problems studied in this paper, we observed that
%%the level repulsion 
%the universal spectral correlations
%% rigidity 
%always appear as soon as the entanglement entropy begins to grow,
%%immediately after the  linear growth regime of the EE ceases,  
%and that they  first 
%develops
%appear
%emerge 
%at the top 
%% in the middle 
%of the spectrum of the reduced density matrix. 

\acknowledgements
We thank  C. Nayak for the discussion on the prethermalized regime. XC was supported by a postdoctoral fellowship from the Gordon and Betty Moore Foundation,
under the EPiQS initiative, Grant GBMF4304, at the Kavli Institute for Theoretical Physics.  
This work was supported by the NSF under Grant No. DMR-1309667 (AWWL).
We acknowledge support from the Center for Scientific Computing from the CNSI, MRL: an NSF MRSEC (DMR-1121053).

\appendix

\section{Spectral Form Factor of the Entanglement Hamiltonian}
\label{SectionAppendixSpectralFormfactorEntanglementHamiltonian}

\begin{figure}%[hbt]
\centering
\includegraphics[width=.5\textwidth]{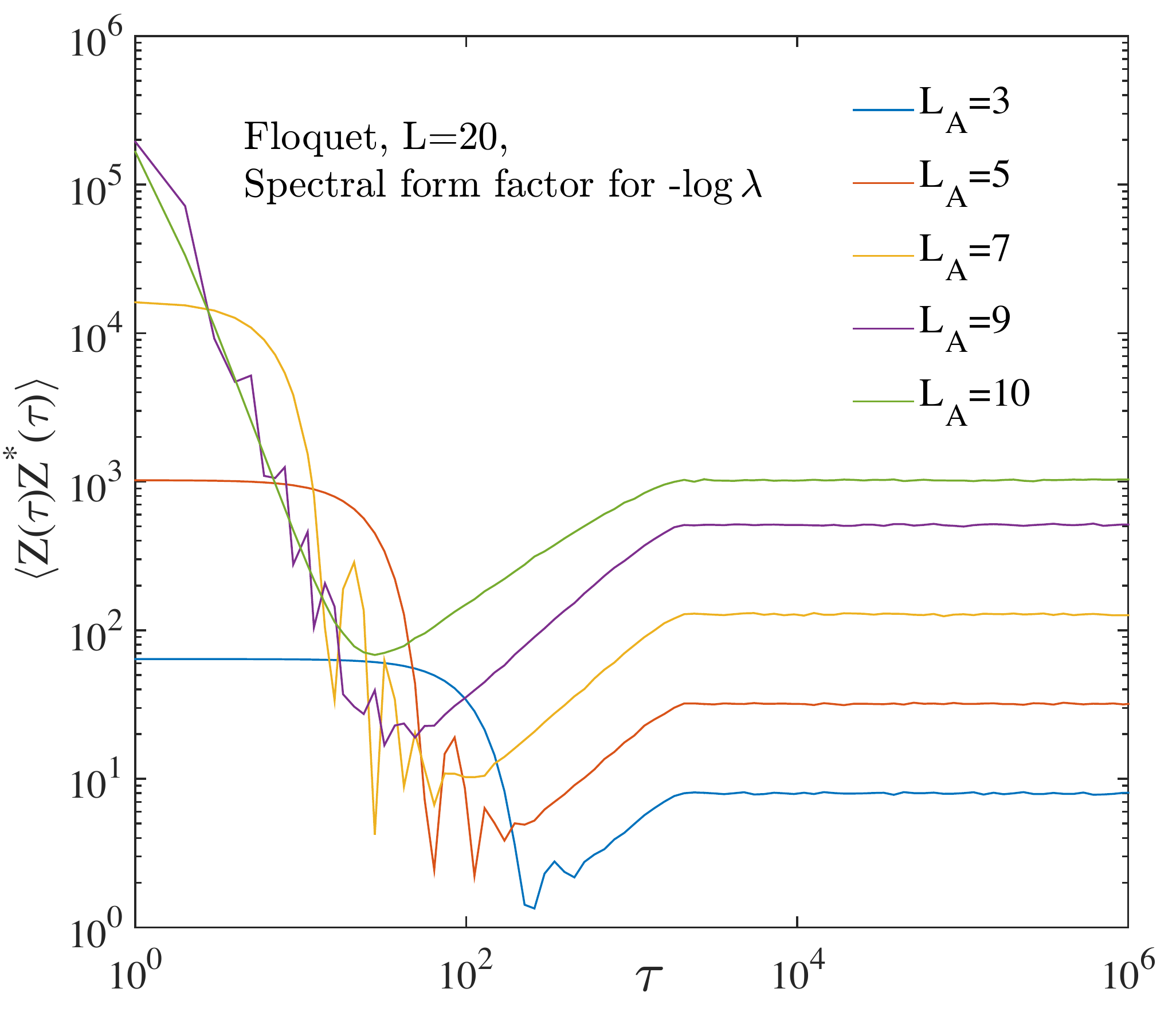}
\caption{Spectral form factor of the entanglement Hamiltonian}
\label{fig:SpectralFormFactorEntanglementHamiltonian}
\end{figure}

Numerical results for the spectral form factor of the entanglement Hamiltonian ${\hat H}_E$
of the ``random pure state''
%, defined in (\ref{DEFEntanglementHamiltonian}) of the main text, 
are displayed in Fig. \ref{fig:SpectralFormFactorEntanglementHamiltonian},
and are to be compared with the spectral form factor of the reduced density matrix ${\hat \rho}_A$ of the
same system, depicted in Fig. \ref{fig:page}.
We see that both spectral form factors exhibit a linear ``ramp'' (unit slope on a the log-log plot), which is the hallmark
of   universal spectral correlations.
% rigidity. 
The form factors
of the entanglement Hamiltonian are shifted by a $N_A$-dependent constant on the log-log plot, which reflects
a $N_A$-dependent prefactor of the linear $\tau$-dependence.

Recall that entanglement Hamiltonian and density matrix are related as in (\ref{DEFEntanglementHamiltonian}),
and that,  as mentioned in the sentence below (\ref{DEFEntanglementHamiltonian}), the spectral form factor of
the former is obtained from that of the latter by letting $\lambda_i \to - \ln \lambda_i$,
where $\lambda_i$ denotes the eigenvalues of the reduced density matix.

%%%%%%%%%%%%%%
%%%%%%%%%%%%%% 

\section{Some Details on Wishart Random Matrix Theory}
\label{AppendixSectionDetailsWishartRandomMatrixTheory}

\subsection{Unscaled Gaussian Probability Weight}
\label{AppendixSubSectionUnscaledGaussian}
 In general, for a Wishart matrix ${\cal W}=YY^{\dag}$ with $Y$ being an arbitrary $N_A\times N_B$ matrix with real $(\beta=1)$, complex $(\beta=2)$ or quaternion $(\beta=4)$ Gaussian entries drawn from the
% joint 
probability distribution
\begin{eqnarray}
\label{WishartGaussianDistributionCal}
{\cal P}(\{Y_{iJ}\})
=
{\cal N}^{-1}
\exp  \{ - {\beta\over 2}  \mbox{Tr}(Y Y^\dagger) \},
\end{eqnarray}
% $P(Y_{ij})\propto \exp\left[ -\frac{\beta}{2}\mbox{Tr}YY^{\dag} \right]$, 
the 
joint probability 
%probability 
 distribution for the $N_A$ eigenvalues  ${\xi}_i$ of  
%the Wishart matrix 
${\cal W}$ is known to be\cite{loggas}
\begin{eqnarray}
\label{JointProbabilityxi}
{\cal P}[\{\xi_i\}]=C_{N_A,N_B}e^{-\frac{\beta}{2}\sum_{i=1}^{N_A}\xi_i}\prod_{i=1}^{N_A} \xi_{i}^{\kappa\beta/2}\prod_{1\leq j<k\leq N_A}
|\xi_j-\xi_k|^{\beta},\quad \xi_i>0,
\end{eqnarray}
where $\kappa=(1+N_B-N_A)-2/\beta$ and $C_{N_A,N_B}$ is a normalization factor. This 
expression
%equation 
can be written in standard Boltzmann form, ${\cal P}[\{\xi_i\}]\propto \exp[-\beta E(\{\xi_i\})]$, where
\begin{align}
E[\{\xi_i\}]=\frac{1}{2}\sum_i^N
%V_i
\left [ 
V(\xi_i) -\frac{1}{2}\log|\xi_i-\xi_j|
\right ]
\label{Appendixlog_gas}
\end{align}
can be
thought of
% identified 
as the energy of a  one-component Coulomb gas of charges with 
%self-
logarithmic interaction 
%and
in an  external potential 
\begin{eqnarray}
\label{WishartPotential}
V(\xi)=(\xi-\kappa \log \xi).
\end{eqnarray}
In the limit $N_A, N_B \to \infty$  with $\alpha=N_A/N_B=$ fixed,
the spectral density can be computed via the saddle point approximation leading to the Marchenko-Pastur (MP)
distribution
\begin{eqnarray}
\label{MPDistributionCal}
\bar{n}(\xi)
%F(\xi)
 \equiv 
{ \langle {\hat \nu}(\xi)\rangle\over N_A}
=\lim_{N_A, N_B \to \infty}
\left \langle {1\over N_A} \sum_i \delta (\xi-\xi_i) \right \rangle=
{1\over 2 \pi \alpha \xi}
\sqrt{ ({\xi\over N_B}
-\alpha_- )
(\alpha_+ - {\xi\over N_B})}, 
\end{eqnarray}
where $\alpha_{\pm} = (1\pm \sqrt{\alpha})^2$
and
$N_B \alpha_- \leq \xi \leq N_B \alpha_+$. This distribution is independent of the Dyson index $\beta$, and by construction satisfies
$\int_{N_B \alpha_-}^{N_B \alpha_+}  d \xi \  \bar{n}(\xi) =1$.

\subsection{Scaled Gaussian - $N_A$-independent Spectral Density}
\label{AppendixSubSectionScaledGaussian}

It is convenient rescale the Wishart random matrix and consequently also its eigenvalues 
\begin{eqnarray}
\label{RescaleW}
{\cal W} \equiv N_B  \ W,
\quad W=Y Y^\dagger,
\qquad \xi_i \equiv N_B \ \mu_i
\end{eqnarray}
so that
\begin{eqnarray}
\label{WishartGaussianDistributionNonCal}
 P(\{Y_{iJ}\})
=
{\cal{\tilde  N}}^{-1}
\exp  \{ - {\beta\over 2} N_B \ \mbox{Tr}(Y Y^\dagger) \}.
\end{eqnarray}
% $P(Y_{ij})\propto \exp\left[ -\frac{\beta}{2}\mbox{Tr}YY^{\dag} \right]$, 
We can think of this as rescaling $\beta \to \beta \ N_B$.
Now, the joint probability probability  distribution for the $N_A$ eigenvalues  ${\mu}_i$ of  the Wishart matrix $W$  
 can be written in standard Boltzmann form, $P[\{\mu_i\}]\propto \exp[-\beta  N_B E(\{\mu_i\})]$, with
$E[\{\mu_i\}]$ the same function as 
% as before
 in (\ref{log_gas}).
In the limit $N_A, N_B \to \infty$, we obtain  from (\ref{MPDistributionCal})
the spectral distribution [noting that   $\bar{n}(\xi) \ d \xi = \bar{n}(\mu) \ d \mu$]
\begin{eqnarray}
\label{MarchenkoPasturDistributionNindependent}
\bar{n}(\mu)
= \lim_{N_A, N_B \to \infty}
{\langle {\hat \nu}(\mu)\rangle \over N_A}
%f(\mu) \equiv \lim_{N_A, N_B \to \infty}
%\left \langle {1\over N_A} \sum_i \delta (\mu-\mu_i) \right \rangle=
={1\over 2 \pi \alpha \mu}
\sqrt{\bigl (
\mu -\alpha_- \bigr)
\bigl (\alpha_+ - \mu\bigr) }.
\end{eqnarray}
In this form the distribution becomes {\it independent} of $N_A, N_B$ in the limit when these are large, and the result depends only on $\alpha
=
N_A/N_B$ which we consider holding fixed.

%Note that $\mu$ has support in the finite interval $\alpha_- \leq \mu \leq \alpha_+$ of $N_A$-independent length
%$(\alpha_+-\alpha_-)= $ $4 \sqrt{\alpha}$. Since the $N_A$ eigenvalues lie in this interval, the average level spacing is
%\begin{eqnarray}
%\label{AverageLevelSpacingMu}
%\overline{\left (\Delta \mu\right )} 
%=
%{4 \sqrt{\alpha}
%\over N_A},
%\qquad ({\rm average \ level \ spacing \ of \ eigenvalues} \ \mu_i).
%\end{eqnarray}
As already discussed in the paragraph surrounding (\ref{ExpectationValueTraceW}),
%We  note the following 
we have the expectation value
\begin{eqnarray}
\langle \mbox{Tr} (Y Y^\dagger)\rangle = 
\sum_{i=1}^{N_A}\sum_{J=1}^{N_B} \ 
\langle |Y_{iJ}|^2\rangle
={(N_A N_B)\over N_B} = N_A.
\end{eqnarray}
%which follows from the Gaussian integral for the matrix elements $Y_{iJ}$ using (\ref{WishartGaussianDistributionNonCal}).
Thus,  in the limit where both $N_A$ and $N_B$ tend to infinity while the ratio $\alpha\equiv N_A/N_B$ remains fixed, 
%the relative fluctuations $f$
%of the random variable $ Tr (Y Y^\dagger)=$ $N_A (1+ f)$ about its mean $N_A$  vanish, and 
we can replace 
\begin{eqnarray}
\label{AppendixLimitTrYYdagNA}
\mbox{Tr} (Y Y^\dagger) \to N_A
\end{eqnarray}
% in this limit
 in the usual sense. We see from 
(\ref{RescaleW}, \ref{WishartGaussianDistributionNonCal})
that the eigenvalues $\lambda_i$ of the reduced density
matrix ${\hat \rho}_A$ are related in the  limit of large $N_A$ and $N_B$ to the eigenvalues $\mu_i$ of the Wishart matrix $W$
via
\begin{eqnarray}
\label{DEFEigenvaluesOfRhoA}
\lambda_i = {\mu_i\over N_A}.
\end{eqnarray}
%Note from (\ref{MarchenkoPasturDistributionNindependent}, \ref{DEFEigenvaluesOfRhoA}, \ref{AverageLevelSpacingMu}) 
%that the eigenvalues $\lambda_i$ of the reduced density matrix lie in the interval
%$\alpha_-/N_A \leq \lambda_i \leq \alpha_+/N_A$ (which becomes small when $N_A$ becomes large), and that their average
%level spacing is
%\begin{eqnarray}
%\label{AverageLevelSpacingLambdai}
%\overline{\left (\Delta \lambda\right )} 
%=
%{4 \sqrt{\alpha}
%\over (N_A)^2},
%\qquad ({\rm average \ level \ spacing \ of \ eigenvalues} \ \lambda_i).
%\end{eqnarray}

\subsection{Some details about  the computation of the  Spectral Form Factor for the (``GUE-type''-) Wishart Random Matrix Ensemble
in (\ref{``ramp''})}
\label{SubSectionAppendixDetailsOnTheSpectralFormFactorWishart}

We can express the spectral form factor (\ref{DEFSpectralFormFactor}) as follows in terms of the density of states (\ref{DEFDensityOfStates})
\begin{eqnarray}\nonumber
&&g(\tau) =  \langle\sum_{i,j} e^{-i \tau (\lambda_i-\lambda_j)}\rangle = \\ \nonumber
&& = \langle\sum_{i,j}
[\int d \lambda \ \delta(\lambda - \lambda_i)]
\ 
[\int d \lambda' \ \delta(\lambda' - \lambda_j)] 
\ \  e^{-i \tau (\lambda_i-\lambda_j)}\rangle= \\  \nonumber
&&=
\int d \lambda \int d \lambda' \ \  e^{-i \tau (\lambda-\lambda')} \ \ 
 \langle\sum_{i,j} \ \delta(\lambda - \lambda_i)
\ \ \delta(\lambda' - \lambda_j)\rangle= \\ 
\label{SpectralFormFactorTwoPointCorrelationFunction}
&& =
\int d \lambda \int d \lambda' \ \  e^{-i \tau (\lambda-\lambda')} \ \ 
 \langle{\hat \nu}(\lambda) \ {\hat \nu}(\lambda')\rangle.
\end{eqnarray}
Using (\ref{DEFDensityOfStates})
%\begin{eqnarray}
%\label{DEFDensityOfStatesAndRescaled}
%&{\hat \nu}(\mu)\equiv \sum_{i=1}^{N_A} \delta(\mu -\mu_i),
%\quad 
%&{\hat \nu}(\lambda)\equiv \sum_{i=1}^{N_A} \delta(\lambda -\lambda_i)
%= N_A {\hat \nu}(\mu)
%\\ \nonumber
%&\nu(\mu)\equiv {1\over N_A} \sum_{i=1}^{N_A} \delta(\mu -\mu_i), 
%\quad 
%&{\nu}(\lambda)\equiv {1\over N_A}  \sum_{i=1}^{N_A} \delta(\lambda -\lambda_i)
%= N_A { \nu}(\mu)
%\end{eqnarray}
we obtain from (\ref{SpectralFormFactorTwoPointCorrelationFunction})
\begin{eqnarray}
\label{SpectralFormFactorTwoPointCorrelationFunctionMu}
&&g(\tau) = 
\int d \mu \int d \mu' \ \  e^{-i (\tau/N_A) (\mu-\mu')} \ \ 
 \langle{\hat \nu}(\mu) \ {\hat \nu}(\mu')\rangle.
\end{eqnarray}
The (2-point) correlation function of the density of states (\ref{DEFDensityOfStates})
appearing in
(\ref{SpectralFormFactorTwoPointCorrelationFunction},\ref{SpectralFormFactorTwoPointCorrelationFunctionMu}) above
can be re-written as follows
\begin{eqnarray}\nonumber
&& \langle{\hat \nu}(\mu) \ {\hat \nu}(\mu')\rangle=
\langle \sum_{i,j} \ \delta(\mu-\mu_i) \  \delta(\mu'-\mu_j)\rangle= \\ \nonumber
&&=
\langle \sum_{i} \ \delta(\mu-\mu_i) \  \delta(\mu'-\mu_i)\rangle
\ \ 
+
\ \ 
\langle \sum_{i \not = j} \ \delta(\mu-\mu_i) \  \delta(\mu'-\mu_j)\rangle \\ \nonumber
&& =
\delta(\mu-\mu') 
\langle \sum_{i} \ \delta(\mu-\mu_i) \rangle
\ \ 
+
\ \ 
\langle \sum_{i \not = j} \ \delta(\mu-\mu_i) \  \delta(\mu'-\mu_j)\rangle = \\ 
\label{DEFLevelLevelTwoPointFunction}
&&=
\delta(\mu-\mu')  \ 
\langle {\hat \nu}(\mu) \rangle
\ \ 
+
\ \ 
\langle \sum_{i \not = j} \ \delta(\mu-\mu_i) \  \delta(\mu'-\mu_j)\rangle,
\end{eqnarray}
%which yields for the connected (2-point) correlation function
%\begin{eqnarray}\nonumber
%&& \langle{\hat \nu}(\lambda_1) \ {\hat \nu}(\lambda_2)\rangle
%\ - \ 
%\langle{\hat \nu}(\lambda_1)\rangle \  \  \langle{\hat \nu}(\lambda_2)\rangle =  \\  \label{ConnectedLevelLevelCorrelator}
%&&=
%\delta(\lambda_1-\lambda_2)  \
%\langle {\hat \nu}(\lambda_1) \rangle
%+
%\langle{\hat \nu}(\lambda_1)\rangle \    \langle{\hat \nu}(\lambda_2)\rangle  \  
 %\left [
%{\langle \sum_{i \not = j}  \delta(\lambda_1-\lambda_i)   \delta(\lambda_2-\lambda_j)\rangle 
%\over \langle{\hat \nu}(\lambda_1)\rangle \    \langle{\hat \nu}(\lambda_2)\rangle} \ 
%%\langle \sum_{i \not = j}  \delta(\lambda_1-\lambda_i)   \delta(\lambda_2-\lambda_j)\rangle 
%- 1
% \right ]
%%\label{ConnectedLevelLevelCorrelator}
%\end{eqnarray}
%We can write this equation for $\nu(\mu)$ from (\ref{DEFDensityOfStatesAndRescaled})
and thus  the connected function reads 
%[upon division by $N_A^2$ - see (\ref{DEFDensityOfStatesAndRescaled})]
\begin{eqnarray}\nonumber
&& \langle{ \hat \nu}(\mu) \ {\hat \nu}(\mu')\rangle_c=
\langle{\hat  \nu}(\mu) \ {\hat \nu}(\mu')\rangle
\ - \ 
\langle{\hat \nu}(\mu)\rangle \  \  \langle{\hat \nu}(\mu')\rangle =  \\  \label{RescaledConnectedLevelLevelCorrelatorMu}
&&=
%{1\over N_A} 
\delta(\mu-\mu')  \
\langle { \hat \nu}(\mu) \rangle
+
\langle{\hat \nu}(\mu)\rangle \    \langle{\hat \nu}(\mu')\rangle  \  
 \left [
{
%{1\over N_A^2} 
\langle \sum_{i \not = j}  \delta(\mu-\mu_i)   \delta(\mu'-\mu_j)\rangle 
\over \langle{\hat \nu}(\mu)\rangle \    \langle{\hat \nu}(\mu')\rangle} \ 
%\langle \sum_{i \not = j}  \delta(\lambda_1-\lambda_i)   \delta(\lambda_2-\lambda_j)\rangle 
- 1
 \right ].
%\label{ConnectedLevelLevelCorrelator}
\end{eqnarray}
Note that
$\langle \sum_{i \not = j}  \delta(\mu-\mu_i)   \delta(\mu'-\mu_j)\rangle $ equals  $N_A (N_A-1)$ times the probability that
one eigenvalue equals $\mu$ and another eigenvalue equals $ \mu' (\not = \mu)$, as computed from (\ref{log_gas}).

When $\mu$ and $\mu'$ are separated by much less than $N_A$ level spacings, 
%When $\mu_1$ and $\mu_2$ are not separated too far 
so that 
%of (\ref{DEFOmegaE})
we can approximate
$\langle{ \nu}(\mu)\rangle \approx$ $ \langle{ \nu}(\mu')\rangle \approx$ $\langle{ \nu}({\cal E})\rangle$,
where
\begin{eqnarray}
{\cal E} \equiv {\mu+\mu'\over 2}, \quad \Omega \equiv (\mu - \mu'),
\end{eqnarray}
the square bracket in (\ref{RescaledConnectedLevelLevelCorrelatorMu})
%the above equation
 is known analytically
 (``sine kernel')
to be \cite{Fox1964,Nagao1992,BREZIN1993}
\begin{eqnarray}
\label{DEFSineSquareKernel}
&& \left [ {\langle{\hat  \nu}(\mu) {\hat \nu}(\mu')\rangle
 \over
\langle {\hat \nu}(\mu)\rangle \ \langle{\hat  \nu}(\mu')\rangle}
-1
\right ]  = \qquad \qquad ({\rm when} \ \Omega=(\mu-\mu') \not = 0)
\\ \nonumber
&&
= \left [
{
%{1\over N_A^2}
\langle \sum_{i \not = j}  \delta(\mu-\mu_i)   \delta(\mu'-\mu_j)\rangle 
\over \langle{\hat \nu}(\mu)\rangle \    \langle{\hat \nu}(\mu')\rangle} \ 
%\langle \sum_{i \not = j}  \delta(\lambda_1-\lambda_i)   \delta(\lambda_2-\lambda_j)\rangle 
- 1
 \right ]
\ =  \ (-1)  \ 
{\sin^2[\pi \langle {\hat \nu}({\cal E})\rangle \Omega]\over
[\pi  \langle {\hat  \nu}({\cal E})\rangle \Omega]^2}.
\end{eqnarray}
%Note that
%$\langle \sum_{i \not = j}  \delta(\mu-\mu_i)   \delta(\mu'-\mu_j)\rangle $ is $N_A (N_A-1)$ times the probability that
%one eigenvalue equals $\mu$ and another eigenvalue equals $ \mu'\not = \mu$.
Analogous to what was  mentioned in the paragraph below (\ref{DEFOmegaE}), the argument of the sine-function is rescaled by
the non-universal
factor $\langle {\hat \nu}({\cal E})\rangle$ which equals the inverse of the local mean level spacing of eigenvalues $\mu_i$ at $\mu={\cal E}$.

%When $\lambda_1$ and $\lambda_2$ are not separated too far so that 
%in the notation of (\ref{DEFOmegaE})
%we can approximate
%$\langle{\hat \nu}(\lambda_1)\rangle \approx$ $ \langle{\hat \nu}(\lambda_2)\rangle \approx$ $\langle{\hat \nu}(E)\rangle$,
%the square bracket in the above equation is known analytically (``sine-square kernel')
%\begin{eqnarray}
%\label{DEFSineSquareKernel}
%\left [
%{\langle \sum_{i \not = j}  \delta(\lambda_1-\lambda_i)   \delta(\lambda_2-\lambda_j)\rangle 
%\over \langle{\hat \nu}(\lambda_1)\rangle \    \langle{\hat \nu}(\lambda_2)\rangle} \ 
%%\langle \sum_{i \not = j}  \delta(\lambda_1-\lambda_i)   \delta(\lambda_2-\lambda_j)\rangle 
%- 1
 %\right ]
%\ =  \
%{\sin^2[\pi \langle {\hat \nu}(E)\rangle \omega]\over
%[\pi \langle {\hat \nu}(E)\rangle \omega]^2}.
%\end{eqnarray}
%Now, (\ref{ConnectedLevelLevelCorrelator}) and (\ref{DEFSineSquareKernel}) together imply (\ref{pair_corr})
%in the main text.

\vspace {.5cm}
We now provide some detailed steps for obtaining (\ref{``ramp''}). The connected spectral form factor on the left hand side of this equation
now
reads explicitly 
\begin{eqnarray}
&& 
%({1\over N_A})^2 \ 
g_c(\tau)
= 
\int d\mu \int d \mu' 
\ \ e^{-i (\mu-\mu')(\tau/N_A)} \ \ 
\langle{\hat \nu}(\mu) \ {\hat \nu}(\mu')\rangle_c
= \\ \nonumber
&& =
\int d{\cal E} \int d\Omega \ 
 e^{-i \Omega (\tau/N_A)} 
\Bigl [
\langle {\hat \nu}({\cal E})\rangle \ 
%{
\delta(\Omega) 
%\over N_A} 
-
\langle{\hat  \nu}({\cal E}+\Omega/2)\rangle \    \langle{ \hat \nu}({\cal E}-\Omega/2)\rangle  \ 
{\sin^2[\pi  \langle {\hat \nu }({\cal E})\rangle \Omega]\over
[\pi \langle  {\hat \nu}({\cal E})\rangle \Omega]^2}
\Bigr ],
\end{eqnarray}
or
\begin{eqnarray}
&& 
\label{ConnectedSpectralFunctionWithSineKernel}
 \ g_c(\tau) =
\int d{\cal E} \int d\Omega \ 
 e^{-i \Omega (\tau/N_A)} 
\Bigl [
%N_A \ 
\langle {\hat \nu}({\cal E})\rangle \ \delta(\Omega) 
-
R({\cal E},\Omega) \ 
{\sin^2[\pi N_A \langle{\hat \nu}({\cal E})\rangle \Omega]\over
[\pi  \Omega]^2}
\Bigr ],
\end{eqnarray}
where
\begin{eqnarray}
R({\cal E},\Omega)
\equiv
{
\langle{ \hat \nu}({\cal E}+\Omega/2)\rangle \    \langle{\hat  \nu}({\cal E}-\Omega/2)\rangle
\over
 \langle {\hat \nu}({\cal E})\rangle^2
}.
\end{eqnarray}
Next, we  implement  a version of an  idea that was  used in Ref.\ \onlinecite{Cotler2016}
for the computation of the spectral form factor of a 
random GUE {\it Hamiltonian}  in the context of the SYK model.
In particular, we limit the integral over $\Omega$ by introducing a cutoff $\Omega_0$,
\begin{eqnarray}
\int d\Omega
\to 
\int_{-\Omega_0/2}^{+\Omega_0/2} d\Omega,
\end{eqnarray}
chosen to satisfy  the requirement that the density of states
$\langle{\hat  \nu}({\cal E}+\Omega/2)\rangle$ does not vary appreciably when $-\Omega_0/2 < \Omega < +\Omega_0/2$.
Then the factor  $R$ in the integrand  in
(\ref{ConnectedSpectralFunctionWithSineKernel}) becomes unity,
 $R({\cal E},\Omega)\to 1$.
[Physically, the cutoff of course implies  that variations of $g_c(\tau)$ on time scales $(\tau/N_A) \lessapprox (1/\Omega_0)$ can no longer be resolved.]
%This means that result for the spectral form factor will  not correctly reproduce the (non-universal) regime of sufficiently small values of $\tau$.
Now we divide the interval $[\alpha_-, \alpha_+]$ in which all eigenvalues $\mu_i$
have support, into a set of 
non-overlapping subintervals  
of length $\Omega_0$ each. 
%Then we  distribute  by dividing  they have support,  over 
The integral
of $\Omega$ over each subinterval  number $I=1, 2, ..., M$ at fixed ${\cal E}_I$ (say at the center of the interval) can now be done
in the limit $N_A \to \infty$ by using 
(\ref{IntegralOfSine}) and  the fact that in that limit $\langle {\hat \nu }({\cal E}_I)\rangle=$
$N_A \ \bar{n}({\cal E}_I)$,  where $\bar{n}({\cal E}_I)$ is a $N_A$-independent constant [see (\ref{mp_dis})]:
% $\langle {\hat \nu }({\cal E}_I)\rangle=$
%$N_A \ {\bar{n}({\cal E}}$
%  is finite  as $N_A\to \infty$:
\begin{eqnarray}
\nonumber
&&
\int_{-\Omega_0/2}^{+\Omega_0/2}
\ e^{-i\Omega (\tau/N_A)} \ \ 
{\sin^2[\pi N_A \  \bar{n}({\cal E}_I) \Omega]\over
[\pi \Omega]^2} \ \ d\Omega =  \\ \nonumber
&&
= 
{N_A \bar{n}({\cal E}_I) \over \pi^2}
\int_{-N_A \bar{n}({\cal E}_I)\Omega_0/2}^{+N_ A\bar{n}({\cal E}_I) \Omega_0/2}
\ e^{-i\Omega'   \tau/  N^2_A\bar{n}({\cal E}_I) }\  \ 
{\sin^2[\pi  \Omega']\over
[\Omega']^2} \  \ d\Omega' \sim \\ \nonumber
&& \sim 
{N_A \bar{n}({\cal E}_I) \over \pi^2}
\int_{-\infty}^{+\infty}
\ e^{-i\Omega' \tau/  N^2_A \bar{n}({\cal E}_I) }\  \ 
{\sin^2[\pi  \Omega']\over
(\omega')^2} \ \ d\Omega'  = \\ \nonumber 
&&
=
{N_A \bar{n}({\cal E}_I)\over \pi^2}
\begin{cases} \pi^2 - {\pi\over 2}  \frac{|\tau|}{N_A^2 \bar{n}({\cal E}_I)}, & 
\frac{|\tau|}{N_A^2 \bar{n}({\cal E}_I)}   < 2\pi \\ 
0, &   \frac{|\tau|}{N_A^2 \bar{n}({\cal E}_I)}   > 2\pi  \end{cases} =
 \\ \nonumber
&&
=
\begin{cases} N_A \bar{n}({\cal E}_I)   - {1\over 2\pi}  \frac{|\tau|}{N_A}, & \frac{|\tau|}{N_A }   < 
2\pi  N_A\bar{n}({\cal E}_I) \\ 
0, &   \frac{|\tau|}{N_A }   > 2\pi  N_A \bar{n}({\cal E}_I)
 \end{cases}.
\end{eqnarray}
 For  subinterval number $I$,  at fixed ${\cal E}_I$, we thus  obtain a ``ramp'',
\begin{eqnarray}
\label{RampNumberI}
g_c^{(I)}(\tau)=
\begin{cases} {1\over 2\pi}  \frac{|\tau|}{N_A}, & {1\over 2 \pi} \frac{|\tau|}{N_A }   <   N_A\bar{n}({\cal E}_I) \\ 
N_A\bar{n}({\cal E}_I), &   {1\over 2\pi} \frac{|\tau|}{N_A }   >   N_A\bar{n}({\cal E}_I)
 \end{cases}
\bigskip
 \ \ = \ \  {\rm min}\{
\  {1\over 2\pi}  \frac{|\tau|}{N_A} \ , \ N_A\bar{n}({\cal E}_I) \  \}.
\end{eqnarray}
Now, doing the integral over ${\cal E}$ as a sum over the subintervals,
\begin{eqnarray}
\int_{\alpha_-}^{\alpha_+} \ d{\cal E} 
 \to \sum_{I=1}^M \ \Omega_0,
\end{eqnarray}
where $\alpha_{\pm}$ were defined immediately below (\ref{mp_dis}),
we obtain from (\ref{RampNumberI})
\begin{eqnarray} \nonumber
&& g_c(\tau)
=
\sum_{i=1}^M \Omega_0 \ g_c^{(I)}(\tau)=
{\rm min}\{
\ \ 
\int_{\alpha_-}^{\alpha_+}  d{\cal E}  \ 
 {1\over 2\pi}  \frac{|\tau|}{N_A}
\ \ , \ \ 
\int_{\alpha_-}^{\alpha_+}  d{\cal E} 
 \ N_A \ \bar{n}({\cal E}) \ 
\ \ \} =  \\  \label{ConnectedSpectralFormFactorComputed}
&&
=
{\rm min}\{
\ \ 
({\alpha_+-\alpha_-\over N_A})\ 
 \frac{|\tau|}{2\pi} 
\ \ , \ \ 
N_A
\ \ \}
=
{\rm min}\{
\ \ 
{
2 \sqrt{\alpha} \  |\tau|\over \pi N_A} 
\ \ , \ \ 
N_A
\ \ \},
\end{eqnarray}
where we used $(\alpha_+-\alpha_-)/N_A=$ ${4 \sqrt{\alpha}\over N_A}=$ ${4 \over \sqrt{N}}$, recalling  $\alpha=N_A/N_B$,
as well as  the normalization 
of $\bar{n}(\mu)$ from (\ref{mp_dis}). We now  obtain (\ref{``ramp''})
from (\ref{ConnectedSpectralFormFactorComputed}), since ${4\over \sqrt{N}} {|\tau_H|\over 2\pi} =N_A$ 
leads to
$\tau_H= $
${\pi \over 2}  N_A \sqrt{N}$. The Heisenberg time is defined to be $2\pi$ times the inverse of the mean level spacing (here of
the reduced density matrix ${\hat \rho}_A$), and 
%we obtain
this yields  (setting $\hbar=1$)
upon using (\ref{AverageLevelSpacingLambdai})
$\tau_H=$
$(2\pi/ \overline{\left (\Delta \lambda\right )} )=$ 
$
{\pi \over 2} (N_A^2/\sqrt{\alpha})=$
${\pi\over 2} N_A \sqrt{N}$ in agreement with the above result. In conclusion
we have obtained the following result for the connected spectral form factor,
\begin{eqnarray}
\label{gConnectedResultOneAppendix}
g_c(\tau)&=&
N_A
\begin{cases}  \  \frac{|\tau|}{\tau_H}, & |\tau| <   \tau_H \\ 
1, &   |\tau| >   \tau_H
 \end{cases},
\qquad \quad {\rm where} \  \ \ \tau_H=
(2\pi/ \overline{\left (\Delta \lambda\right )} )=
{\pi\over 2} N_A \sqrt{N} \\ \label{gConnectedResultTwoAppendix}
&=&\begin{cases}  \   {2 \over \pi} {1\over \sqrt{N}} \ |\tau|, & |\tau| <   \tau_H \\ 
N_A, &   |\tau| >   \tau_H
 \end{cases},
\qquad \quad {\rm where} \  \ \ \tau_H=
(2\pi/ \overline{\left (\Delta \lambda\right )} )=
{\pi\over 2} N_A \sqrt{N}.
\end{eqnarray}
The last equation, displaying explicitly the $N_A$-independence of the prefactor of the linear growth in $\tau$,  is  the result 
%displayed 
shown in (\ref{``ramp''}).

%\bibliography{level_repulsion_biblio3}

%\bibliography{level_repulsion_biblio}

\bibliography{level_repulsion_biblio-new1}

\begin{thebibliography}{60}
\bibitem [{foo()}]{footnote_1}%
  \BibitemOpen
  \href@noop {} {}\bibinfo {note} {We actually studied $\langle ZZ^*\rangle$
  for eigenstates in both Floquet and Ising models with $L=14$ and we can
  observe a smaller linear ramp in it}\BibitemShut {NoStop}%
\end{thebibliography}
\end{document}